\documentclass[preprint,10pt]{aastex}

\usepackage[utf8]{inputenc}
\usepackage[english]{babel}
\usepackage{natbib}
\setcitestyle{authoryear,open={(},close={)}}

\usepackage{array}
\usepackage{wrapfig}
\usepackage{multirow}
\usepackage{tabu}
\usepackage{graphicx}
\usepackage{import}
\usepackage{subfiles}
\usepackage{pdfpages}
\usepackage{pgffor}
\usepackage{comment}
\usepackage{xfrac}
\usepackage{xifthen}
\usepackage{color}
\usepackage[normalem]{ulem}
\usepackage{csquotes}

\newcommand{\Sensitivityf}{$16.3$}
\newcommand{\Sensitivityp}{$16.1$}
\newcommand{\vsspeed}{$0.1$} 

\newcommand{\ftsearchstep}{20} 
\newcommand{\ftsearchiniamp}{500} 
\newcommand{\ftsearchsecamp}{1000} 
\newcommand{\spiralgrowthfactor}{2}
\newcommand{\numtosemilock}{3} 
\newcommand{\locktosearch}{3} 

\newcommand{\be}{\begin{eqnarray*}}
\newcommand{\ee}{\end{eqnarray*}}
\newcommand{\bs}{\begin{small}}
\newcommand{\es}{\end{small}}

\newcommand{\tpdelayline}{$73\%$} 
\newcommand{\tpfeedsystem}{$75\%$} 
\newcommand{\tpspecfrng}{$62\%$}  
\newcommand{\tpqefrng}{$83\%$} 
\newcommand{\tpspecphot}{$62\%$} 
\newcommand{\tpqephot}{$83\%$} 
\newcommand{\tpthreeinchflatmirror}{$96\%$} 
\newcommand{\tpoap}{$98\%$} 

\newcommand{\tptwoinchflatmirror}{$98\%$} 
\newcommand{\tplenslet}{$98\%$}  
\newcommand{\tplensletphot}{$98\%$} 
\newcommand{\tpfrngfocuslens}{$97\%$} 
\newcommand{\tpcylindricallens}{$98\%$} 

\newcommand{\tpfiberfresnel}{$92\%$}

\newcommand{\tpfrngtheo}{$4.2\%$}
\newcommand{\tpphottheo}{$4.4\%$}
\newcommand{\tpfrngobs}{$0.45\%$}
\newcommand{\tpphotobs}{$0.22\%$}
\newcommand{\tptheo}{$8.6\%$} 
\newcommand{\tpobs}{$0.67\%$}

\newcommand{\tpfibercoupling}{$55\%$~} 

\newcommand{\zetapa}{223.9\pm1.0} 
\newcommand{\zetasep}{40.6\pm1.8} 
\newcommand{\zetafratio}{2.18\pm0.13} 
\newcommand{\hummelfratio}{2.2\pm0.1}
\newcommand{\hummelsep}{40.1\pm1.0}
\newcommand{\hummelpa}{223.2\pm2.3}

\newcommand{\vsqrerr}{4.5} 
\newcommand{\cphaseerr}{1.9} 
\newcommand{\baserange}{10.77-25.92} 

\title{VISION: A Six-Telescope Fiber-Fed Visible Light Beam Combiner for the Navy Precision Optical Interferometer}

\author{Eugenio V. Garcia\altaffilmark{1,2}, 
	   Matthew W. Muterspaugh\altaffilmark{3,4}, 
	   Gerard van Belle\altaffilmark{1}, 
	   John D. Monnier\altaffilmark{5},
	   Keivan G. Stassun\altaffilmark{2},
	   Askari Ghasempour\altaffilmark{6},
	   James H. Clark\altaffilmark{7}, 
	   R. T. Zavala\altaffilmark{8},
	   James A. Benson\altaffilmark{8},
	   Donald J. Hutter\altaffilmark{8},
	   Henrique R. Schmitt\altaffilmark{7}, 
	   Ellyn K. Baines\altaffilmark{7},
	   Anders M. Jorgensen\altaffilmark{10},
	   Susan G. Strosahl\altaffilmark{1},
	   Jason Sanborn\altaffilmark{1},
	   Stephen J. Zawicki\altaffilmark{1},
	   Michael F. Sakosky\altaffilmark{1},
	   Samuel Swihart\altaffilmark{9}}

\altaffiltext{1}{Lowell Observatory, Flagstaff, AZ 86001, USA, eugenio.v.garcia@gmail.com}
\altaffiltext{2}{Department of Physics\& Astronomy, Vanderbilt University, 6301 Stevenson Science Center, Nashville, TN, 37212}
\altaffiltext{3}{Department of Mathematical Sciences, College of Life and Physical Sciences, Tennessee State University, Boswell Science Hall, Nashville, TN 37209, United States}
\altaffiltext{4}{Center of Excellence in Information Systems Engineering and Management, Tennessee State University, 3500 John A. Merritt Blvd., Box No.~9501, Nashville, TN 37209-1561}
\altaffiltext{5}{University of Michigan (Astronomy), 500 Church St, Ann Arbor, MI 48109}
\altaffiltext{6}{Optical Spectroscopsy Division, HORIBA Scientific, 3880 Park Ave. Edision, NJ 08820, USA}
\altaffiltext{7}{Remote Sensing Division, Naval Research Laboratory, 4555 Overlook Avenue SW, Washington, DC 20375, USA}
\altaffiltext{8}{U.S. Naval Observatory, Flagstaff Station, 10391 W. Naval Observatory Road, Flagstaff, AZ 86005, USA}
\altaffiltext{9}{Michigan State University, 3265 Biomedical and Physical Sciences Building, Okemos, MI 48864, USA}
\altaffiltext{10}{Electrical Engineering Department,New Mexico Institute of Mining and Technology, Socorro, NM, USA}


\begin{document}

\begin{abstract}
Visible-light long baseline interferometry holds the promise of advancing a number of important applications in fundamental astronomy, including the direct 
measurement of the angular diameters and oblateness of stars, 
and the direct measurement of the orbits of binary and multiple star systems. 
To advance, the field of visible-light interferometry requires development of instruments capable of combining light from 15 baselines (6 telescopes) simultaneously. The Visible Imaging System for Interferometric Observations at NPOI (VISION) is a new visible light beam combiner for the Navy Precision Optical Interferometer (NPOI)
that uses single-mode fibers to coherently combine light from up to 
six telescopes simultaneously with an image-plane combination scheme. 
It features a photometric camera for calibrations and spatial filtering from single-mode fibers 
with two Andor Ixon electron multiplying CCDs. This paper presents the VISION system, results of laboratory tests, and results of commissioning on-sky observations.  
A new set of corrections have been determined for 
the power spectrum and bispectrum by taking into account non-Gaussian 
statistics and read noise present in 
electron-multipying CCDs to enable measurement of visibilities 
and closure phases in the VISION post-processing pipeline.  
The post-processing pipeline has been verified via new on-sky observations of the O-type supergiant binary $\zeta$ Orionis A, obtaining a flux ratio of $\zetafratio$  with a position angle of $\zetapa^{\circ}$ and separation $\zetasep$ mas over 570-750 nm, in good agreement with expectations from the previously published orbit. 

\end{abstract}

\section{Introduction}

A beam combiner coherently combines 
the starlight from the multiple telescopes of the interferometer to form interference patterns (fringes). 
These fringes are the Fourier components of the image of the object being observed and thus allow for the measurement of the angular diameters of stars, the orbits of binary and multi-star systems
with milliarcsecond separations, and the direct observation of stellar surface features. 
The advantage of optical interferometry is high angular resolution typically in the milliarcsecond or sub-milliarcsecond range. 

Over the past two decades, visible-light beam combiners have been commissioned for 
three interferometers: the Navy Precision Optical Interferometer \citep[NPOI,][]{Arm98},
the Center for High Angular Resolution Astronomy (CHARA) Array
\citep{tenBrummelaar05}, and the Sydney University Stellar Interferometer \citep[SUSI,][]{SUSIA,SUSIB}.
Currently, visible-light beam combiners being commissioned or recently made operational are
the Precision Astronomical Visible Observations (PAVO) for the CHARA Array and SUSI \citep{PAVO08,PAVO12,PAVO13}, 
the Micro-arcsecond University of Sydney Companion Astrometry instrument (MUSCA) for SUSI \citep{Kok12},
and the Visible spEctroGraph and polArimeter \citep[VEGA,][]{Vega08,Vega09,Vega11,Vega12} for CHARA. In addition, at NPOI 
there is the ``Classic" pupil-plane combiner \citep{Mozurk94}, which is having its fringe engine upgraded \citep{Sun14,Landavazo14}, and is named the New Classic fringe engine. All of these beam combiners have 
provided insights into rapidly rotating stars \citep{Ohishi04,Peterson06,Jamial15}, 
direct measurements of stellar radii \citep{Arm01,Witt06,North07,North09,Bazot11,Arm12,Baines13,Baines14,Challouf2014,Jorg14}, 
validation of transiting exoplanets \citep{Huber12b,Huber12a}, and 
binary and multiple star systems \citep{Hummel98,Hummel01,Hummel03,Patience08,Schmitt09,Zavala10,Tango09,Hummel13,Wang15}. 

However, with the exception of the upgraded ``Classic" fringe engine for the NPOI, none of these are yet capable of simultaneous 
measurement of fringes on all available baselines. Dense coverage of the UV plane is critical for making the first 
direct stellar surface images at visible wavelengths, and this is best accomplished by simultaneously observing 
a star with as many baselines as possible. This is one of main advantages of the Visible Imaging System for Interferometric Observations at NPOI (VISION), which has recently been commissioned and is described in this paper. 

VISION's design was derived from the six-telescope 
Michigan InfraRed Combiner \citep[MIRC,][]{Monnier04,Monnier06,Monnier07}. VISION is a six-telescope, all-in-one beam combiner using single-mode fibers and visible 
light electron-multiplying charge coupled devices (EMCCDs). Prior to VISION and MIRC, the IONIC (Integrated Optics Near-infrared Interferometric Camera) instrument for the Infrared Optical Telescope Array (IOTA) \citep{Rouss99,Rouss00,Berger03,Traub04} to filter and guide light also used single mode fibers to measure closure phases \citep{Ragland04}.

VISION was built by Tennessee State University in collaboration with Lowell Observatory, the United States Naval Observatory, 
and the Naval Research Laboratory \citep{Ghasempour12}. It monitors individual telescope 
throughputs and fiber coupling efficiencies in real time for visibility calibration. VISION operates from $580-850$ nm, whereas MIRC operates in the near 
infrared (IR) at $1490-1750$ nm ($H$-band). It is capable of simultaneously measuring 15 visibilities, 20 triple amplitudes, and 
20 closure phases, allowing for dense UV plane coverage and image 
reconstruction. 

VISION's six-telescope simultaneous beam combination 
allows for multi-pixel images across the surface of a target star via 
image reconstruction. VISION 
is intended to deliver complementary visible-light observations to MIRC's near-IR observations of
rapidly rotating stars, binary stars, and red super giants 
\citep{Monnier07,Zhao08,Che11,Monnier12,White13,Baron14,Klop15,Roet15a,Roet15b}. 
Early science targets for VISION include imaging the surfaces of rapidly rotating stars and red supergiants
for testing of 2D and 3D stellar models. 
It will be used to test 2D models of rapidly rotating stars \citep{Esp11,Esp13} and 
3D radiative-hydrodynamic models of red supergiants 
\citep{Freytag02,Freytag08,Chiavassa09,Chiavassa10,Chiavassa11,Chiavassa12}; to 
date, only a few rapidly rotating stars and 
several red supergiants have been 
observed via interferometry \citep[see reviews by][]{vanBelle12,DeSouza03,
Peterson06,Monnier07,Zhao08,Che11,Monnier12,DeSouza12, Haubois09,Baron14}.

Furthermore, VISION has high spatial resolution imaging coupled with a large field of view 
from moderate spectral resolution. This allows VISION to 
study hierarchical triple star systems, where one
of the two components of a relatively wide pair of stars is itself a much more narrowly separated binary.
VISION can measure the relative astrometry between the different components 
of the triple or quadruple system. There are only a handful of fully characterized orbits of multi-star systems
\citep{Hummel03,Muterspaugh05,Muterspaugh06a,Muterspaugh06c,Muterspaugh06b,Muterspaugh08,Muterspaugh10}. 

This paper presents the design of the VISION instrument, laboratory tests evaluating the system performance, and validation of the 
data-processing pipeline from new on-sky resolved measurements of the 
O-type supergiant binary $\zeta$ Orionis A. 
In \S\ref{sec:ins}, 
the VISION optical design and light path is detailed. In \S\ref{sec:dataaqc} 
the data acquisition sequence is described. In \S\ref{sec:char} the
throughput, cameras, system visibility, and fringe crosstalk are evaluated. 
In \S\ref{sec:post} 
the adaptation of the MIRC data-processing pipeline for VISION is described 
and the theoretical bispectrum and power spectrum bias subtraction equations for an  EMCCD in the photon counting regime are evaluated. 
The post-processing pipeline is validated in \S\ref{sec:zeta} using on-sky commissioning observations of $\zeta$ Orionis A and new resolved astrometric measurements of the flux ratio, separation, and position angle of this benchmark binary star system are reported. 
Finally, in \S\ref{sec:conclude} a summary and a brief description of planned future work are discussed.

\section{The VISION Instrument \label{sec:ins}} 

\subsection{Optical design \label{sec:overalldesign}} 

\begin{figure}[ht]
\begin{center}
\includegraphics[width=\textwidth]{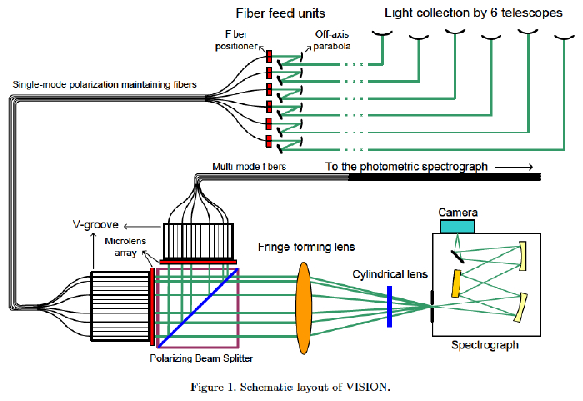}  
\end{center}
\caption{\label{fig:schematic}
A diagram of light path for the VISION beam combiner. Light from up to six 
telescopes is coupled to single-mode polarization-maintaining fibers via 
off-axis parabolas and picomotor fiber positioners. Beam 6 has not yet been commissioned 
due to observatory maintenance. The single-mode fibers 
are placed on a V-groove array to ensure unique fringe frequencies for each beam pair. 
Finally, the light from each beam is split by a polarizing beam splitter: 50\% is transmitted and
focused onto the fringing camera, which records the interferograms, and 50\% is reflected by the polarizing beam splitter 
and focused onto the photometric camera, which monitors real-time fluxes of each beam for visibility 
and triple amplitude calibration.  
}
\end{figure}

\begin{figure}[ht]
\begin{center}
\includegraphics[width=\textwidth]{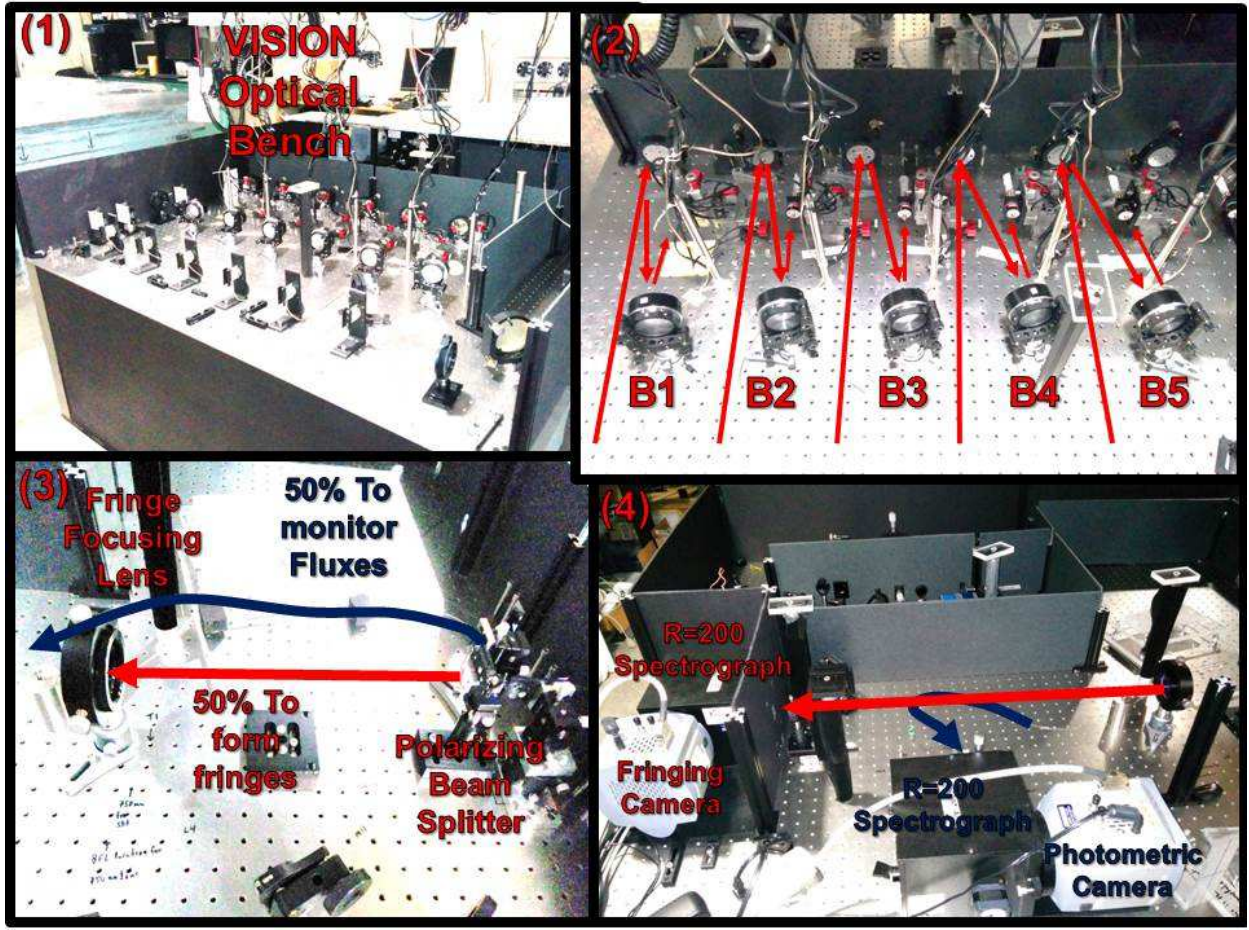}
\end{center}
\caption{\label{fig:ins}
The light path through the VISION beam combiner. (1) The entire VISION 
optical bench. (2) Each $35$ mm diameter beam is collapsed to a $4-8$~$\mu$m spot size, focused 
onto each single-mode fiber tip held in place by a fiber positioner. (3) The light from each single-mode fiber 
is positioned on a V-groove array, where $50\%$ is sent to the photometric camera via multi-mode fibers
and $50\%$ is sent to the fringing camera after passing through a fringe focusing lens and cylindrical lens.  
The photometric and fringing cameras and the identical $R=200$ spectrographs attached to these cameras 
are shown in (4). 
}
\end{figure}

\begin{figure}[ht]
\begin{center}
\includegraphics[width=\textwidth]{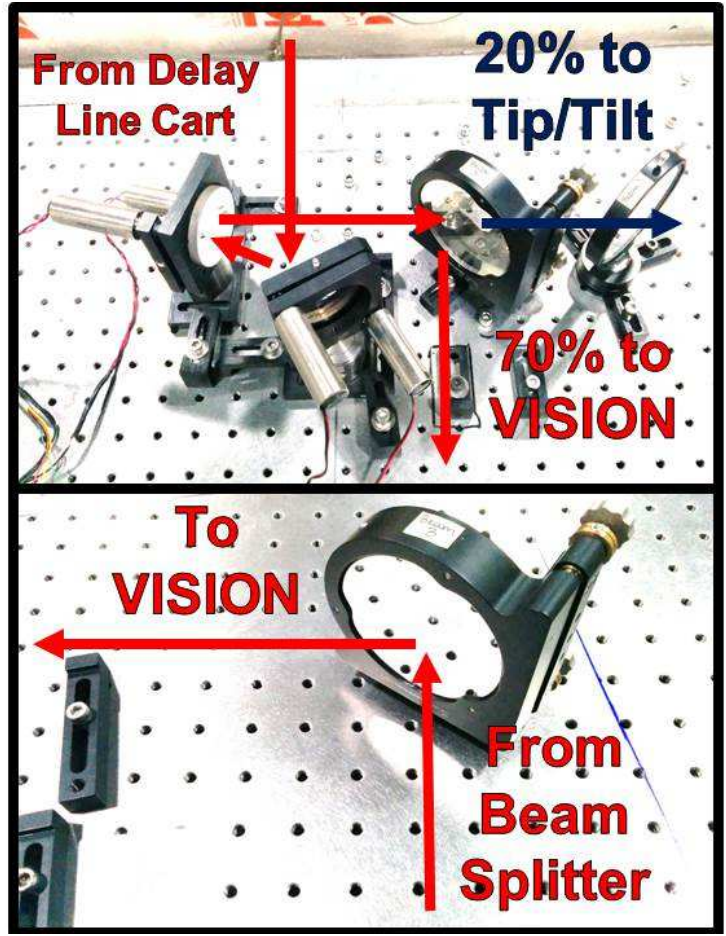}  
\end{center}
\caption{\label{fig:routing}
The light path towards the VISION beam combiner. (TOP) A minority ($20\%$) of the light 
passes through the beam splitter in transmission to the tip/tilt mirrors for 1st order correction 
of the atmospheric turbulence. The majority (70\%) of the light is reflected towards VISION. 
(BOTTOM) The light is re-routed towards the VISION beam combiner.
}
\end{figure}

\begin{figure}[ht]
\begin{center}
\includegraphics[angle=90,width=\textwidth]{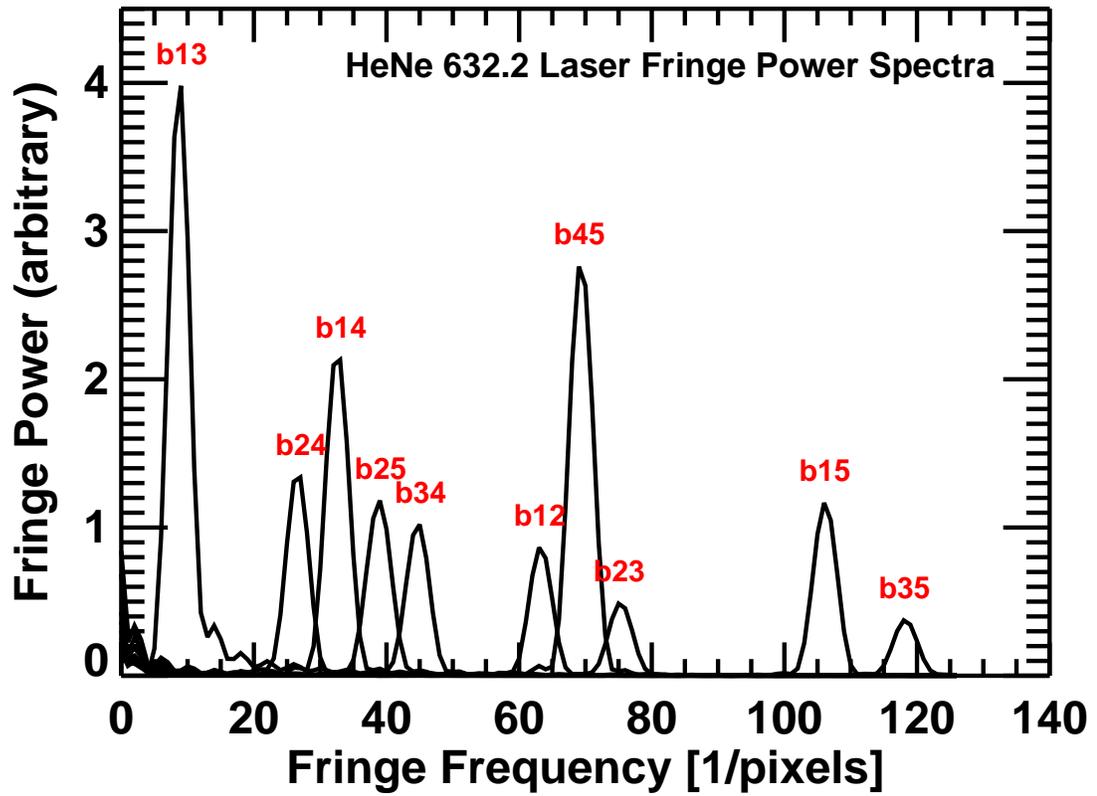}  
\end{center}
\caption{\label{fig:frngfreq}
The unique fringe frequencies for each beam pair combination for beams 1-5 for VISION as measured 
using an in-laboratory 632.8 nm HeNe laser source. At fixed wavelength, the
fringe frequency increases with increased distance between the fibers on the V-groove array.  The peaks 
in the power spectra for each beam pair are well isolated resulting in very low cross talk between the fringes. The label 
bAB is the power spectrum peak for a fringe resulting from combining beams A and B. 
}
\end{figure}

The VISION optical design is shown
in Figure~\ref{fig:schematic}, and the beam combiner itself in Figure~\ref{fig:ins}.  
Similar to MIRC, VISION uses single-mode optical fibers that spatially
filter incoming starlight, enabling precise visibility and closure phase measurements \citep{Shaklan92}. Unlike MIRC, the VISION fibers are polarization-maintaining.  
The six outputs of VISION's single mode fibers are arranged 
in a non-redundant pattern using a 
V-groove array. The polarization of the 
starlight parallel to the optical bench is reflected by a polarizing beam splitter, focused into multimode fibers, which are reconfigured into linear arrangement with equal spacing, and imaged onto a EMCCD to 
monitor the fluxes of each beam in real-time (the ``photometric camera'' hereafter). 
The polarization of the light perpendicular to the optical bench is imaged onto an identical EMCCD to 
form 15 unique sets of fringes (the ``fringing camera'' hereafter).
The EMCCDs feature sub-electron, but non-negligible, effective read noise. Light is spectrally dispersed using 
identical optical ($570-850$ nm) slit spectrographs attached to photometric and fringing cameras.
Each spectrograph has a low resolution ($R\approx200$) and medium resolution ($R\approx1000$) option.   
Below is a sequential description of each optical system, in order from the siderostats where light is collected 
to the EMCCD detectors: 

\begin{itemize}
\item {\bf Light Gathering:} NPOI uses a 12.5 cm circular beam of starlight gathered by 50 cm siderostats. 
The light is guided to the central beam combining facility in vacuum pipes that include 125:35 beam reducers.
After beam reduction, the light is passed to delay line carts with mirrors that receive feedback from the 
VISION fringe tracker to match optical path lengths. 
The result is six coherently phased, collimated circular beams of light with 35-mm diameters.

\item {\bf Routing Light to the VISION Optical Bench:} The routing of the six 35-mm collimated beams
 from NPOI delay lines to the optical bench is shown in Figure~\ref{fig:routing}. The 35-mm beams are reflected 
 by six UV fused silica broadband plate $\sfrac{70}{20}$ beam splitters placed at a $45^{\circ}$ angle to each beam (top panel, 
 Figure~\ref{fig:routing}). Some light is lost in transmission and scattering through our custom made broadband beam splitters which results in the $\approx70$ reflectivity and $\approx20$ transmissivity across $570-850$ nm. The $\sfrac{70}{20}$ beam splitter reflects 70\% of the light to VISION and transmits $\approx$20\% of the light to NPOI's tip/tilt quad cells. The six beams are then reflected to the optical bench
 via 3-inch flat Newport Zerodur Broadband Metallic Mirrors with 
 silver coating\footnote{http://www.newport.com/Broadband-Metallic-Mirrors/141088/1033/info.aspx\#tab\_Specifications}. 
 The 3-inch flat mirrors are placed $45^{\circ}$ to the beam (bottom panel, Figure~\ref{fig:routing}). At the front of the VISION optical bench are shutters that can be controlled either manually or by computer.  These 
can block light for individual beams (panel 1, Figure~\ref{fig:ins}). The six beams are then reflected off 2-inch flat mirrors to the VISION off-axis parabolas. 

\item {\bf Coupling Light to Single-Mode Fibers:} Each 
35-mm beam is coupled to a single-mode polarization-maintaining 
fiber using 2-inch Nu-tek 480-900 nm 
silver coated off-axis parabolas (OAPs). The 
165 mm focal lengths of the OAPs (f$/4.7$ optics) were calculated by \cite{Ghasempour12} 
for typical $r_{0}=9$ cm site seeing using the method of \cite{Shaklan88} (panel 2, Figure~\ref{fig:ins}). The OAPs 
collapse the beams from 35-mm to 4-8~$\mu$m over 165 mm. 
Newport closed-loop picomotors and drivers are used to move the fibers vertically and horizontally relative 
to the VISION optical bench to maximize the coupling 
efficiency of the starlight into each fiber. This alignment is both manually and computer controllable. Except 
for the X-axis of the beam 5 driver and all of the beam 6 drivers, the picomotor drivers operate in closed-loop
mode, which corrects for fiber positioning hysteresis. The fibers can be positioned to better than 2~$\mu$m, and the X and Y axis of the
fibers can be aligned using an automated fiber alignment algorithm written in C++ and Python. The Newport picomotors and
drivers occasionally fail with error status on startup; 
the picomotor drivers often require restarting a few times until they operate normally. This is an issue that will eventually be addressed but does not significantly affect performance. 


\item {\bf Single-Mode Polarization-Maintaining Fibers:} In order to increase fringe contrast, VISION uses single-mode 
fibers that filter the atmospheric turbulence, removing residual wavefront errors for each beam. VISION uses Nufern PM630-HP single-mode polarization
maintaining fibers that are operational over 570-900 nm. The fibers are multimode
at wavelengths $\lesssim570$ nm. The single-mode fibers spatially 
filter wavefront errors by transmitting only the fundamental transverse mode of incoming light (LP$_{01}$). 
This filtering enhances fringe contrast by partially removing spatial but not temporal atmospheric turbulence. 
The polarization of starlight through the fibers is maintained 
via strong birefringence due to stress rods along the slow axis of the fiber.  
The mode field diameter of the fibers is $4.5~\mu$m at 630 nm, with a core size of $3.5~\mu$m and 
numerical aperture of $0.12$. The beam exiting each single mode fiber can be described by a Gaussian model. 

\item {\bf Non-redundant Spacing with V-groove Array:} VISION produces a unique fringe frequency for 
each telescope pair by arranging the outputs of the fibers in a 
non-redundant linear pattern on a silicon OZ-optics V-groove array (Table~ \ref{table:frngfreq}). The V-groove array has a base spacing of 250~$\mu$m. 
Non redundant fiber positions of 0-2-8-13-17-20 were chosen. After passing the light through a lens with long focal length, interference fringes are imaged.  
The fringe frequency, or number of pixels per fringe is:
\begin{equation}
\frac{\rm Pixels}{\rm Fringe}=\frac{f \lambda}{p (d_{A}-d_{B})}
\end{equation}
where $\lambda$ is the wavelength of the fringe, $d_{A}$ and $d_{B}$ 
are the location of fiber $A$ and $B$ on the V-groove array, $f$ is the 
focal length ($f=750$ mm), and $p=24$~$\mu$m is the size of the pixels of 
the camera on which the fringes are formed. Given that each fiber pair $AB$ has a
unique physical separation $d_{A}-d_{B}$ on the V-groove array, the corresponding telescope pair have a unique fringe frequency on the VISION cameras and therefore each of VISION's 15 telescope pairs has a unique signature 
in a power spectrum of the image (Figure~\ref{fig:frngfreq}). The VISION geometry minimizes the overlap between the peaks in power spectra for
each telescope pair (the ``fringe cross talk"). The fast axis of all fibers on output of the V-groove array are aligned vertically. A lenslet array glued to the polarizing beam splitter re-collimates the beams to 250~$\mu$m diameter after they exit the fibers. 

\item {\bf Output to Photometric and Fringing Cameras:} VISION splits 
the light between a camera to record interferograms (the ``fringing'' camera), 
and a camera to record individual beam fluxes (the ``photometric'' camera), using 
a polarizing beam splitter. The polarization of the six beams 
parallel to the optical bench is reflected through the beam splitter at 90$^{\circ}$ 
and is coupled to six multimode fibers positioned on a second V-groove array (blue arrow, panel 3, Figure~\ref{fig:ins}). It is then guided to the spectrograph attached the photometric camera (blue arrow, Figure~\ref{fig:ins}, panel 4). 
Each beam output has a unique spatial location on the photometric camera due to a V-groove array that positions 
the multi-mode output for each fiber onto the photometric camera. The photometric camera monitors the real-time wavelength-dependent flux of each beam. 

The polarization of the light perpendicular to the optical bench is transmitted directly through the polarizing 
beam splitter (red arrow, panel 3, Figure~\ref{fig:ins}), and is focused by a 750 mm focal length Thorlabs antireflection coated
achromatic cemented doublet to a $\sim 3.5$ mm diameter spot size. Next, to condense the image in the non-fringing direction, VISION uses a 50 mm focal length cylindrical lens to collapse the light in the horizontal direction (red arrow, panel 4, Figure~\ref{fig:ins}) to 24~$\mu$m, the size of a single pixel on the EMCCDs.  This results in an image 24~$\mu$m by 3.5 mm, with fringing in the long direction that  
is passed to a spectrograph attached to the fringing camera, which in turn results in 128 unique spectral channels of fringes with a height of 3.5 mm in the fringing direction. The combination of the microlens array, polarizing beam splitter, and 750 mm achromatic cemented doublet achieve the desired overlap of the six Gaussian beam profiles from the six single-mode fibers along the fringing direction on the VISION camera, which produces the fringes (see the fringe forming lens to cylindrical lens ray tracing in Figure~\ref{fig:schematic}). The fringes are produced at the entrance to the slit of the spectrograph. 

\item {\bf Spectrographs:} VISION spectrally disperses the incoming light 
using two identical Princeton Instruments SP-2156 Acton spectrographs,
attached to the fringing and photometric cameras. The spectrographs 
are 1:1, i.e. there is no magnification of the Gaussian beam profiles 
at the entrance to the slit. VISION has two observing modes, one low resolution ($R=200$) 
and one medium resolution ($R=1000$). Switching between 
the $R=200$ and $R=1000$ grating for each spectrometer 
is remotely controllable and can be accomplished in a few seconds. The 
wavelength solution for the spectrographs was initialy derived using 
a Ne-Ar lamp source. This wavelength solution was verified 
with the pixel locations of the H$\alpha$ feature from 
on-sky observations of Vega, and an in-lab HeNe 
laser source on the fringing and photometric cameras. 
The resulting wavelength solution for the 
$R=200$ mode is: 
\begin{center} 
$\lambda(\rm i) = (\lambda_{\rm cent} - 232.294)+2.91(\rm i+1) - 1.30 \rm~nm,~fringing~camera$\\
$\lambda(\rm i) = (\lambda_{\rm cent} - 214.986)+2.91(\rm i+1) + 4.66 \rm~nm,~photometric~camera$
\end{center}
where $i=0-127$ is the pixel number, $\lambda_{\rm cent}$ is the user-chosen central 
wavelength of the spectrograph in nanometers, and $\lambda(i)$ is the wavelength 
in nanometers corresponding to pixel $i$. In this paper, 
only the commissioning of the low resolution observing 
mode is described, as 
the medium resolution mode has not yet been fully tested on sky. 

\item {\bf Andor Ixon EMCCDs:} VISION features two identical 128$\times$128 pixel Andor Ixon DU 860 
EMCCDs, with 24~$\mu$m square pixels and quantum efficiencies of 70-85\% over 550-850 nm
at $-50^{\circ}$ C and dark current of 0.002 electrons/pixel/second. 
For recording stellar interferograms, the EMCCDs are operated at $-50^{\circ}$ C, with 6 ms exposure times, electron multiplying 
gains of 300, and fast readout rates of 10 MHz with vertical clock speeds of \vsspeed~$\mu$s to minimize 
clock induced charge (CIC) noise. The typical CIC event 
rate for the Andor Ixon EMCCDs was found to be $0.08-0.11$ events/pixel/frame. Longer exposure times of 10-12 ms were tested on 
sky resulting in interferograms with significantly reduced fringe contrast due to atmospheric turbulence. 
Custom C++ and Python code controls data acquisition, the fringe searching, the fringe tracking, the spectrographs, the shutters, and the single-mode 
fiber positioners using a computer running Ubuntu Linux OS $12.04$.


\end{itemize}

\subsection{\bf Raw Interferograms \& Calibration Data \label{sec:overalldata}}
\begin{figure}[ht]
\begin{center}
\includegraphics[width=\textwidth]{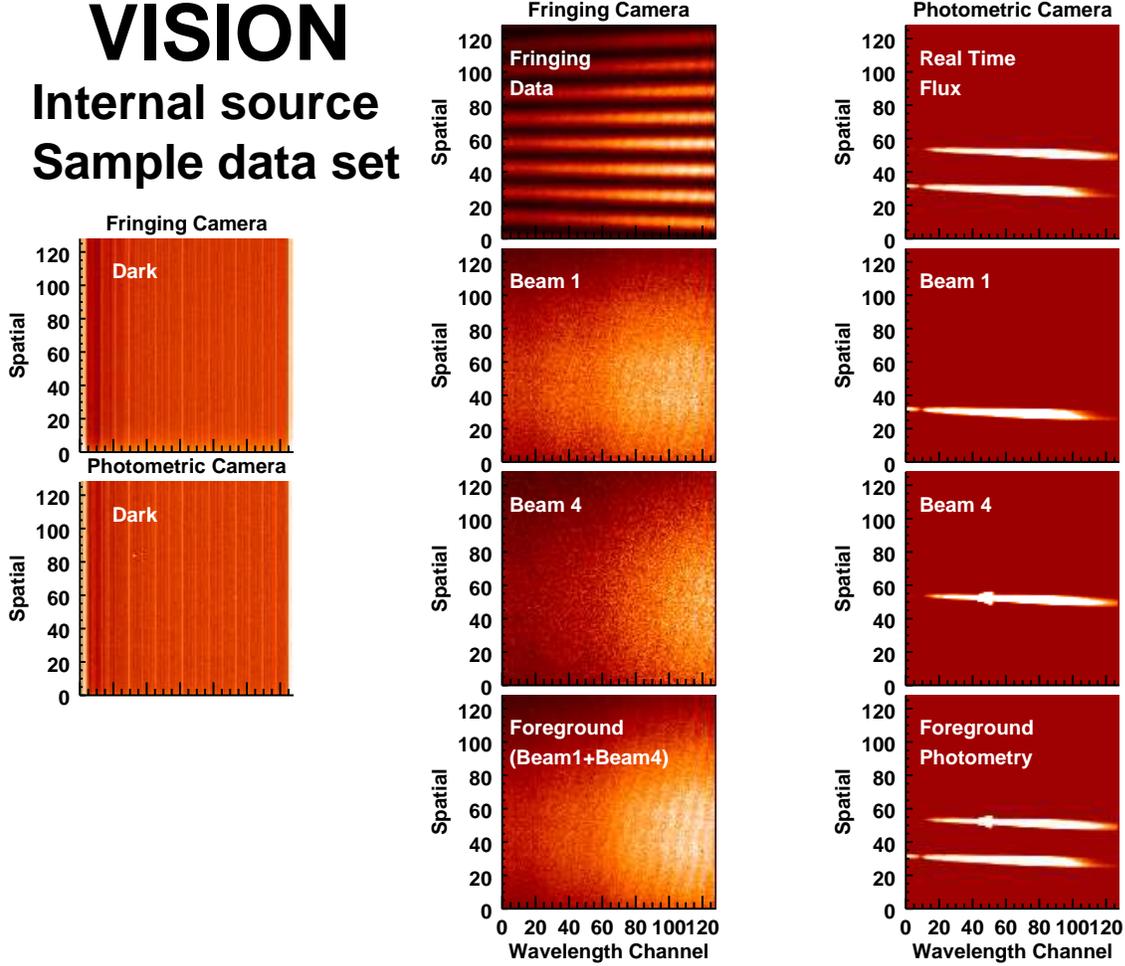}  
\end{center}
\caption{\label{fig:dataex}
Sample VISION data set using an laboratory white light source with an $R=1000$ spectrograph. A complete VISION data set to measure visibilities and closure 
phases includes (1) an average dark frame for both the fringing and photometric cameras to characterize the bias count 
structure across the EMCCDs, (2) the fringing data and simultaneous real-time flux measurements, (3) frames with light from each beam individually 
on the fringing and photometric cameras to measure the split of the light from the polarizing beam splitter, and (4) the foreground data to characterize 
the bias in the power spectrum (visibility bias) and the bispectrum. The small fringing seen in the foreground data is due to pixel-to-pixel sensitivity of the EMCCD chips. This pixel-to-pixel sensitivity only appears strongly after averaging at least 10 minutes of frames. The small point sources of light on the photometric camera frames are due to leakage of the laser metrology used by the delay line carts onto the EMCCDs.  
}
\end{figure}

VISION requires large amounts of support measurements to calibrate the raw interferograms
of a given star. A complete VISION data set for extracting photometrically calibrated visibilities 
and closure phases is shown in Figure~\ref{fig:dataex} for beams 1 and 4. For 
both the fringing and photometric cameras, the spatial direction 
is vertical in the figure, and the wavelength direction is horizontal. 

A sample averaged interferogram is shown in the top center
panel of Figure~\ref{fig:dataex} from the combined light of beams 1 and 4
using a laboratory white-light source. To produce the interferogram, 
the light path length difference between beams 1 and 4 was minimized using the delay line carts. The averaged interferogram was constructed from several hundred co-added, dark subtracted, 20 ms frames with a gain of 300 in medium ($R\sim 1000$) resolution mode. 
The low resolution observing mode is not typically used for measurements illuminated by internal light sources because the laboratory 
light source path passes through the $\sfrac{70}{20}$ beam splitters 
in transmission, leading to significant dispersion. 
The different types of data necessary to extract calibrated squared visibilities and closure phases from any given set of interferograms are:

\begin{itemize}

\item {\bf Darks} are frames with no starlight on the detector. For each data set, $\approx5\times10^{3}-10^{4}$ six-ms dark frames are recorded
(requiring 30--60 seconds of real time). Darks 
are recorded by blocking incoming starlight by closing all shutters at the front of the VISION optical bench.  Darks are recorded semi-hourly throughout the night for typical observing to carefully 
characterize the clock induced charge and bias count levels. The sky background is not included in the darks, as it is 
not a significant source of photons given that VISION observes bright ($R_{\rm mag} <4$) stars. Darks are used to estimate the EMCCD read noise, gain, and CIC rate, which are necessary parameters for extracting calibrated squared visibilities, bias-corrected closure phases, and triple amplitudes from raw interferograms.  Sample average dark frames are shown in the left panels of Figure~\ref{fig:dataex}
for the fringing and photometric cameras.


\item {\bf Foregrounds} are frames with incoherent light from all beams on the detector. Incoherent light is obtained when the delay line carts are moved many coherence lengths away from the fringing position. A sample foreground for beams 1 and 4 is shown in the bottom-center panel of Figure~\ref{fig:dataex}.
Foregrounds are used to characterize the Gaussian profiles from the single-mode fibers on the fringing camera. This is needed to compute the power spectrum and bispectrum biases, which are in turn needed for calibrating squared visibilities and triple amplitudes.

\item{\bf Real-Time Flux Estimation.} The flux for each beam is recorded simultaneously with the interferograms. 
A sample image from the photometric camera with beams 1 and 4 is shown 
in the bottom right and top right panels of Figure~\ref{fig:dataex}. The real-time flux is used to estimate the system visibility due to beam intensity mismatch. The photometric imbalance between 
two beams with intensities $I_A$, $I_B$ will reduce the visibility 
by $\frac{2\sqrt{I_A I_B}}{{I_A+I_B}}$\citep{Coude97}. The fringing camera beam fluxes $I_{f,A}$ and $I_{f,B}$ are estimated from the photometric camera beam fluxes, as described next.


\item {\bf Single Telescope Data.} A sample set of single telescope data for beams 1 and 4 is shown 
in the center panels of Figure~\ref{fig:dataex}. The precise ratio of fluxes between the fringing and photometric cameras using
the polarizing beam splitter can deviate from an exact $\sfrac{50}{50}$ split. 
This is the result of slightly varying polarization of light from the several 
telescopes, which is due to the 
siderostats' motions when tracking stars.  
Thus, the fringing-to-photometric light flux ratio 
can vary from star to star and night to night at different wavelengths. In order to calibrate this effect, the time and
wavelength dependent flux ratio $\alpha(\lambda)$ between the fringing and photometric cameras are measured for each beam and
for each star observed. This flux ratio is then used to correlate the real-time flux for each beam on the photometric camera to 
the fringing camera. The flux for each beam on the fringing camera is given as:
\begin{equation}
\label{eqn:if}
I_{f,i}(\lambda,t) = \alpha_i(\lambda,t) I_{p,i}(\lambda,t).
\end{equation}
$I_{f,i}(\lambda,t)$ is the flux at wavelength $\lambda$ and time $t$ on the fringing camera for beam $i$, and
$\alpha_i(\lambda,t)$ is the measured wavelength- and time-dependent flux ratio between the fringing and photometric cameras using single beam data for beam $i$: 
\begin{equation} 
\label{eqn:alf}
\alpha_i(\lambda) = \frac{I_{f, i}(\lambda)}{I_{p, i}(\lambda)} 
\end{equation}
where the fringing-to-photometric camera flux ratio $\alpha_i(\lambda)$ is measured using single beam data (center panels, Figure~\ref{fig:dataex}).  
Equations~\ref{eqn:if} and~\ref{eqn:alf} are used to estimate the flux on the fringing camera for each beam, $i$, separately. The estimated fluxes for each beam on the fringing camera are used to correct observed squared visibilities and triple amplitudes for beam intensity mismatch. 

\end{itemize}

\subsection{Daily Alignments} 
A series of daily alignments are performed for each of the six beams to maximize the starlight throughput for
on-sky observations, using
a 632.8 nm HeNe laser source. The procedure is as follows: 
\begin{enumerate} 
\item Align the ``switchyard" mirror (top panel, second mirror in the light path of Figure~\ref{fig:routing})
to the $\sfrac{70}{20}$ beam splitters in order to route each beam towards the VISION ``switchyard'' table. 
\item Mount the $\sfrac{70}{20}$ beam splitters and accompanying beam-shear compensating windows.  The beam splitters route light from the feed system to the VISION optical bench. 
\item Place an auto-collimation mirror directly in front of each of the $\sfrac{70}{20}$ beam splitters, 
thus retro-reflecting the HeNe laser back on itself and to the VISION optical bench. This is necessary because
the HeNe laser light path is opposite that of the feed system. 
\item Align the $\sfrac{70}{20}$ beam splitters (top panel, Figure~\ref{fig:routing}) to place the laser spot
as close as possible to the fiber tip of each single-mode fiber. 
\item Align each fiber with an automated algorithm 
that directs the fiber to move horizontally 
and vertically to the optical bench until the laser light coupled to the fiber is at maximum.  
This alignment algorithm typically is repeated twice, once as a rough pass with total grid search size of 32$\times$32~$\mu$m, and 
once with a smaller grid search size of 8$\times$8~$\mu$m. Occasionally, 
the fiber focus for 
each beam is determined using the fiber alignment algorithm. 
\item Remove the auto-collimation mirrors, and after acquiring a star, 
re-align the fibers to maximize the coupling of starlight to each fiber. 
\end{enumerate}

\section{Data Acquisition \label{sec:dataaqc}}

\subsection{Fringe Searching \label{sec:frngsearch}}

The fringe search algorithm acquires fringes by automatically 
stepping the delay line carts back and forth until the fringes are found. For the first 
star observed each night, this procedure typically takes several minutes. After that, 
offsets of 1--3 mm in the cart positions generally remain fixed throughout the night, and the fringe
searching on subsequent stars can take less than a few seconds. 
VISION takes advantage of the roughly equal spacing of the NPOI array to 
use the shortest baselines to phase the long baselines via baseline bootstrapping \citep{Arm98spie,Jorg06}. 
The fringe search algorithm uses up to 5 baselines for fringe searching and fringe tracking. For single stars 
that are resolved with NPOI baselines, the shortest 
baselines are typically used to fringe search, 
since the visibility is highest on the
first peak of the visibility function. For binary stars, 
the baselines used for fringe tracking are strategically chosen 
based on the observed 
fringe SNR on each baseline, and this can be done in real time. 

For example, for 
3-way beam combination on a star using 
beams 2, 4, and 5, beam pairs 2-4 and 4-5 can be selected for fringe tracking 
while the longer baseline with beam pair 2-5 is also phased 
without additional delay line feedback. 
Each of the fringe searching and fringe tracking parameters are adjustable. 

Next the fringe search 
algorithm is detailed for beams 2, 4, and 5 with the nominal settings:    
\begin{enumerate} 
\item First, the algorithm begins searching for fringes between beams 2 and 4 with an increasing search pattern around the nominal delay point, stepping delay line cart 4 by 
$\sim + \ftsearchstep$~$\mu$m.  If the fringe SNR is greater 
than the semi-lock SNR at least $\sim\numtosemilock$  consecutive
times, then the fringe is considered ``found''. The 
fringe SNR is estimated in the control system code after 
the sum of $\sim 20$ power spectra co-adds: 
\begin{equation}
\label{eqn:SNR}
{\rm SNR} = \frac{\rm Peak~of~power~spectrum}{\rm Average~power~spectrum~noise}. 
\end{equation}
If the fringe is not found by the time 
the delay line cart has reached $\sim\ftsearchiniamp$~$\mu$m, the cart 
stepping reverses direction with steps of $\sim - \ftsearchstep$~$\mu$m
until the cart has reached $\sim - \ftsearchiniamp$~$\mu$m. 
\item If no fringe is found, the search range is repeated 
and increased by $\sim\spiralgrowthfactor\times$, from $\sim - \ftsearchsecamp$~$\mu$m to 
$+\sim\ftsearchsecamp$~$\mu$m about the nominal delay point. 
This search range is increased continually until the fringe is found, 
which is typically within $1-3$ mm of the nominal delay point for the first 
star observed that night. 

\item Once the fringe is found, if the fringe SNR is greater than the track SNR, tracking begins on
beam pair 2-4, with delay line feedback sent to cart 4 to correct 
for atmospheric piston errors. If the fringe SNR is lower than a separate ``fringe-lost'' SNR
at least $\sim\locktosearch$ times, then the fringe searching for cart 4 resumes, with 
a small, fixed delay range of $\pm\sim 5$~$\mu$m. 

\item The fringe search algorithm then repeats the above process 
to search for fringes between beams 4 and 5. Searching for fringes using 
delay line cart 5 is relative to any delay line feedback sent to cart 4. The fringe 
search algorithm automatically keeps the cart of the first beam given as the stationary 
cart. In this example, the algorithm was given tracking beam pairs 2-4 and 4-5 and thus cart 
2 was kept stationary. If the algorithm is tracking beam pairs 4-2 and 2-5, it would keep 
cart 4 stationary.

\end{enumerate}

\subsection{Fringe Tracking \label{sec:frngtrack}}


Due to atmospheric turbulence, the path length that light travels from each star to the telescopes often changes on 10-500 ms timescales. To correct for this effect the fringes are tracked 
in real time using a fringe tracking algorithm. The fringe tracking code is written in C++ and installed on the VISION control system computer, which sends feedback directly to the delay line carts. 

VISION forms spatially dispersed fringes on the fringing camera in real time, and thus avoids the need for modulation of the delay line mirrors to create temporal fringes. This design was chosen to avoid possible non-linear modulations in the shapes of the delay line modulations observed in NPOI classic, which can lead to cross talk between the fringe amplitudes and phases, when multiple baselines are observed with the same detector pixel. 



A fringe-fitting approach is used to estimate group delay in the fringe tracking algorithm. 
The delay to move each of the five operational delay line carts is evaluated 
on $\approx100$ ms timescales to minimize the path length differences between the carts, ideally to within a few hundred nm or better. The fringe tracking approach uses a Fourier transform along 
the wavelength direction, and a direct fit to the data along the fringing (spatial) 
direction for each 6 ms frame of data. The theoretical fringe model 
for an image plane combiner adopted from Equation 10.1 on page 569 of \cite{Born1999} is:
\begin{equation} 
\label{eqn:physical}
I(y) = I_1+I_2+2\sqrt{I_1 I_2}|{\gamma}|\cos\bigg(fy+{\rm arg(\gamma)}\bigg)
\end{equation}
where, $I_1$ and $I_2$ are the fluxes for beams $1$ and $2$ respectively.  The coherence between the two 
beams $\gamma$ has both an amplitude $|\gamma|$ and a phase arg$(\gamma)$, and the fringes are modulated by frequency $f$.  
Note that the phase can be instead represented by replacing the cosine 
function with independent cosine and sine functions each with independent 
amplitudes using the identity 
$$
\cos(\theta_1-\theta_2)=\cos(\theta_1)\cos(\theta_2)+\sin(\theta_1)\sin(\theta_2);
$$
the phase would be found as the inverse tangent of the ratio of 
sine to cosine amplitudes.  This allows the nonlinear phase parameter to be 
replaced by linear coefficients, simplifying model fitting procedures.  

If only a single telescope pair were being used, and if there were no further 
modulation of the fringe amplitude, the intensity pattern could be fit using a 
Fourier Transform, and implemented efficiently in real time using, for 
example, the Fast Fourier Transform algorithm (with attention to zero-pad the array for better sampling of the fringe frequencies $f$, since the fringe wavelengths will not necessary be integer fractions of the number of pixels).

In practice this theoretical fringe model is modulated by both a coherence 
envelope (as a 
function of distance from zero differential optical path length) and detector 
illumination pattern, and multiple fringe signals are present simultaneously.  
The first can be modeled according to the instrument and 
source bandpasses (though only with nonlinear parameters), 
and the second according to 
incoherent illumination pattern measurements.  For fringe tracking, a model 
combining the effects of the illumination pattern and fringing is used as follows (however, the coherence envelope is presently ignored in real-time analysis due to nonlinear parameter complexities).

Fringes are dispersed horizontally ($x$), with the fringing (delay) direction vertical ($y$).  For a given spectral channel $x$, the interferogram is modeled as
\begin{equation}
\label{eqn:frngmulti}
I(x, y) =  e^{-(y-P_{1})^{2}/P_{2}^{2}} \left( P_{3} + \sum_{k=1} \left[P_{4,k} \cos \left(P_{6, x, k} \left(y - 64 - P_{7, x, k}\right)\right) + P_{5,k}\sin \left(P_{6, x, k} \left(y - 64- P_{7, x, k}\right)\right)\right]\right)
\end{equation}
where $k$ is the index of each beam pair, $P_1$ and $P_2$ describe the approximately Gaussian illumination pattern on the detector, $P_{3}$ is the overall intensity, $P_{4, k}$ and $P_{5, k}$ are the cosine and sine amplitudes for each pair (with the phase $\phi$ given by $\tan \phi = P_{5, k}/P_{4, k}$ and the total fringe amplitude $\sqrt{P_{4, k}^2 + P_{5, k}^2}$), and $P_{6, x, k}$ is the (wavelength dependent) fringe frequency.  More generally, the form can be written as
$$
I(y) = \sum_{m=0}^{m=2N} A_m g_m
$$
where
$$
g_m(x, y) = \left\{ \begin{array}{cc}
1 & m = 0\\
e^{-(y-P_{1})^{2}/P_{2}^{2}} \cos \left(P_{6, x, k} \left(y - 64 - P_{7, x, k}\right)\right) & m = 2n-1, \, m \, {\rm odd}\\
e^{-(y-P_{1})^{2}/P_{2}^{2}} \sin \left(P_{6, x, k} \left(y - 64 - P_{7, x, k}\right)\right) & m = 2n, \, m \, {\rm even}\\
\end{array}\right.
$$
and these functions can be precomputed based on laboratory evaluations of the Gaussian profile parameters $P_{1}$ and $P_{2}$, fringe frequencies $P_{6, x, k}$, and internal differential dispersion $P_{7, x, k}$ (presently set to zero).  The remaining coefficients $A_m$ are all linear, allowing for a single matrix inversion to solve the best $\chi^2$ fit.  This is efficient to implement in real-time, whereas an iterative nonlinear fitting procedure would be prohibitively slow.  It is also trivial to parallelize the fit computations, as each spectral channel's fit is evaluated independently.

The typical fringe tracking parameters are given in Table~\ref{table:fringesearchtrack} and are optimized with on-sky observations in median seeing conditions. 
An exposure time of 6 ms is commonly used for the EMCCD. The typical fringe search step sizes are $\sim 12.5$~$\mu$m. During fringe tracking, two 6 ms frames are added together for a total on-sky coherent integration time of 12 ms (2 coherent co-adds). While a 12 ms effective exposure time on sky does reduce the fringe SNR (see Equation~\ref{eqn:SNR} in \S\ref{sec:frngsearch}) due to the atmospheric fluctuations, the added flux more than makes up for the lost fringe SNR. Fringe fitting is done as described above using Equation~\ref{eqn:frngmulti} to determine $P_4$ and $P_5$. The group delay for each telescope pair is estimated by treating $P_4$ and $P_5$ as the real and imaginary components of a 1D Fourier Transform (FT) 
for that frequency, and another 1D FT along the wavelength direction using the FFTW\footnote{http://www.fftw.org/} program in C++ \citep{FFTW} is performed. The resulting power spectrum is coadded over thirty 12 ms co-added frames, for an effective in-coherent exposure time of $360$ ms (30 incoherent co-adds) to generate a total power spectrum. The location of the peak of the total power spectrum corresponds to the delay that is sent to the delay line carts. 




\subsection{Observing Sequence \label{sec:obsseq}}
A complete observing sequence
for a target or calibrator star is detailed in Table~\ref{table:obsseq}. After acquiring the star, 
the fibers are aligned to maximize the light coupled
using an automatic fiber alignment algorithm. This step is typically required 
several times a night depending upon whether the observed fluxes are lower than expected. 
Next the different types of VISION data necessary to calibrate the observed 
interferograms in post-processing are recorded: darks, interferograms while fringe tracking, foregrounds, and
single beam data, as detailed in~\S\ref{sec:overalldata}. 

Target star and calibrator star observation sequences are interleaved. 
Calibrator stars are selected that are typically 
1--4 mas (depending upon whether longer (30--80 m) or shorter (8--12 m) baselines are used) to 
correct for the system visibility drift and bispectrum bias in the data. For typical $\approx2$ hour observations of a given star, a calibrator-target-calibrator pattern is alternated on $\approx20$ minute timescales, given that single beam and foreground data are required. A typical VISION observation of a star produces 20--50 GB of raw data, and consists of $\approx10^{6}$ individual frames with $6$ ms exposures; an observing run typically produces 200--400 GB of raw data per night.

\subsection{Faint Magnitude Limit} 

Fringe detection has been demonstrated with current hardware at
apparent magnitude $R_{\rm mag} = 4.5$, in excellent seeing. Funds from the Office of Naval Research DURIP competition
have recently been received to replace the existing Andor DU-860 EMCCDs used for fringe detection and
intensity mismatch monitoring with new N{\"u}v{\"u} EMCCDs. The new
EMCCDs feature $\times10$ less CIC noise (0.005 events pixels$^{-1}$ per frame) as compared to our measured CIC of $0.08-0.11$ events pixels$^{-1}$ per frame. This reduced noise is expected to greatly improve our faint magnitude limit.

\section{Characterizing the VISION instrument \label{sec:char}}
 
\subsection{System Throughput \label{sec:throughput}} 
An observed throughput of~\tpobs~was measured using the average total flux of
of $\gamma$ Orionis on the night of March 16, 2015, in median seeing. 
The total observed 
photons per second $F_{\rm obs}$ for $\gamma$ Orionis was estimated as:
\begin{equation}
\label{eqn:tp}
F_{\rm obs} ({\rm phot~s^{-1}}) = \sum_{i=0}^{i=127}\frac{\pi}{4}D^{2} <F_{0}> 10^{-0.4( R_{\rm mag}+kz)} \frac{\lambda_{i}}{hc} \Delta\lambda <T>
\end{equation}
where $i$ is the pixel index on the VISION fringing and photometric cameras, 
$D=12.5$ cm is the effective collecting area diameter, $k=0.11$ is the extinction in
$R$-band in magnitudes, $z=1.43$ is the airmass during the observations, $<F_{0}> = 2.25\times10^{-9}$
ergs cm$^{-2}$ s$^{-1}$ \AA$^{-1}$ is the zero magnitude $R$ band flux, $R_{\rm mag}=1.73\pm0.1$ \citep{Ducati02} is the 
magnitude $\gamma$ Orionis from SIMBAD\footnote{http://simbad.u-strasbg.fr/simbad/} \citep{Wenger00}, 
$\lambda_{i}$ is wavelength of 
light at pixel $i$ on the cameras, $h$ is Planck's constant, $c$ is the speed of light,
$\Delta\lambda\approx1860$ \AA~is the wavelength range over the entire filter, 
and finally $<T>$ is the average throughput for the observations. $F_{\rm obs}$ for $\gamma$ Orionis was measured on both the fringing and photometric cameras on March 16, 2015 in median seeing. We computed the observed throughput, for both the fringing and photometric cameras by solving Equation~\ref{eqn:tp} above for $<T>$. 

\begin{figure}[ht]
\begin{center}
\includegraphics[angle=90,width=\textwidth]{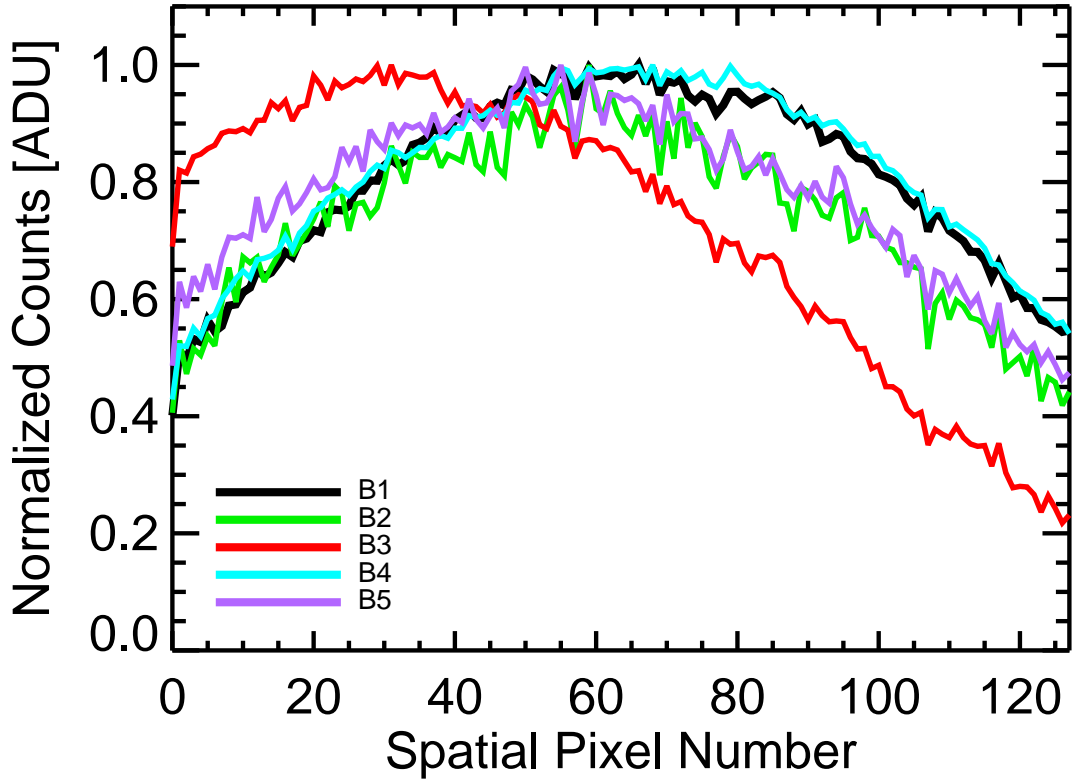}  
\end{center}
\caption{\label{fig:gaussprof}
The wavelength averaged ($550-850$ nm) shape of the Gaussian beam profile outputs from the single 
mode fibers on the fringing camera (see \S\ref{sec:gaussprof}). All five beams overlap well. A Gaussian shape is expected 
for the LP$_{01}$ mode of electric field that propagates through each single-mode fiber used 
for VISION. Beam 3 is offset from the center of fringing camera chip due the single mode fiber 
being slighlty misplaced in the V-groove array. 
}
\end{figure}

Accounting for all optical surfaces from the telescopes to the beam combiner, a
total theoretical throughput of~$\approx$\tptheo\ was computed 
by multiplying the reflectivity and transmission of all 
optical surfaces from the telescope to the VISION cameras including filter response 
and quantum efficiency as detailed in Table~\ref{table:throughput}.
The total observed throughput was 
$\approx$\tpobs, which is nearly $13$ times lower than the theoretical throughput. This significant difference could be due to
the adopted throughput of the NPOI feed system and the delay 
line carts of $\approx$\tpfeedsystem~and~$\approx$\tpdelayline~respectively,
as measured 9 years ago by \citet{Zhang06}. The loss of light 
in the feed system could be much larger due to 9 additional years of optical coating degradation. Similarly, 
the delay line cart mirrors have drifted out of focus, meaning the adopted theoretical fiber coupling 
efficiency of~\tpfibercoupling is 
likely overestimated. An additional cause of lost light is that the fringing camera 
does not sample the full Gaussian profile from each fiber on the detector as 
shown in Figure~\ref{fig:gaussprof}, which leads to $\approx30$\% loss in light. Finally, the misalignment of 
the focusing optics for light from the multi-mode fiber output to the photometric camera 
also could lead to an additional loss of light for beams 3 and 5, which are on the edges of the photometric 
camera chip.

\subsection{Beam Overlap\label{sec:gaussprof}} 
Fringes can only exist where the Gaussian beam profiles
of each single-mode fiber overlap on the fringing camera. 
The Gaussian beam profile was measured 
for each beam individually on the fringing camera using a laboratory 
white light source, 20 ms exposure times, a
gain of 300, and the low resolution $R=200$ observing mode. 
Ten minutes of frames were recorded on the fringing camera for each beam 
to build high-SNR Gaussian beam profiles. 
Figure \ref{fig:gaussprof} illustrates the overlap for all 5 beams averaged over
$570-850$ nm.  With the exception of beam 3, 
the percent flux overlap for each beam with each other beam is $>90\%$. 
The percent flux for beam 3 that overlaps with beams 1, 2, 4, and 5 is $72\%-79\%$. The lower 
overlap for beam 3 is likely due to a slight 
misplacement of the fiber for beam 3 in the V-groove array. 
The lower overlap fraction in beam 3 can lead to slightly lower fringe contrast 
between beam 3 and the other beams. These systematic 
differences in visibility can be removed by observing a calibrator star. 

\subsection{Laboratory Fringe Model \label{sec:frngmodel}}

Given that VISION uses single-mode fibers, 
high visibilities of $>80\%$ are expected for all beam pairs under ideal 
laboratory conditions, similar to commissioning 
tests of other beam combiners such as MIRC and AMBER \citep{Petrov07} that 
use optical fibers. This maximum possible visibility measured for a given beam 
pair under ideal conditions is the ``system visibility". 
In order to verify that the system visibility is $>80\%$,  
a fringe model was chosen to match sets of 
high signal-to-noise laboratory fringes that do not suffer from visibility loss due to 
beam intensity mismatch and atmospheric turbulence or CIC noise.

Equation~\ref{eqn:frngmulti} was parameterized making the basic fringe model 
directly comparable to VISION interferograms, at a single wavelength channel: 
\begin{equation} 
\label{eqn:final}
I(x, y) =  e^{-(y-P_{1})^{2}/P_{2}^{2}} \left(P_{3} + \sum_{k=1} \frac{\sin(P_8 y + P_9)}{P_8 y + P_9} \left[P_{4,k} \cos \left(P_{6, x, k} \left(y - 63.5 -P_{5, k}\right)\right)\right]\right)
\end{equation}
where parameters $P_{4,k}$ and $P_{5, k}$ are redefined as overall fringe amplitude and phase, and new parameters $P_8, P_9$ are introduced modulation by the coherence envelope due to the finite spectral resolution of the detector, which can be modeled by integrating over several wavelengths.

The integration is done by evaluating the fringe model at $0.01$ pixel steps, or $1.28\times10^4$ evenly spaced points from
pixels 0 to 127, and integrating the resulting model fringe to 128 pixels. By integrating, the pixelation of each interferogram by the finite spectral resolution is modeled.  The fringe model (Equation~\ref{eqn:final}) 
assumes a rectangular bandpass, due to the use of a sinc function. The model is an 
approximation given that the VISION bandpass is likely closer to a Gaussian and not square (as dictated by a sinc function)
for each resolution element, but was sufficient for the present study. 

The system visibilities for all 10 available beam pairs (beams 1--5)
were measured under laboratory conditions using the fringe model above. 
The fringe parameters for 
each beam pair were evaluated by fitting the fringe model in Equation~\ref{eqn:final} 
to interferograms from a laboratory HeNe 632.8 nm laser source. 
The visibility was estimated as the ratio between total flux on the detector and the amplitude of the coherence: 
\begin{equation}
V=\frac{P_4}{P_3}=2\frac{\sqrt{I_1 I_2}}{I_1+I_2}|\gamma|. 
\end{equation} 

The procedure to obtain the interferograms for each beam pair was 
as follows. First, a 632.8 nm HeNe laser source was coupled 
to the single-mode fiber of each beam. 
The delay line carts were positioned 
to minimize light path difference between the 
two beams to within the coherence length ($\lesssim1300$~$\mu$m), 
maximizing the visibility for a given
pair of beams. 
Several minutes of 2-ms exposure time raw interferograms were recorded 
for each beam pair and the gain set to zero. 
Electron multiplying gain was not used in order 
to avoid additional multiplication noise. 
A time averaged dark was subtracted from each of the raw interferograms to 
remove the bias counts on the EMCCD. 
A median selected dark was not used because there was 
negligible CIC on the detector when running 
the EMCCD with zero gain. 

\begin{figure}[ht]
\begin{center}
\includegraphics[angle=90.0,width=\textwidth]{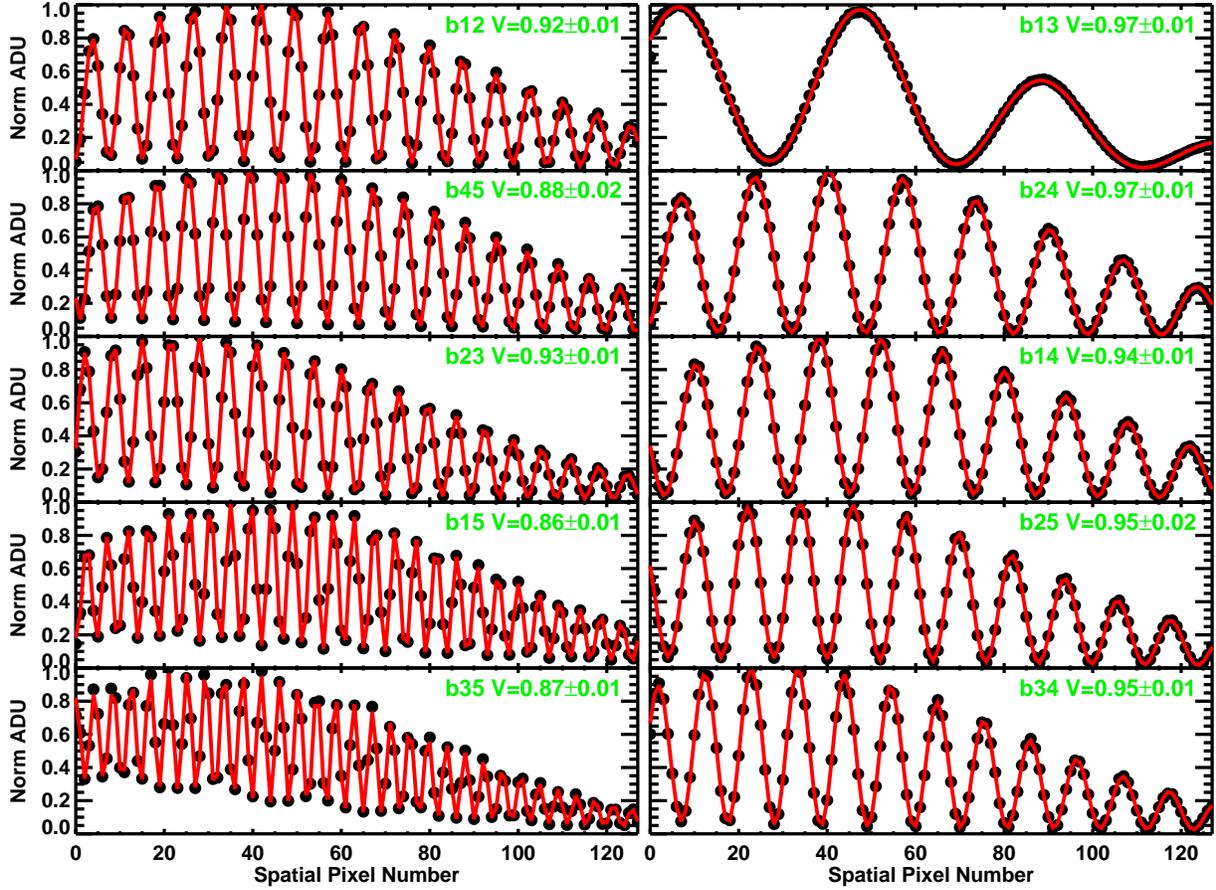}
\end{center}
\caption{\label{fig:labvis} 
Interferograms using a 632.8 nm HeNe laser source and matched light paths for each 
beam pair for beams 1-5 (see \S\ref{sec:frngmodel}). A fringe model was fit to each set of laser fringes for each beam pair using the fringe model Equation~\ref{eqn:final}, 
and measure visibilities of 85-97\%, which implies that the majority of light is interfering. The small visibility 
loss could be the result of polarization effects, beam intensity mismatch.
}
\end{figure}

\begin{figure}[ht]
\begin{center}
\includegraphics[angle=90.0,width=\textwidth]{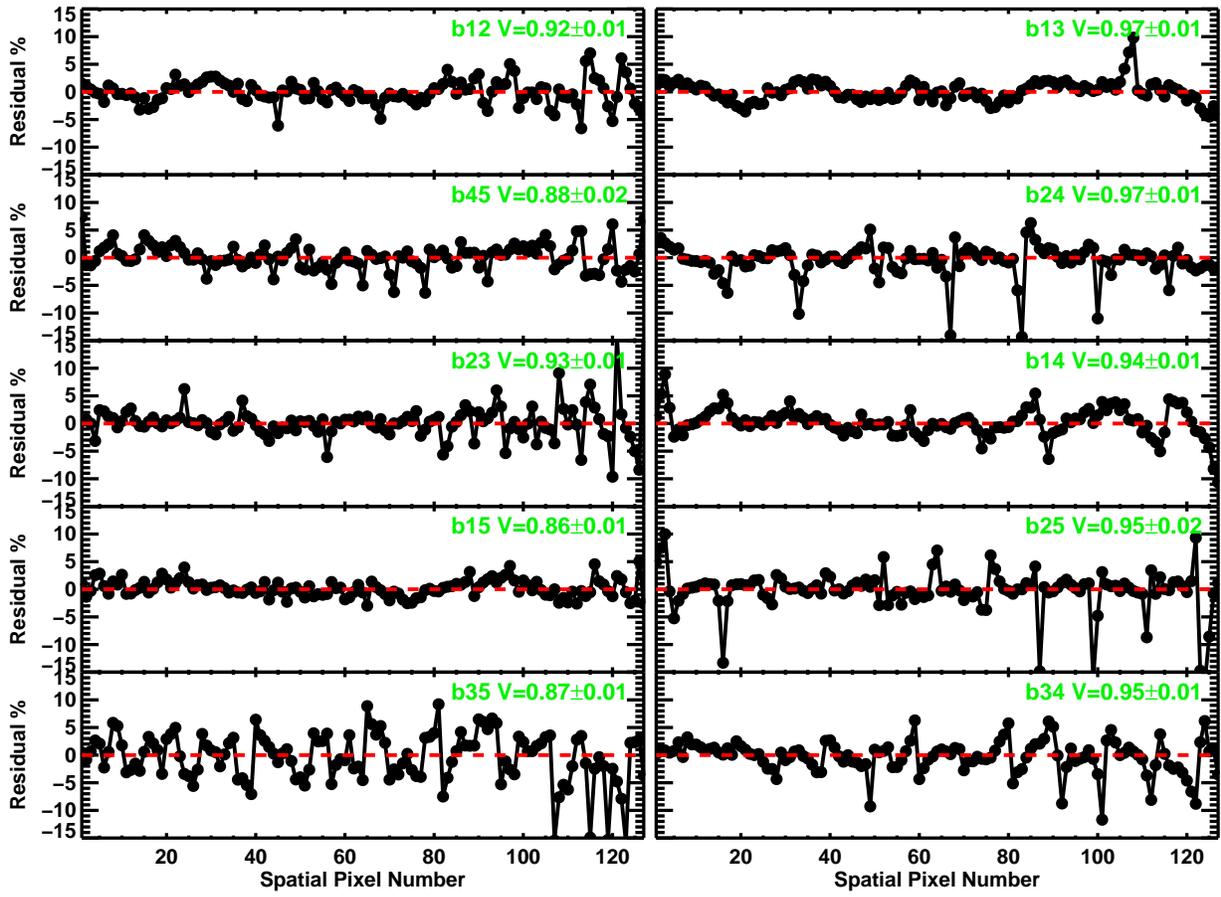}
\end{center}
\caption{\label{fig:labvisresid} 
The residuals for figure~\ref{fig:labvis}. 
}
\end{figure}

The system visibility was 
$V_{\rm laser}\approx85-97\%$, as expected when using single-mode fibers.  
Fitting was performed using IDL's MPFIT\footnote{http://www.physics.wisc.edu/$\sim$craigm/idl/down/mpfit.pro} which employs 
a modified Levenberg-Marquardt $\chi^{2}$ minimization to fit
the fringe model (Equation~\ref{eqn:final}) to the interferograms. 
Sample interferograms for each beam 
and model fits are shown in 
Figure~\ref{fig:labvis}. The residuals between 
the fringe model and the interferograms 
are 1--5\% as shown in Figure~\ref{fig:labvisresid}, as expected 
given the imperfect fringe model (e.g.~sinc envelope instead of Gaussian). 
Nevertheless, the laboratory system 
visibilities for VISION were $> 80\%$ for all beam pairs, as expected. 


\subsection{Fringe Crosstalk \label{sec:crosstalk}}

Fringe analysis based on Fourier Transforms shows slight overlap among the peaks in the power spectra from each beam pair (see Figure~\ref{fig:frngfreq}).  This is because the fringes do not fit along the 128 pixels of the fringe direction an exact integer number of times.  While the fringe frequencies were selected to be unique and produce orthogonal intensity pattern functions, sinusoids are only orthogonal on domains in which both sinusoids have an exact integer number of waves.  (Even if a perfect integer number of waves fit across the detector for one wavelength, this would not be the case for other wavelengths.)  As a result, the fringe sinusoids are not strictly orthogonal functions on the domain of 128 pixels, though they may be unique and orthogonal overall.  This results in perceived cross-talk between channels even in the case of a perfect setup (instrumental alignment inaccuracies will add to the effect).  This is independent of the method used to evaluate the fringe models and instead related to the non-orthogonality of the basis functions on the 128 pixel restricted domain.

As expected, the observed crosstalk between 
peaks in the power spectra were found to occur between pairs of beams that are 
closest in fringe frequency such as beam pairs 1-4 and 2-5. 
The crosstalk percentage was calculated as the total power in the power 
spectrum for each beam pair at the location of the peak of each other beam pair.
The magnitude of the fringe crosstalk using the Fourier Transform method was found to be $\approx1-8\%$ of the power in the power spectra
for the laboratory fringes. Fringe fitting with better fringe models (including illumination profile, coherence reduction far from zero optical path delay, dispersion, etc.), as described in \S \ref{sec:frngtrack} and below, can improve but not entirely eliminate these effects.

In an attempt to further understand the crosstalk, a multi-fringe model based on Equation~\ref{eqn:final},
was used to fit fringes from multiple beam pairs on the detector. 
Laser fringes from beam pairs 1-4 and 2-5 were added together and 
fit for parameters of both beam pairs simultaneously. 

\begin{figure}[ht]
\begin{center}
\includegraphics[angle=90.0,width=\textwidth]{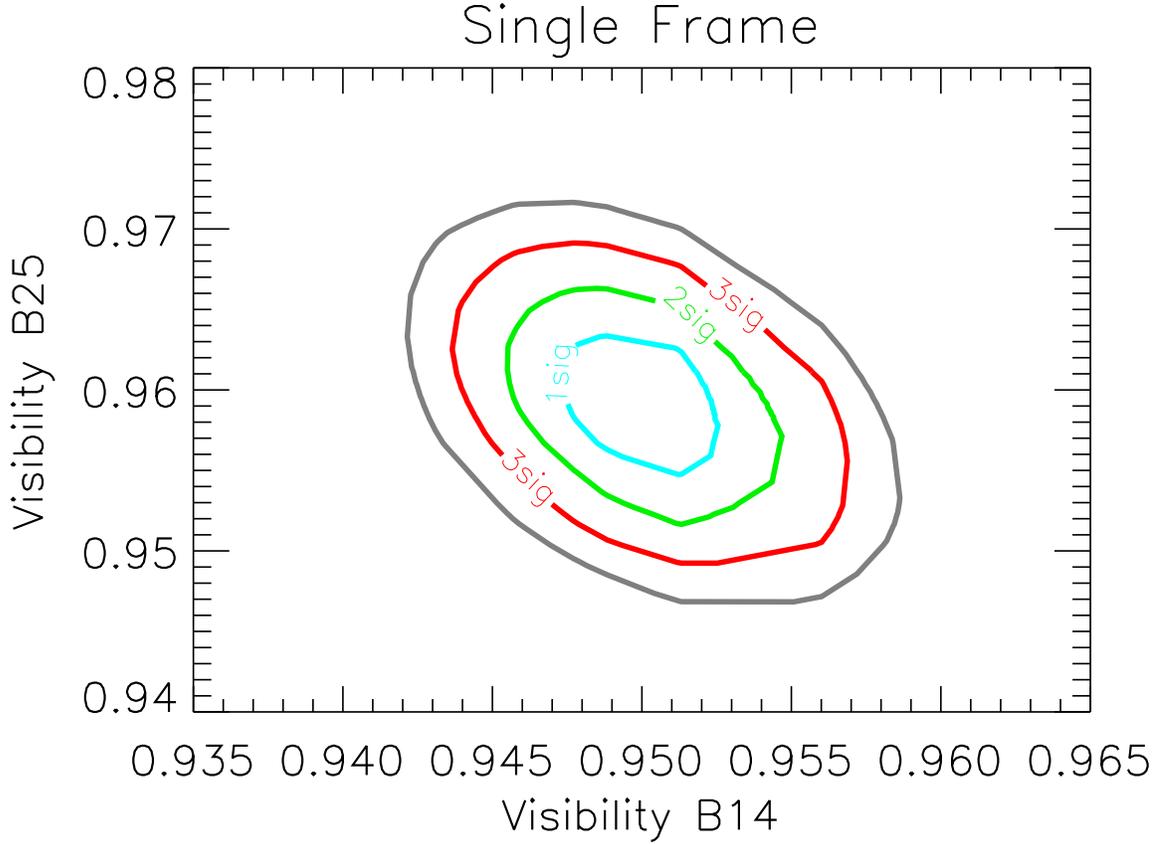}
\end{center}
\caption{\label{fig:contour} 
A map of $\chi^2$ space for the correlation between the visibility of beam pair 1-4 
and beam pair 2-5, using co-added laser fringes. The confidence intervals 
correspond to $\Delta\chi^{2}= \chi^{2}-\chi_{\rm min}^{2} = 2.30$, $6.17$ and $11.8$ for $1\sigma$, $2\sigma$ 
and $3\sigma$ respectively for 2 parameters of interest \citep[see \S\ref{sec:crosstalk}, and also][]{NR}. The visibility parameters 
for either beam pair are correlated on the $\approx2\%$ level for the $3\sigma$ confidence interval even for these very high signal-to-noise laser fringes. This suggests crosstalk is inherent to the instrument set up.  
}
\end{figure}

In Figure~\ref{fig:contour} 
the $\chi^2$ space is mapped by varying the amplitude of the fringes, parameter $P_{4,k}$ 
for both beam pairs 1-4 and 2-5 in Equation~\ref{eqn:final} above. For two parameters of interest, 
$\Delta\chi^{2} = 2.30$, $6.17$, and $11.8$ for $1\sigma$, $2\sigma$, and $3\sigma$ confidence intervals 
respectively \citep{NR}. Even for ideal, high SNR fringes, a correlation in the visibilities is seen 
of both beam pairs at the $\approx2\%$ level for the $3\sigma$ confidence interval. This could also be interpreted as a form of crosstalk. This test suggests that the $\chi^{2}$ minimization between a multi-fringe model and 
multi-fringe data likely leads to correlations between the visibility parameters and 
is thus a manifestation of crosstalk, similar to the overlapping peaks in the power spectra. 



\section{The VISION Data-Processing Pipeline \label{sec:post}}

\subsection{Dark Subtraction Algorithm \label{sec:darksub}}

Dark frames are used to subtract the bias counts 
for all raw calibration data and interferograms.  
The bias counts are important to characterize given that average signal counts 
can be as low as $\approx0.1$ photons/pixel/frame 
for $R_{\rm mag}\approx3.5$ stars. 
Bias level subtracted interferograms 
and calibration data are used in the data processing pipeline. 
Following \citet{Harpsoe12a} 
the bias count levels on the photometric and fringe camera chips 
were characterized by the sum of time-dependent (frame to frame) and spatially 
dependent bias counts: 
\begin{equation} 
b(x,y,t) = b(x,y) + b(t)
\end{equation}
where $b(x,y,t)$ are the bias counts at pixel $(x,y)$, and time $t$. Andor's ``baseline clamp'' 
software stabilizes $b(t)$ for each frame by subtracting off a 
running average of a $128$ pixel overscan region of the chip and then adding 
back in $100$ Analog-to-Digital Units (ADU). 
The output probabilities of ADU for the 
EMCCDs are characterized using an analytic model that is a
convolution of the probability that a pixel will have a CIC event or just a bias count \citep[Equation 8,][]{Harpsoe12a}.

This analytic EMCCD model was modified 
to include the probability that two CIC events occur 
in the same pixel, given that the probability of a CIC electron is
 $\approx8-11\%$ for the EMCCDs, and thus the probability 
of two concurrent CIC electrons was $\approx0.6-1.2\%$, which is
significant for characterizing the high end tail of pixels with $>200$ ADU: 
\begin{equation} 
\label{eqn:emccd}
P(Z=n) = \int_{0}^{n} \Bigg[(1-p-p^{2})\Bigg(\delta(x) + \Big(\frac{p}{G} e^{-x/G} + \frac{p^{2}}{G^2} x e^{-x/G}\Big)H(x)\Bigg)\Bigg]\times N(n-x,\sigma_{\rm RN}) dx
\end{equation}
where $P(Z=n)$ is the normalized probability that a given pixel will have output $n$ ADU. 
Furthermore, $p$ is the probability of CIC, $G$ is the gain, $\sigma_{\rm RN}$ is the 
read noise in ADU, $H(x)$ is the heaviside step function, and $N(n-x,\sigma_{\rm RN})$ is the 
normal distribution that describes the bias counts $b(x,y)$ at each pixel: 
\begin{equation}
N(n-x,\sigma_{\rm RN}) = \frac{1}{\sqrt{2\pi\sigma_{\rm RN}^{2}}} e^{-(n-x)^2/(2\sigma_{\rm RN}^2)}
\end{equation}
where $N(n-x,\sigma_{\rm RN})$ describes the bias counts (which has Gaussian read noise) 
when no CIC event occurs. If a CIC event occurs, the output counts $n$ from the 
EMCCD go as $\frac{e^{-x/G}}{G}$ \citep{Basden04}.

\begin{figure}[ht]
\begin{center}
\includegraphics[angle=90.0,width=\textwidth]{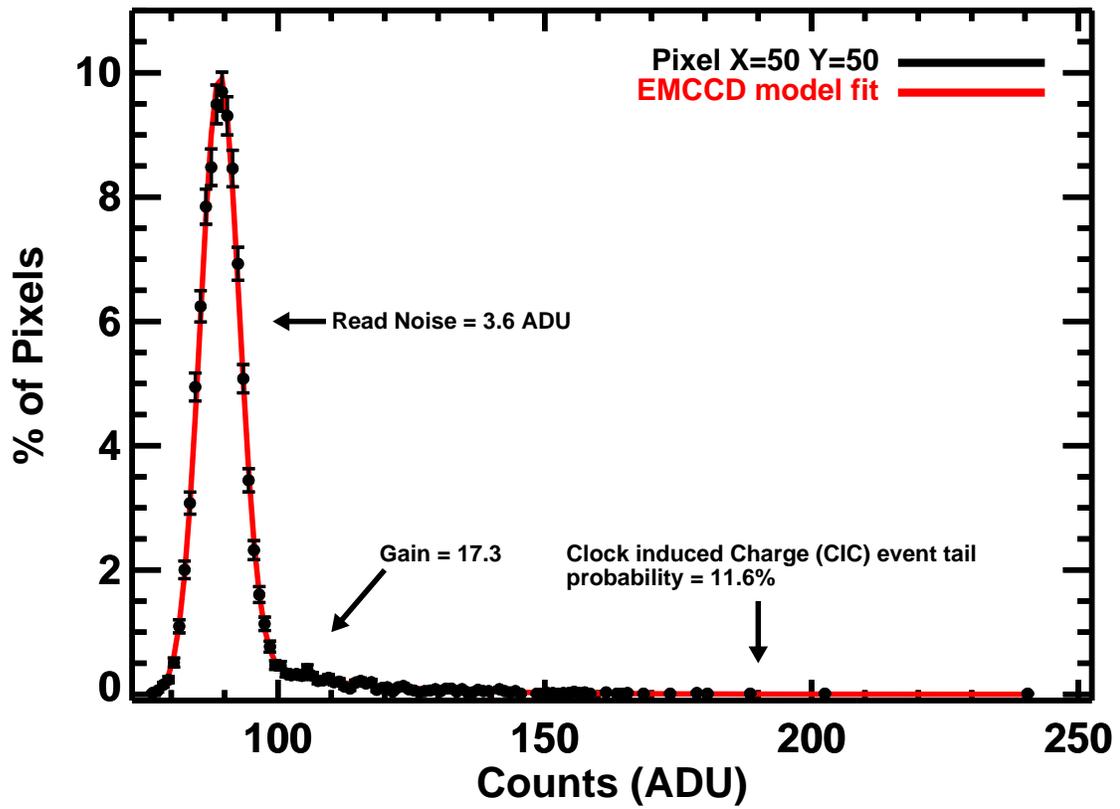}
\end{center}
\caption{\label{fig:emccd} 
Sample fit to a histogram of a time series of $\approx10^{5}$ dark frames for pixel $(50,50)$, (black circles). The 
gain, read noise, and clock induced charge rate are derived by fitting an analytic EMCCD model to the data (red line). The analytic 
EMCCD model (see \S\ref{sec:darksub}) is the convolution of the probability that a pixel only has read noise, or has read noise and a clock induced 
charge event. 
}
\end{figure}

This EMCCD model was tested by fitting a histogram of 
10$^{5}$ raw dark frames from the fringing camera
on a pixel by pixel basis to derive a bias count level $b(x,y)$, read noise $\sigma_{\rm RN}(x,y)$, gain $G(x,y)$, 
and CIC probability $p(x,y)$. The darks were recorded with $6$ ms exposures, and a gain 
setting of $300$ on the camera. 
The modified Levenberg least squares minimization algorithm MPFIT
\citep{Markwardt09} was used to fit the EMCCD model to the darks. A sample 
fit to pixel $(x,y) = (50,50)$ is shown in Figure~\ref{fig:emccd}. 
The EMCCD model fit the data well with small residuals.  
The typical read noise was $\sigma_{\rm RN,obs}\approx3.5$ 
ADU\footnote{Following the literature convention, counts instead of electrons are used given that the gain on an EMCCD is stochastic and varies as a function of time and location on the detector. Therefore there is no exact conversion between electrons and ADU for an EMCCD.}, gain of $\approx16-18$, and clock induced charge probability of $\approx11-13\%$ averaged over the entire chip for the fringing camera. The observed gain of $\approx16-19$ is $258-305$ when multiplied by the e$^{-}$/ADU conversion 
of \Sensitivityf~and~\Sensitivityp~for the fringing and photometric camera, respectively. This gain is comparable to 
the camera gain setting of 300 for these darks. The read noise, gain, and CIC were similar for the photometric camera. 
Andor's listed CIC rate of $0.05$ events/pixel/frame for the cameras is likely 
underestimated since it does not include CIC events buried within the read noise. 
A higher CIC rate of $0.08-0.11$ events/pixel/frame was measured and accounted 
for CIC within the read noise using the EMCCD model above.  

With a derived CIC $p(x,y)$ and gain $G(x,y)$ from fitting dark frames using Equation~\ref{eqn:emccd}, 
the dark subtraction algorithm of \citet{Harpsoe12a} was used to subtract the bias count level. 
First, the mean bias level for each raw frame was computed using pixel 127 to further stabilize 
the frame to frame bias count variability:  
\begin{equation} 
b_{o}(t) = {\rm avg} \Big[c(127,y,t)-b(127,y)\Big]
\end{equation}
where $c(127,y,t)$ are pixel values for the raw frame at pixel $x=127$. 
Finally, 
the bias counts from each raw frame are subtracted as: 
\begin{equation}
\label{eqn:darksub}
c_{b}(x,y,t) = c(x,y,t)-[b(x,y)+b_{o}(t)]-G(x,y)p(x,y)
\end{equation}
where $c_{b}(x,y,t)$ is the bias count subtracted frame. 

\subsection{Pre-Processing Raw Interferograms \label{sec:preprocess}}

Prior to extracting squared visibilities, closure phases, and triple amplitudes from raw interferograms, the following steps are performed: 
\begin{enumerate}
\item estimate the gain, read noise, and clock induced charge for darks, 
\item perform dark subtraction, 
\item remove poor quality interferograms, and 
\item bin the data spectrally to increase SNR. 
\end{enumerate}

First, a series of dark frames taken during the observing sequence (see \S\ref{sec:obsseq})
is used to estimate the gain, clock induced charge rate, and read noise for the raw interferograms 
as detailed in \S\ref{sec:darksub} above. Next, dark subtraction is performed using Equation~\ref{eqn:darksub} from 
\S\ref{sec:darksub} above. 
Raw interferograms for which the fringe tracking 
SNR is so low that no fringes were identified are removed.  
This loss of coherence can occur often due to the turbulence 
of the atmosphere at visible wavelengths. With this step 
1--5\% of the raw interferograms that have the lowest fringe SNR (see \S\ref{sec:frngtrack}) are removed. Next, 
spectral channels with very little light are removed. For stars with spectral types 
O--F, typically only spectral channels with wavelengths $580-750$~nm are used, given that 
the single-mode fibers enter multimode at $<580$~nm and spectral channels with wavelengths 
$>750$~nm had too little SNR. Next, these spectral channels are binned in the wavelength by a factors of 8, 9, or 10. For a typical 9 spectral channels, this results in an overall read noise of $\sigma_{\rm RN,binned} = \sqrt{9}\sigma_{\rm RN,obs} \approx 10.5$ ADU, and $\approx19$ nm per spectral channel. Finally, 
the raw interferograms are divided by the mean gain as derived above, typically $15-18$ ADU/e$^{-}$.


\subsection{Adapting the MIRC Data-Processing Pipeline for VISION \label{sec:postprocess}}

Since MIRC and VISION are nearly identical in design, 
the MIRC data-processing pipeline \citep{Monnier04,Monnier07} 
was modified to estimate calibrated squared visibilities, closure phases, and triple amplitudes 
from interferograms pre-processed as detailed in \S\ref{sec:preprocess} above. 
Briefly, the MIRC pipeline measures
uncalibrated squared visibilities and triple amplitudes from raw interferograms 
after a series of Fourier transformations
and foreground subtractions. The MIRC pipeline then 
calibrates the squared visibilities and triple amplitudes using 
fluxes measured simultaneously with fringes. The last step of the MIRC pipeline is to
use calibrators with known sizes to compensate for 
system visibility drift. 

A significant change 
to the MIRC pipeline for processing VISION interferograms is the use of single  
$6$ ms frames to estimate the uncalibrated squared visibilities. By contrast, MIRC 
pipeline coherently co-adds several frames of data before estimating 
uncalibrated squared visibilities. 
In adapting the MIRC pipeline, 
the complex bispectrum bias must also be corrected on a frame by frame basis, since VISION data 
are photon noise limited whereas MIRC data are read-noise limited; the MIRC pipeline 
only partially implements this correction, since MIRC operates in 
a regime where photon noise bias in the bispectrum is rarely important. 

The uncorrected bispectrum is given as
\begin{equation} 
B_{0,ijk} = C_{ij} C_{jk} C_{ki}
\end{equation}
where $B_{0,ijk}$ is the bispectrum for beams $i,j,k$, and $C_{ij}$ is the complex discrete 
Fourier transform for beam pair $i,j$. The triple 
product bias was adopted from EMCCD simulations of \cite{Basden04}, with an additional read noise term from 
\cite{Gordon12} that was modified for the EMCCDs:   
\begin{equation} 
\label{eqn:tpbias}
B_{1,ijk} = B_{0,ijk} - 2\bigg(|C_{ij}|^{2} + |C_{jk}|^{2} + |C_{ki}|^{2}\bigg) + 6N + 6N_{\rm pix}\sigma_{\rm RN,binned}^2. 
\end{equation}
where $|C_{ij}|^{2}$ is the power spectra, $N$ is the total number of counts in the frame, $N_{\rm pix} = 128$ 
is the total number of pixels in the spectral channel, $B_{1,ijk}$ is the bias-corrected 
bispectrum, and $\sigma_{\rm RN,binned}$ is the summed read noise in quadrature over 
the wavelength-binned pixels in Equation~\ref{eqn:tpbias}. 

Equation~\ref{eqn:tpbias} was derived by attempting to recover the correct closure phases extracted from simulated fringes of a binary star with an input orbit, 
a simulated EMCCD gain register, and added read-noise and Poisson noise. 
The EMCCD simulations of \citet{Basden04} were replicated, and their Equation 4 was extended by adapting the \cite{Gordon12} treatment 
of read noise, yielding an extra term of $6N_{\rm pix} \sigma_{\rm RN,binned}^2$. \cite{Gordon12} provide equations to correct 
the bispectrum in the presence of read noise, but their equations only apply for Poission statistics, 
and the output of the EMCCDs is non-Poissonian due to the stochastiscity of the electron multiplying gain. As detailed in \cite{Basden04}, the coefficients of 2 and 6 multiplied by the power spectra $|C_{ij}|^{2}$ and total counts $N$ in Equation~\ref{eqn:tpbias} also differ from the traditional \citet{wirnitzer85} coefficients of 1 and 2 
given that the output from an EMCCD is not Poissonian. The VISION implementation uses 
the bispectrum bias subtraction in Equation~\ref{eqn:tpbias} on a frame-by-frame basis as recommended by both \cite{Basden04} and \cite{Gordon12} for extracting unbiased closure phases and triple amplitudes from VISION raw interferograms.  

\begin{figure}[ht]
\includegraphics[angle=90,width=\textwidth]{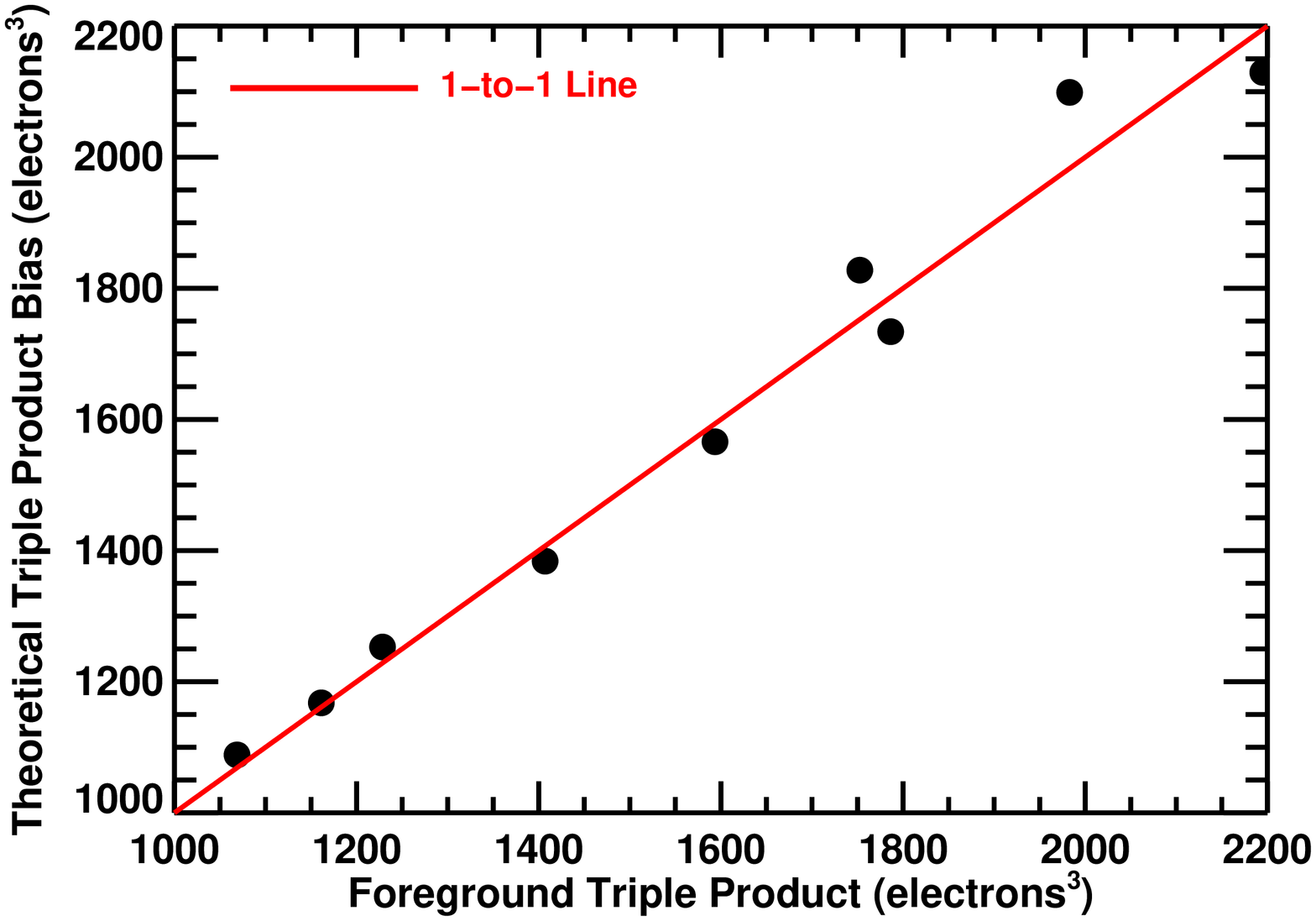}
\caption{The theoretical bispectrum matches the amplitude of foreground
bispectrum for $\gamma$ Orionis, further verifying 
that the bias correction procedures in the data-processing pipeline are accurate. The foreground 
amplitude of the bispectrum and power spectrum were averaged over 15.36 seconds, using 6 ms exposure times.}
\label{fig:tpbias}
\end{figure}

The corrected bispectrum $B_{1,ijk}$ should be zero within error for 
foreground data, given that these data have no fringes. 
Therefore, Equation~\ref{eqn:tpbias} was further tested by comparing the 
uncorrected bispectrum $B_{0,ijk}$ of a sample set of foregrounds to the theoretical bispectrum bias: 
\begin{equation}
\label{eqn:tpbiasq}
B_{0,ijk} {\rm~of~Foreground} = - 2\bigg(|C_{ij}|^{2} + |C_{jk}|^{2} + |C_{ki}|^{2}\bigg) + 6N + 6N_{\rm pix}\sigma_{\rm RN,binned}^2. 
\end{equation}
$B_{0,ijk}$ was computed for foreground data from 
observations of $\gamma$ Orionis, and it closely matched 
the right side of Equation~\ref{eqn:tpbiasq} above, as shown in Figure~\ref{fig:tpbias}. This further 
validated that the derived bispectrum bias correction (Equation~\ref{eqn:tpbias}) for the EMCCDs was correct. 

\begin{figure}[ht]
\includegraphics[angle=90,width=\textwidth]{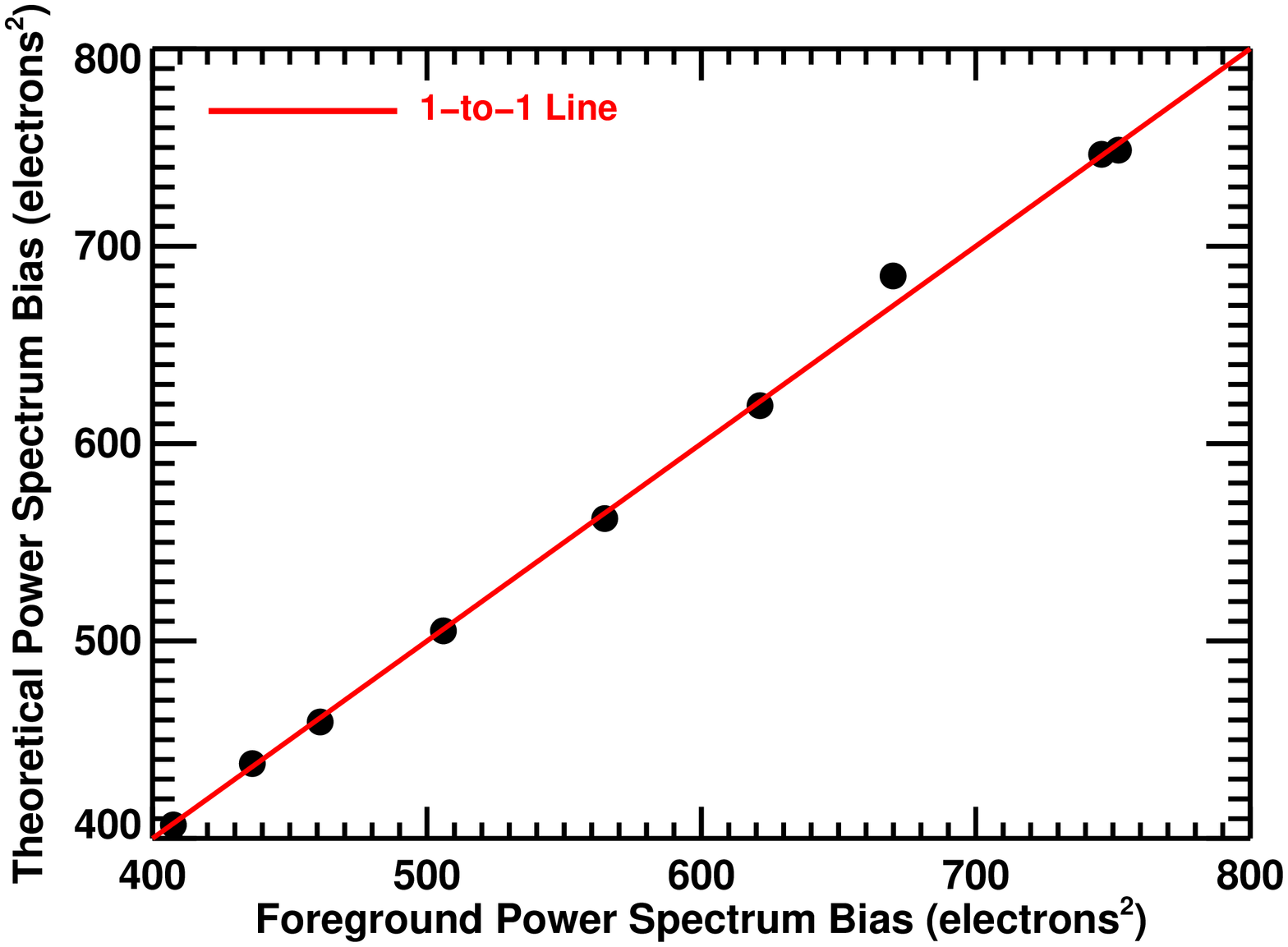}
\caption{The theoretical power spectrum bias matches the foreground power spectrum for $\gamma$ Orionis, further verifying 
that the bias correction procedures in the data-processing pipeline are accurate. The foreground 
amplitude of the bispectrum and power spectrum were averaged over 15.36 seconds, using 6 ms exposure times. }
\label{fig:bias}
\end{figure}

The theoretical prediction for the power spectrum bias was also derived as a modified version of the \cite{Gordon12} 
power spectrum bias adapted for EMCCDs, in the presence of read noise:
\begin{equation} 
|C_{1,ij}|^{2} = |C_{0,ij}|^{2} - 2N+N_{\rm pix} \sigma_{\rm RN,binned}^{2}
\end{equation}
where $|C_{1,ij}|^{2}$ is the corrected power spectrum and $|C_{0,ij}|^{2}$ is 
the uncorrected power spectrum. Similar to the previous approach, foreground data
contains no fringes and therefore no peaks in the power spectrum. Thus, the corrected 
power spectrum $|C_{1,ij}|^{2}$ should be zero. 
The power spectrum bias correction matches the uncorrected power 
spectrum $|C_{0,ij}|^{2}$ to within $1\%$ as shown in Figure~\ref{fig:bias}. 

\section{The Orbit and Flux Ratio of $\zeta$ Orionis A\label{sec:zeta}}

To both provide a first on-sky science result and to
verify the VISION data-processing pipeline from \S\ref{sec:preprocess} and \S\ref{sec:postprocess}, 
VISION was used to obtain new resolved observations of the massive binary star $\zeta$ Orionis A. The orbit and flux ratio of $\zeta$ Orionis A have previously 
been measured by \citet{Hummel13}, therefore the new observations serve as an established test of the VISION system and provide an additional epoch of constraint on the orbit of this benchmark astrometric binary. 

$\zeta$ Orionis A was observed on March 16th, 2015, with stations 
AC, AE and N3, with baselines between $\baserange$~m using the 
observation sequence in Table~\ref{table:obsseq}.  Fringe searching and 
tracking were performed as described in 
\S\ref{sec:frngsearch} and 
\S\ref{sec:frngtrack}. The calibrator star
$\gamma$ Orionis \citep[$\theta_{\rm UD}=0.701\pm0.005$;][]{Challouf2014}
was observed immediately after, using the same observing sequence to compensate for any visibility drift. The wavelength solutions for the cameras 
from \S\ref{sec:overalldesign} was used. 

\begin{figure}[ht]
\begin{center}
\includegraphics[angle=90.0,width=\textwidth]{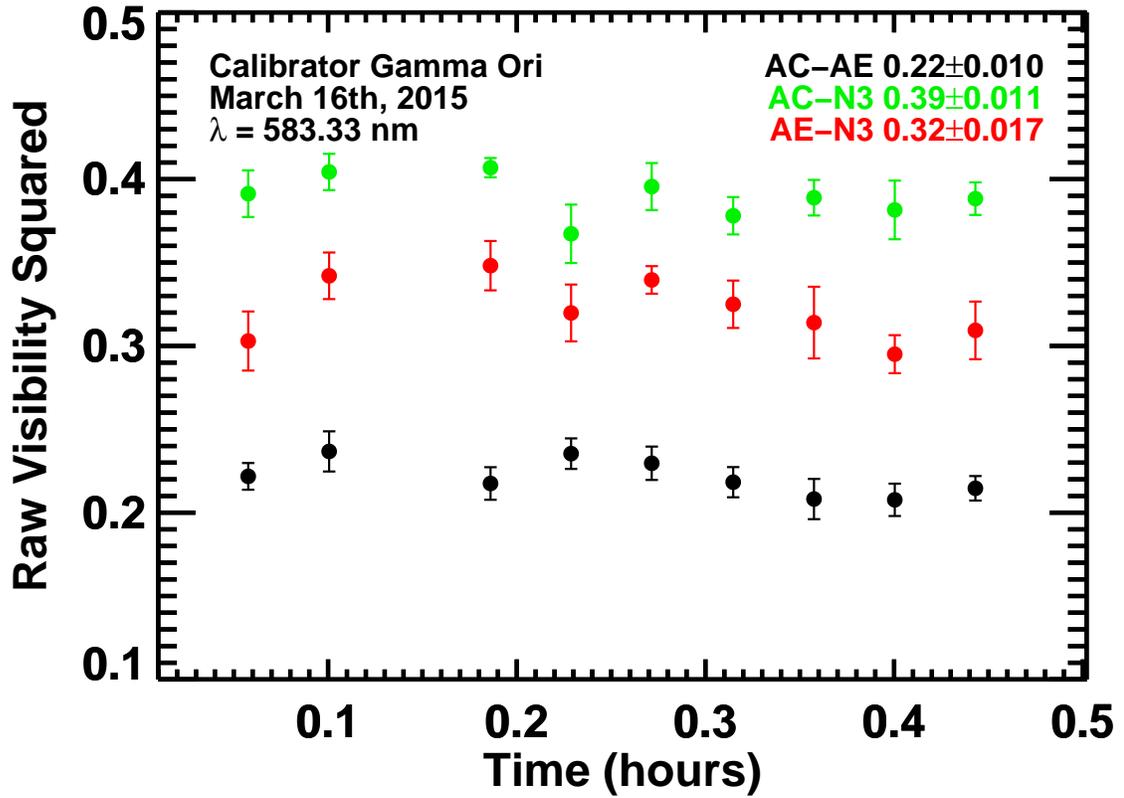}
\end{center}
\caption{\label{fig:gamori} 
Beam intensity mismatch corrected 
squared visibilities vs time for observations of calibrator $\gamma$ Orionis during 
commissioning on March 16th, 2015 with stations AC, AE and N3. 
The system visibility drift is at max $0.01-0.02$ 
over half an hour for $\gamma$ Orionis. 
}
\end{figure}

Sample squared visibilities versus time for 
$\gamma$ Orionis are shown in Figure~\ref{fig:gamori}. 
The squared visibility drift for VISION as 
measured using $\gamma$ Orionis is $0.01-0.02$ over 30 minutes. 
Dark subtraction was performed on all the raw frames of
$\zeta$ Orionis and the calibrator $\gamma$ 
Orionis as described in \S\ref{sec:darksub}. 
The data were pre-processed as described in \S\ref{sec:preprocess} and 
calibrated squared visibilities and 
bias-corrected closure phases were extracted 
as described in \S\ref{sec:postprocess}.


From the orbit of \citet{Hummel13}, 
$\zeta$ Orionis A is predicted to have a separation of $\hummelsep$~mas and a position angle of 
$\hummelpa^{\circ}$ at the epoch of the observations, with a flux ratio of $\hummelfratio$ mag. The 
$1\sigma$ errors on the predicted separation and position angle were calculated from distributions of 
separation and position angle from $10^{7}$ uniformly 
randomly selected orbits from the reported $1\sigma$ errors of the orbital elements from \cite{Hummel13} and 
then projected on sky.  

\begin{figure}[ht]
\begin{center}
\includegraphics[angle=90.0,width=\textwidth]{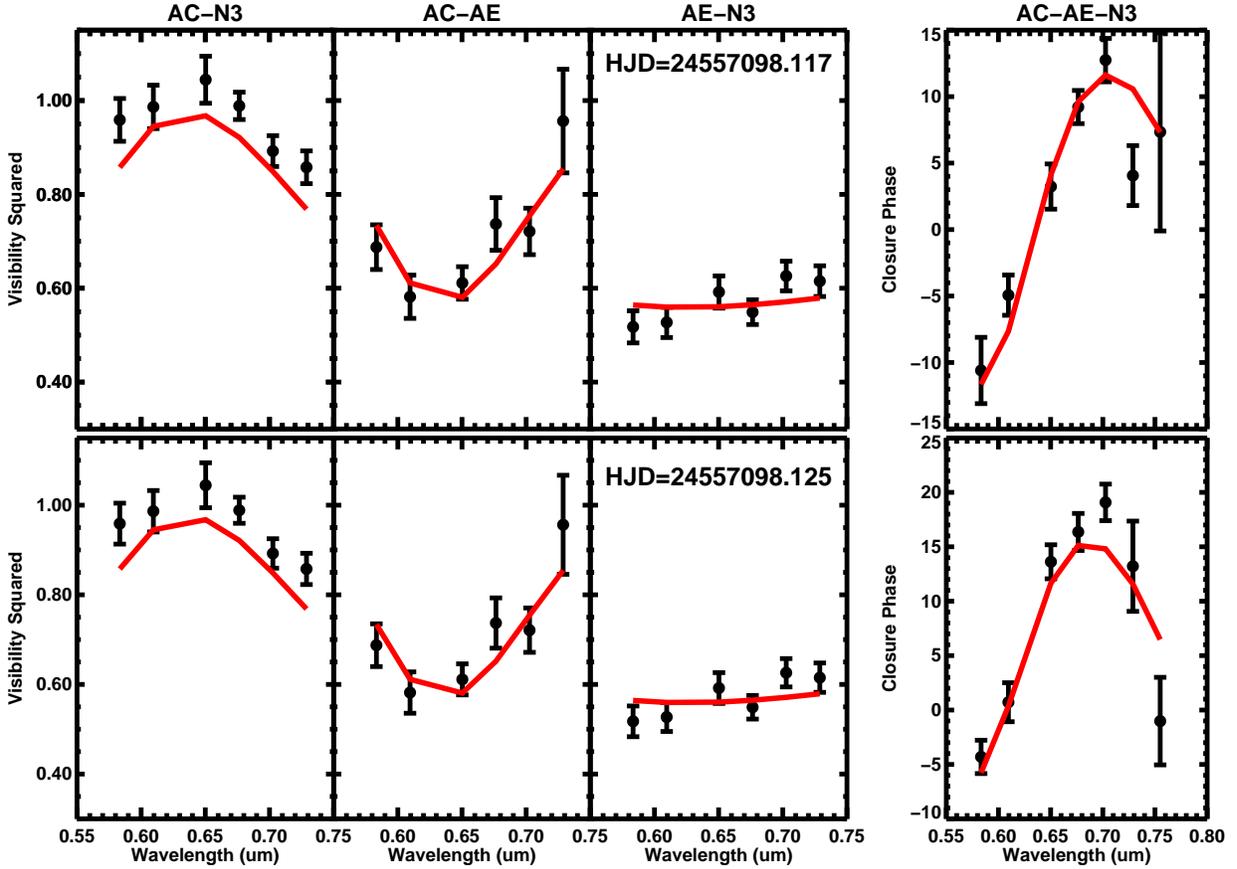}
\end{center}
\caption{\label{fig:zetori} 
Sample calibrated squared visibilities and closure phase observations of $\zeta$ Orionis on UT 03\/15\/2016 at two different times 
of the night. The red solid line is the best fit model to the data. the best fit model for this binary star at the epoch 
the observations yields a separation of $\zetasep$~mas, a position angle of $\zetapa^{\circ}$ for observations from $580-750$ nm, 
in good agreement with the predicted separation of $\hummelsep$~mas and 
position angle $\hummelpa$~mas from \citet{Hummel13}. The observed flux ratio of $\zetafratio$ mag is also in good agreement 
with the flux ratio $\hummelfratio$ mag from \citet{Hummel13}. The AC-N3 visibilities have a small bias likely due possibly to imperfect photometric calibration.}
\end{figure}

\begin{figure}[ht]
\begin{center}
\includegraphics[angle=90.0,width=\textwidth]{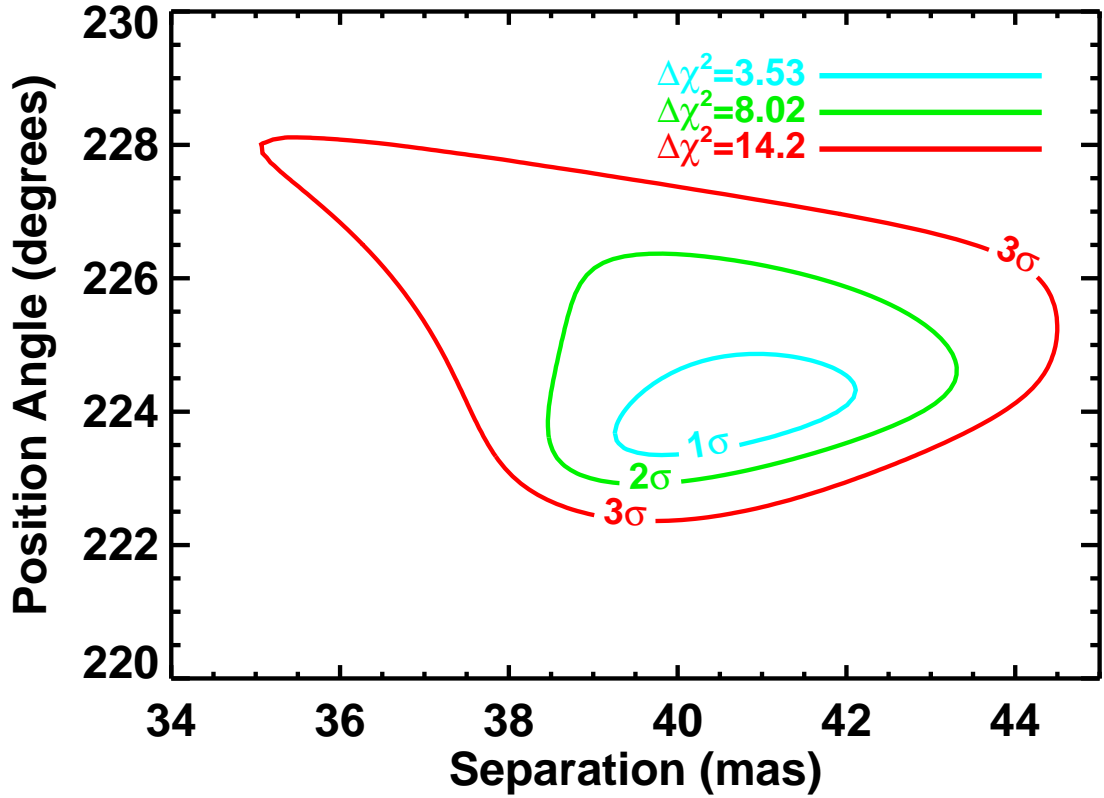}
\end{center}
\caption{\label{fig:cont} 
Confidence intervals for $1\sigma$ (blue line), $2\sigma$ (green line) and $3\sigma$ (red line) 
errors on the separation and position angle corresponding to $\Delta\chi^{2}=\chi^{2}-\chi_{\rm min}^{2}$ of 3.53, 8.02 and 14.03 respectively for 3 parameters of interest \citep{NR}. The minimum $\chi_{\rm min}^{2}$ corresponds to the best fit model (red solid line)
in Figure~\ref{fig:zetori}. 
}
\end{figure}

\begin{figure}[ht]
\begin{center}
\includegraphics[angle=90.0,width=\textwidth]{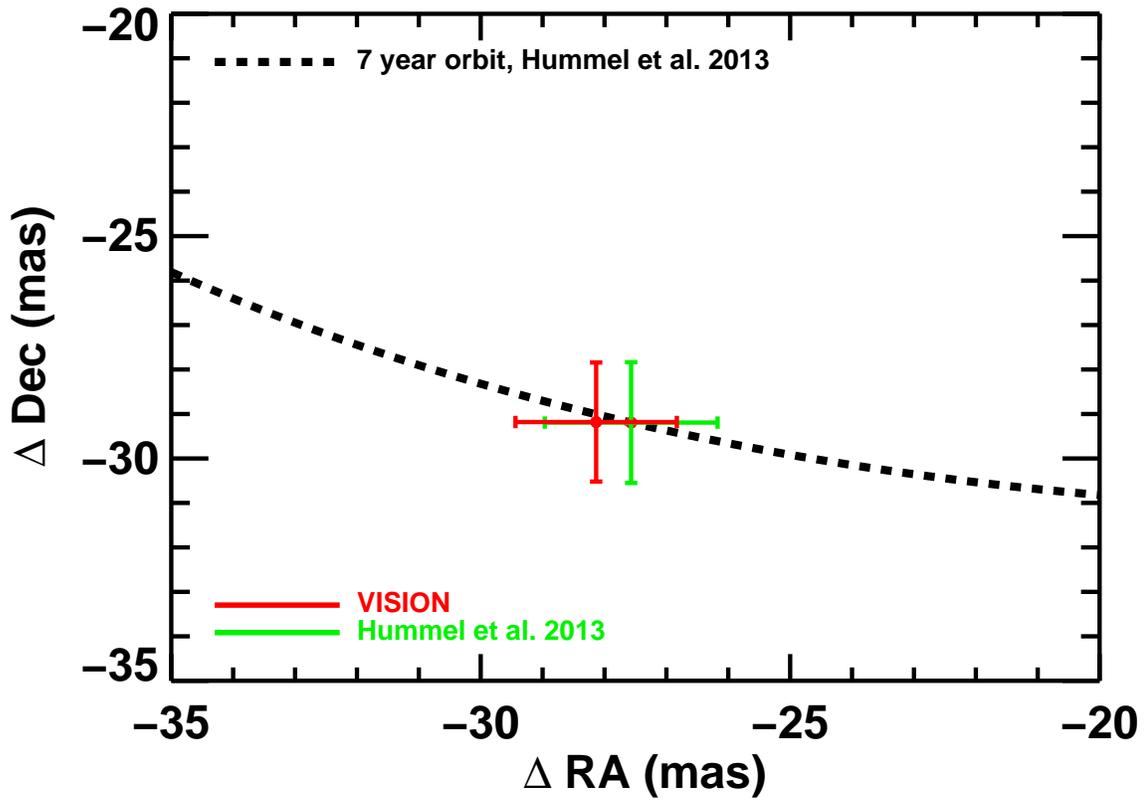}
\end{center}
\caption{\label{fig:orbit} 
The observed position angle and separation for the binary $\zeta$ Orionis A (red) match the known 
orbit \citet{Hummel13} (black line). The errors on the predicted 
$\Delta$RA and $\Delta$Dec (green) was calculated 
from the error on the orbital elements from \citet{Hummel13}.}
\end{figure}

The new observations of $\zeta$ Orionis A with VISION yield a best-fit separation of $\zetasep$~mas, 
a position angle of $\zetapa^{\circ}$, and flux ratio of $\zetafratio$ mag at $580-750$ nm and are 
listed in Table~\ref{table:orbit}. 
A sample of the extracted squared visibilities and closure phases along 
with the best fit model are shown in Figure~\ref{fig:zetori}. The median 
error on the closure phase is $\cphaseerr^{\circ}$ and 
on $V^{2}$ is $\vsqrerr$\%. The error increases towards redder 
wavelengths due to decreased SNR, which is in part due to a decrease in 
quantum efficiency of $83\%$ to $73\%$ from 600 to 750 nm. 
The formal $1\sigma$ 
errors were plotted on the fitted orbit and flux ratio using $1\sigma$ confidence 
intervals with $\Delta\chi^{2}=\chi^{2}-\chi_{\rm min}^{2}=3.53$, which 
corresponds to $1\sigma$ for 3 parameters of interested \citep{NR}. 
The $1\sigma$,  $2\sigma$, and $3\sigma$ confidence intervals for the fitted orbit to $\zeta$ Orionis are shown in Figure~\ref{fig:cont}. 
Finally, as shown in Figure \ref{fig:orbit},
there is excellent agreement between the observed separation and position angle for $\zeta$ Orionis A as observed by VISION versus that predicted by the previously published orbit \citep{Hummel13}.

\section{Conclusions \& Future Work\label{sec:conclude}}

This paper introduces the VISION beam combiner for NPOI: a six-telescope
image plane combiner featuring optical fibers, electron multiplying CCDs, and real-time 
photometric monitoring of each beam for visibility calibration. 
The VISION cameras, the fringe crosstalk, and the system visibility have been characterized, and a version of 
the MIRC data-processing pipeline has been adapted and verified for VISION with an observation of the benchmark astrometric binary star $\zeta$ Orionis A. 

Future work on the instrument includes installation of new N{\"u}v{\"u} cameras with $10$ times less clock-induced charge noise.  Recently, the 
750 mm fringe-forming lens was replaced with a 500 mm lens to fully sample the Gaussian profile on the fringing camera, with early indications showing 
a gain of $30\%$ flux, as well as a significant reduction in fringe crosstalk. 
Finally, the control system code will be updated from a text user interface to a graphical user interface. 

With the capabilities demonstrated here, 
we anticipate now being able to use VISION to make the first 5- or 6-telescope 
reconstructed images at visible wavelengths of the main sequence stars Altair and Vega, 
as well observations of triple star systems and the TiO lines on the surfaces of spotted red supergiant stars. 

\acknowledgments 
EVG would like to acknowledge the gracious support of his Lowell Pre-doctoral Fellowship by the BF foundation. This work was supported by NSF-AST 0958267 ``MRI-R$^2$ Consortium:  Development of VISION: The Next Generation Science Camera for the Navy Prototype Optical Interferometer''.  We gratefully acknowledge the VISION instrument PI team at TSU, led by Matthew Muterspaugh. This research has made use of the SIMBAD database, operated at CDS, Strasbourg, France. The work done with the NPOI was performed through a collaboration between the Naval Research Lab and the US Naval Observatory, in association with Lowell Observatory, and was funded by the Office of Naval Research and the Oceanographer of the Navy. 

\bibliography{bib}

\begin{thebibliography}{91}
\providecommand{\natexlab}[1]{#1}
\providecommand{\url}[1]{\texttt{#1}}
\expandafter\ifx\csname urlstyle\endcsname\relax
  \providecommand{\doi}[1]{doi: #1}\else
  \providecommand{\doi}{doi: \begingroup \urlstyle{rm}\Url}\fi

\bibitem[{Armstrong} et~al.(1998{\natexlab{a}}){Armstrong}, {Mozurkewich},
  {Pauls}, and {Hajian}]{Arm98spie}
J.~T. {Armstrong}, D.~{Mozurkewich}, T.~A. {Pauls}, and A.~R. {Hajian}.
\newblock {Bootstrapping the NPOI: keeping long baselines in phase by tracking
  fringes on short baselines}.
\newblock In R.~D. {Reasenberg}, editor, \emph{Astronomical Interferometry},
  volume 3350 of \emph{Society of Photo-Optical Instrumentation Engineers
  (SPIE) Conference Series}, pages 461--466, July 1998{\natexlab{a}}.

\bibitem[{Armstrong} et~al.(1998{\natexlab{b}}){Armstrong}, {Mozurkewich},
  {Rickard}, {Hutter}, {Benson}, {Bowers}, {Elias}, {Hummel}, {Johnston},
  {Buscher}, {Clark}, {Ha}, {Ling}, {White}, and {Simon}]{Arm98}
J.~T. {Armstrong}, D.~{Mozurkewich}, L.~J. {Rickard}, D.~J. {Hutter}, J.~A.
  {Benson}, P.~F. {Bowers}, N.~M. {Elias}, II, C.~A. {Hummel}, K.~J.
  {Johnston}, D.~F. {Buscher}, J.~H. {Clark}, III, L.~{Ha}, L.-C. {Ling}, N.~M.
  {White}, and R.~S. {Simon}.
\newblock {The Navy Prototype Optical Interferometer}.
\newblock \emph{\apj}, 496:\penalty0 550--571, Mar. 1998{\natexlab{b}}.
\newblock \doi{10.1086/305365}.

\bibitem[{Armstrong} et~al.(2001){Armstrong}, {Nordgren}, {Germain}, {Hajian},
  {Hindsley}, {Hummel}, {Mozurkewich}, and {Thessin}]{Arm01}
J.~T. {Armstrong}, T.~E. {Nordgren}, M.~E. {Germain}, A.~R. {Hajian}, R.~B.
  {Hindsley}, C.~A. {Hummel}, D.~{Mozurkewich}, and R.~N. {Thessin}.
\newblock {Diameters of {$\delta$} Cephei and {$\eta$} Aquilae Measured with
  the Navy Prototype Optical Interferometer}.
\newblock \emph{\aj}, 121:\penalty0 476--481, Jan. 2001.
\newblock \doi{10.1086/318007}.

\bibitem[{Armstrong} et~al.(2012){Armstrong}, {Jorgensen}, {Neilson},
  {Mozurkewich}, {Baines}, and {Schmitt}]{Arm12}
J.~T. {Armstrong}, A.~M. {Jorgensen}, H.~R. {Neilson}, D.~{Mozurkewich}, E.~K.
  {Baines}, and H.~R. {Schmitt}.
\newblock {Precise stellar diameters from coherently averaged visibilities}.
\newblock In \emph{Society of Photo-Optical Instrumentation Engineers (SPIE)
  Conference Series}, volume 8445 of \emph{Society of Photo-Optical
  Instrumentation Engineers (SPIE) Conference Series}, page~3, July 2012.
\newblock \doi{10.1117/12.926508}.

\bibitem[{Baines} et~al.(2013){Baines}, {Armstrong}, and {van Belle}]{Baines13}
E.~K. {Baines}, J.~T. {Armstrong}, and G.~T. {van Belle}.
\newblock {Navy Precision Optical Interferometer Observations of the Exoplanet
  Host {$\kappa$} Coronae Borealis and Their Implications for the Star's and
  Planet's Masses and Ages}.
\newblock \emph{\apjl}, 771:\penalty0 L17, July 2013.
\newblock \doi{10.1088/2041-8205/771/1/L17}.

\bibitem[{Baines} et~al.(2014){Baines}, {Armstrong}, {Schmitt}, {Benson},
  {Zavala}, and {van Belle}]{Baines14}
E.~K. {Baines}, J.~T. {Armstrong}, H.~R. {Schmitt}, J.~A. {Benson}, R.~T.
  {Zavala}, and G.~T. {van Belle}.
\newblock {Navy Precision Optical Interferometer Measurements of 10 Stellar
  Oscillators}.
\newblock \emph{\apj}, 781:\penalty0 90, Feb. 2014.
\newblock \doi{10.1088/0004-637X/781/2/90}.

\bibitem[{Baron} et~al.(2014){Baron}, {Monnier}, {Kiss}, {Neilson}, {Zhao},
  {Anderson}, {Aarnio}, {Pedretti}, {Thureau}, {ten Brummelaar}, {Ridgway},
  {McAlister}, {Sturmann}, {Sturmann}, and {Turner}]{Baron14}
F.~{Baron}, J.~D. {Monnier}, L.~L. {Kiss}, H.~R. {Neilson}, M.~{Zhao},
  M.~{Anderson}, A.~{Aarnio}, E.~{Pedretti}, N.~{Thureau}, T.~A. {ten
  Brummelaar}, S.~T. {Ridgway}, H.~A. {McAlister}, J.~{Sturmann},
  L.~{Sturmann}, and N.~{Turner}.
\newblock {CHARA/MIRC Observations of Two M Supergiants in Perseus OB1:
  Temperature, Bayesian Modeling, and Compressed Sensing Imaging}.
\newblock \emph{\apj}, 785:\penalty0 46, Apr. 2014.
\newblock \doi{10.1088/0004-637X/785/1/46}.

\bibitem[{Basden} and {Haniff}(2004)]{Basden04}
A.~G. {Basden} and C.~A. {Haniff}.
\newblock {Low light level CCDs and visibility parameter estimation}.
\newblock \emph{\mnras}, 347:\penalty0 1187--1197, Feb. 2004.
\newblock \doi{10.1111/j.1365-2966.2004.07283.x}.

\bibitem[{Bazot} et~al.(2011){Bazot}, {Ireland}, {Huber}, {Bedding},
  {Broomhall}, {Campante}, {Carfantan}, {Chaplin}, {Elsworth}, {Mel{\'e}ndez},
  {Petit}, {Th{\'e}ado}, {Van Grootel}, {Arentoft}, {Asplund}, {Castro},
  {Christensen-Dalsgaard}, {Do Nascimento}, {Dintrans}, {Dumusque}, {Kjeldsen},
  {McAlister}, {Metcalfe}, {Monteiro}, {Santos}, {Sousa}, {Sturmann},
  {Sturmann}, {ten Brummelaar}, {Turner}, and {Vauclair}]{Bazot11}
M.~{Bazot}, M.~J. {Ireland}, D.~{Huber}, T.~R. {Bedding}, A.-M. {Broomhall},
  T.~L. {Campante}, H.~{Carfantan}, W.~J. {Chaplin}, Y.~{Elsworth},
  J.~{Mel{\'e}ndez}, P.~{Petit}, S.~{Th{\'e}ado}, V.~{Van Grootel},
  T.~{Arentoft}, M.~{Asplund}, M.~{Castro}, J.~{Christensen-Dalsgaard}, J.~D.
  {Do Nascimento}, B.~{Dintrans}, X.~{Dumusque}, H.~{Kjeldsen}, H.~A.
  {McAlister}, T.~S. {Metcalfe}, M.~J.~P.~F.~G. {Monteiro}, N.~C. {Santos},
  S.~{Sousa}, J.~{Sturmann}, L.~{Sturmann}, T.~A. {ten Brummelaar},
  N.~{Turner}, and S.~{Vauclair}.
\newblock {The radius and mass of the close solar twin 18 Scorpii derived from
  asteroseismology and interferometry}.
\newblock \emph{\aap}, 526:\penalty0 L4, Feb. 2011.
\newblock \doi{10.1051/0004-6361/201015679}.

\bibitem[{Berger} et~al.(2003){Berger}, {Haguenauer}, {Kern},
  {Rousselet-Perraut}, {Malbet}, {Gluck}, {Lagny}, {Schanen-Duport}, {Laurent},
  {Delboulbe}, {Tatulli}, {Traub}, {Carleton}, {Millan-Gabet}, {Monnier},
  {Pedretti}, and {Ragland}]{Berger03}
J.-P. {Berger}, P.~{Haguenauer}, P.~Y. {Kern}, K.~{Rousselet-Perraut},
  F.~{Malbet}, S.~{Gluck}, L.~{Lagny}, I.~{Schanen-Duport}, E.~{Laurent},
  A.~{Delboulbe}, E.~{Tatulli}, W.~A. {Traub}, N.~{Carleton},
  R.~{Millan-Gabet}, J.~D. {Monnier}, E.~{Pedretti}, and S.~{Ragland}.
\newblock {An integrated-optics 3-way beam combiner for IOTA}.
\newblock In W.~A. {Traub}, editor, \emph{Interferometry for Optical Astronomy
  II}, volume 4838 of \emph{Society of Photo-Optical Instrumentation Engineers
  (SPIE) Conference Series}, pages 1099--1106, Feb. 2003.

\bibitem[{Born} and {Wolf}(1999)]{Born1999}
M.~{Born} and E.~{Wolf}.
\newblock \emph{{Principles of optics : electromagnetic theory of propagation,
  interference and diffraction of light}}.
\newblock 1999.

\bibitem[{Challouf} et~al.(2014){Challouf}, {Nardetto}, {Mourard}, {Graczyk},
  {Aroui}, {Chesneau}, {Delaa}, {Pietrzy{\'n}ski}, {Gieren}, {Ligi},
  {Meilland}, {Perraut}, {Tallon-Bosc}, {McAlister}, {ten Brummelaar},
  {Sturmann}, {Sturmann}, {Turner}, {Farrington}, {Vargas}, and
  {Scott}]{Challouf2014}
M.~{Challouf}, N.~{Nardetto}, D.~{Mourard}, D.~{Graczyk}, H.~{Aroui},
  O.~{Chesneau}, O.~{Delaa}, G.~{Pietrzy{\'n}ski}, W.~{Gieren}, R.~{Ligi},
  A.~{Meilland}, K.~{Perraut}, I.~{Tallon-Bosc}, H.~{McAlister}, T.~{ten
  Brummelaar}, J.~{Sturmann}, L.~{Sturmann}, N.~{Turner}, C.~{Farrington},
  N.~{Vargas}, and N.~{Scott}.
\newblock {Improving the surface brightness-color relation for early-type stars
  using optical interferometry}.
\newblock \emph{\aap}, 570:\penalty0 A104, Oct. 2014.
\newblock \doi{10.1051/0004-6361/201423772}.

\bibitem[{Che} et~al.(2011){Che}, {Monnier}, {Zhao}, {Pedretti}, {Thureau},
  {M{\'e}rand}, {ten Brummelaar}, {McAlister}, {Ridgway}, {Turner}, {Sturmann},
  and {Sturmann}]{Che11}
X.~{Che}, J.~D. {Monnier}, M.~{Zhao}, E.~{Pedretti}, N.~{Thureau},
  A.~{M{\'e}rand}, T.~{ten Brummelaar}, H.~{McAlister}, S.~T. {Ridgway},
  N.~{Turner}, J.~{Sturmann}, and L.~{Sturmann}.
\newblock {Colder and Hotter: Interferometric Imaging of {$\beta$} Cassiopeiae
  and {$\alpha$} Leonis}.
\newblock \emph{\apj}, 732:\penalty0 68, May 2011.
\newblock \doi{10.1088/0004-637X/732/2/68}.

\bibitem[{Chiavassa} et~al.(2009){Chiavassa}, {Plez}, {Josselin}, and
  {Freytag}]{Chiavassa09}
A.~{Chiavassa}, B.~{Plez}, E.~{Josselin}, and B.~{Freytag}.
\newblock {Radiative hydrodynamics simulations of red supergiant stars. I.
  interpretation of interferometric observations}.
\newblock \emph{\aap}, 506:\penalty0 1351--1365, Nov. 2009.
\newblock \doi{10.1051/0004-6361/200911780}.

\bibitem[{Chiavassa} et~al.(2010){Chiavassa}, {Haubois}, {Young}, {Plez},
  {Josselin}, {Perrin}, and {Freytag}]{Chiavassa10}
A.~{Chiavassa}, X.~{Haubois}, J.~S. {Young}, B.~{Plez}, E.~{Josselin},
  G.~{Perrin}, and B.~{Freytag}.
\newblock {Radiative hydrodynamics simulations of red supergiant stars. II.
  Simulations of convection on Betelgeuse match interferometric observations}.
\newblock \emph{\aap}, 515:\penalty0 A12, June 2010.
\newblock \doi{10.1051/0004-6361/200913907}.

\bibitem[{Chiavassa} et~al.(2011){Chiavassa}, {Pasquato}, {Jorissen}, {Sacuto},
  {Babusiaux}, {Freytag}, {Ludwig}, {Cruzal{\`e}bes}, {Rabbia}, {Spang}, and
  {Chesneau}]{Chiavassa11}
A.~{Chiavassa}, E.~{Pasquato}, A.~{Jorissen}, S.~{Sacuto}, C.~{Babusiaux},
  B.~{Freytag}, H.-G. {Ludwig}, P.~{Cruzal{\`e}bes}, Y.~{Rabbia}, A.~{Spang},
  and O.~{Chesneau}.
\newblock {Radiative hydrodynamic simulations of red supergiant stars. III.
  Spectro-photocentric variability, photometric variability, and consequences
  on Gaia measurements}.
\newblock \emph{\aap}, 528:\penalty0 A120, Apr. 2011.
\newblock \doi{10.1051/0004-6361/201015768}.

\bibitem[{Chiavassa} et~al.(2012){Chiavassa}, {Bigot}, {Kervella}, {Matter},
  {Lopez}, {Collet}, {Magic}, and {Asplund}]{Chiavassa12}
A.~{Chiavassa}, L.~{Bigot}, P.~{Kervella}, A.~{Matter}, B.~{Lopez},
  R.~{Collet}, Z.~{Magic}, and M.~{Asplund}.
\newblock {Three-dimensional interferometric, spectrometric, and planetary
  views of Procyon}.
\newblock \emph{\aap}, 540:\penalty0 A5, Apr. 2012.
\newblock \doi{10.1051/0004-6361/201118652}.

\bibitem[{Coud{\'e} du Foresto} et~al.(1997){Coud{\'e} du Foresto}, {Ridgway},
  and {Mariotti}]{Coude97}
V.~{Coud{\'e} du Foresto}, S.~{Ridgway}, and J.-M. {Mariotti}.
\newblock {Deriving object visibilities from interferograms obtained with a
  fiber stellar interferometer}.
\newblock \emph{\aaps}, 121:\penalty0 379--392, Feb. 1997.
\newblock \doi{10.1051/aas:1997290}.

\bibitem[{Davis} et~al.(1999{\natexlab{a}}){Davis}, {Tango}, {Booth}, {ten
  Brummelaar}, {Minard}, and {Owens}]{SUSIA}
J.~{Davis}, W.~J. {Tango}, A.~J. {Booth}, T.~A. {ten Brummelaar}, R.~A.
  {Minard}, and S.~M. {Owens}.
\newblock {The Sydney University Stellar Interferometer - I. The instrument}.
\newblock \emph{\mnras}, 303:\penalty0 773--782, Mar. 1999{\natexlab{a}}.
\newblock \doi{10.1046/j.1365-8711.1999.02269.x}.

\bibitem[{Davis} et~al.(1999{\natexlab{b}}){Davis}, {Tango}, {Booth},
  {Thorvaldson}, and {Giovannis}]{SUSIB}
J.~{Davis}, W.~J. {Tango}, A.~J. {Booth}, E.~D. {Thorvaldson}, and
  J.~{Giovannis}.
\newblock {The Sydney University Stellar Interferometer - II. Commissioning
  observations and results}.
\newblock \emph{\mnras}, 303:\penalty0 783--791, Mar. 1999{\natexlab{b}}.
\newblock \doi{10.1046/j.1365-8711.1999.02270.x}.

\bibitem[{Domiciano de Souza} et~al.(2003){Domiciano de Souza}, {Kervella},
  {Jankov}, {Abe}, {Vakili}, {di Folco}, and {Paresce}]{DeSouza03}
A.~{Domiciano de Souza}, P.~{Kervella}, S.~{Jankov}, L.~{Abe}, F.~{Vakili},
  E.~{di Folco}, and F.~{Paresce}.
\newblock {The spinning-top Be star Achernar from VLTI-VINCI}.
\newblock \emph{\aap}, 407:\penalty0 L47--L50, Aug. 2003.
\newblock \doi{10.1051/0004-6361:20030786}.

\bibitem[{Domiciano de Souza} et~al.(2012){Domiciano de Souza}, {Hadjara},
  {Vakili}, {Bendjoya}, {Millour}, {Abe}, {Carciofi}, {Faes}, {Kervella},
  {Lagarde}, {Marconi}, {Monin}, {Niccolini}, {Petrov}, and
  {Weigelt}]{DeSouza12}
A.~{Domiciano de Souza}, M.~{Hadjara}, F.~{Vakili}, P.~{Bendjoya},
  F.~{Millour}, L.~{Abe}, A.~C. {Carciofi}, D.~M. {Faes}, P.~{Kervella},
  S.~{Lagarde}, A.~{Marconi}, J.-L. {Monin}, G.~{Niccolini}, R.~G. {Petrov},
  and G.~{Weigelt}.
\newblock {Beyond the diffraction limit of optical/IR interferometers. I.
  Angular diameter and rotation parameters of Achernar from differential
  phases}.
\newblock \emph{\aap}, 545:\penalty0 A130, Sept. 2012.
\newblock \doi{10.1051/0004-6361/201218782}.

\bibitem[{Ducati}(2002)]{Ducati02}
J.~R. {Ducati}.
\newblock {VizieR Online Data Catalog: Catalogue of Stellar Photometry in
  Johnson's 11-color system.}
\newblock \emph{VizieR Online Data Catalog}, 2237:\penalty0 0, 2002.

\bibitem[{Espinosa Lara} and {Rieutord}(2011)]{Esp11}
F.~{Espinosa Lara} and M.~{Rieutord}.
\newblock {Gravity darkening in rotating stars}.
\newblock \emph{\aap}, 533:\penalty0 A43, Sept. 2011.
\newblock \doi{10.1051/0004-6361/201117252}.

\bibitem[{Espinosa Lara} and {Rieutord}(2013)]{Esp13}
F.~{Espinosa Lara} and M.~{Rieutord}.
\newblock {Self-consistent 2D models of fast-rotating early-type stars}.
\newblock \emph{\aap}, 552:\penalty0 A35, Apr. 2013.
\newblock \doi{10.1051/0004-6361/201220844}.

\bibitem[{Freytag} and {H{\"o}fner}(2008)]{Freytag08}
B.~{Freytag} and S.~{H{\"o}fner}.
\newblock {Three-dimensional simulations of the atmosphere of an AGB star}.
\newblock \emph{\aap}, 483:\penalty0 571--583, May 2008.
\newblock \doi{10.1051/0004-6361:20078096}.

\bibitem[{Freytag} et~al.(2002){Freytag}, {Steffen}, and {Dorch}]{Freytag02}
B.~{Freytag}, M.~{Steffen}, and B.~{Dorch}.
\newblock {Spots on the surface of Betelgeuse -- Results from new 3D stellar
  convection models}.
\newblock \emph{Astronomische Nachrichten}, 323:\penalty0 213--219, July 2002.
\newblock \doi{10.1002/1521-3994(200208)323:3/4<213::AID-ASNA213>3.0.CO;2-H}.

\bibitem[Frigo and Johnson(2005)]{FFTW}
M.~Frigo and S.~G. Johnson.
\newblock The design and implementation of {FFTW3}.
\newblock \emph{Proceedings of the IEEE}, 93\penalty0 (2):\penalty0 216--231,
  2005.
\newblock Special issue on ``Program Generation, Optimization, and Platform
  Adaptation''.

\bibitem[{Ghasempour} et~al.(2012){Ghasempour}, {Muterspaugh}, {Hutter},
  {Monnier}, {Benson}, {Armstrong}, {Williamson}, {Fall}, {Harrison}, and
  {Sergeyous}]{Ghasempour12}
A.~{Ghasempour}, M.~W. {Muterspaugh}, D.~J. {Hutter}, J.~D. {Monnier}, J.~A.
  {Benson}, J.~T. {Armstrong}, M.~H. {Williamson}, S.~{Fall}, C.~{Harrison},
  and C.~{Sergeyous}.
\newblock {Building the next-generation science camera for the Navy Optical
  Interferometer}.
\newblock In \emph{Society of Photo-Optical Instrumentation Engineers (SPIE)
  Conference Series}, volume 8445 of \emph{Society of Photo-Optical
  Instrumentation Engineers (SPIE) Conference Series}, page~1, July 2012.
\newblock \doi{10.1117/12.924995}.

\bibitem[{Gordon} and {Buscher}(2012)]{Gordon12}
J.~A. {Gordon} and D.~F. {Buscher}.
\newblock {Detection noise bias and variance in the power spectrum and
  bispectrum in optical interferometry}.
\newblock \emph{\aap}, 541:\penalty0 A46, May 2012.
\newblock \doi{10.1051/0004-6361/201117335}.

\bibitem[{Harps{\o}e} et~al.(2012){Harps{\o}e}, {J{\o}rgensen}, {Andersen}, and
  {Grundahl}]{Harpsoe12a}
K.~B.~W. {Harps{\o}e}, U.~G. {J{\o}rgensen}, M.~I. {Andersen}, and
  F.~{Grundahl}.
\newblock {High frame rate imaging based photometry. Photometric reduction of
  data from electron-multiplying charge coupled devices (EMCCDs)}.
\newblock \emph{\aap}, 542:\penalty0 A23, June 2012.
\newblock \doi{10.1051/0004-6361/201219059}.

\bibitem[{Haubois} et~al.(2009){Haubois}, {Perrin}, {Lacour}, {Verhoelst},
  {Meimon}, {Mugnier}, {Thi{\'e}baut}, {Berger}, {Ridgway}, {Monnier},
  {Millan-Gabet}, and {Traub}]{Haubois09}
X.~{Haubois}, G.~{Perrin}, S.~{Lacour}, T.~{Verhoelst}, S.~{Meimon},
  L.~{Mugnier}, E.~{Thi{\'e}baut}, J.~P. {Berger}, S.~T. {Ridgway}, J.~D.
  {Monnier}, R.~{Millan-Gabet}, and W.~{Traub}.
\newblock {Imaging the spotty surface of <ASTROBJ>Betelgeuse</ASTROBJ> in the H
  band}.
\newblock \emph{\aap}, 508:\penalty0 923--932, Dec. 2009.
\newblock \doi{10.1051/0004-6361/200912927}.

\bibitem[{Huber} et~al.(2012{\natexlab{a}}){Huber}, {Ireland}, {Bedding},
  {Brand{\~a}o}, {Piau}, {Maestro}, {White}, {Bruntt}, {Casagrande},
  {Molenda-{\.Z}akowicz}, {Silva Aguirre}, {Sousa}, {Barclay}, {Burke},
  {Chaplin}, {Christensen-Dalsgaard}, {Cunha}, {De Ridder}, {Farrington},
  {Frasca}, {Garc{\'{\i}}a}, {Gilliland}, {Goldfinger}, {Hekker}, {Kawaler},
  {Kjeldsen}, {McAlister}, {Metcalfe}, {Miglio}, {Monteiro}, {Pinsonneault},
  {Schaefer}, {Stello}, {Stumpe}, {Sturmann}, {Sturmann}, {ten Brummelaar},
  {Thompson}, {Turner}, and {Uytterhoeven}]{Huber12b}
D.~{Huber}, M.~J. {Ireland}, T.~R. {Bedding}, I.~M. {Brand{\~a}o}, L.~{Piau},
  V.~{Maestro}, T.~R. {White}, H.~{Bruntt}, L.~{Casagrande},
  J.~{Molenda-{\.Z}akowicz}, V.~{Silva Aguirre}, S.~G. {Sousa}, T.~{Barclay},
  C.~J. {Burke}, W.~J. {Chaplin}, J.~{Christensen-Dalsgaard}, M.~S. {Cunha},
  J.~{De Ridder}, C.~D. {Farrington}, A.~{Frasca}, R.~A. {Garc{\'{\i}}a}, R.~L.
  {Gilliland}, P.~J. {Goldfinger}, S.~{Hekker}, S.~D. {Kawaler}, H.~{Kjeldsen},
  H.~A. {McAlister}, T.~S. {Metcalfe}, A.~{Miglio}, M.~J.~P.~F.~G. {Monteiro},
  M.~H. {Pinsonneault}, G.~H. {Schaefer}, D.~{Stello}, M.~C. {Stumpe},
  J.~{Sturmann}, L.~{Sturmann}, T.~A. {ten Brummelaar}, M.~J. {Thompson},
  N.~{Turner}, and K.~{Uytterhoeven}.
\newblock {Fundamental Properties of Stars Using Asteroseismology from Kepler
  and CoRoT and Interferometry from the CHARA Array}.
\newblock \emph{\apj}, 760:\penalty0 32, Nov. 2012{\natexlab{a}}.
\newblock \doi{10.1088/0004-637X/760/1/32}.

\bibitem[{Huber} et~al.(2012{\natexlab{b}}){Huber}, {Ireland}, {Bedding},
  {Howell}, {Maestro}, {M{\'e}rand}, {Tuthill}, {White}, {Farrington},
  {Goldfinger}, {McAlister}, {Schaefer}, {Sturmann}, {Sturmann}, {ten
  Brummelaar}, and {Turner}]{Huber12a}
D.~{Huber}, M.~J. {Ireland}, T.~R. {Bedding}, S.~B. {Howell}, V.~{Maestro},
  A.~{M{\'e}rand}, P.~G. {Tuthill}, T.~R. {White}, C.~D. {Farrington}, P.~J.
  {Goldfinger}, H.~A. {McAlister}, G.~H. {Schaefer}, J.~{Sturmann},
  L.~{Sturmann}, T.~A. {ten Brummelaar}, and N.~H. {Turner}.
\newblock {Validation of the exoplanet Kepler-21b using PAVO/CHARA
  long-baseline interferometry}.
\newblock \emph{\mnras}, 423:\penalty0 L16--L20, June 2012{\natexlab{b}}.
\newblock \doi{10.1111/j.1745-3933.2012.01242.x}.

\bibitem[{Hummel} et~al.(1998){Hummel}, {Mozurkewich}, {Armstrong}, {Hajian},
  {Elias}, and {Hutter}]{Hummel98}
C.~A. {Hummel}, D.~{Mozurkewich}, J.~T. {Armstrong}, A.~R. {Hajian}, N.~M.
  {Elias}, II, and D.~J. {Hutter}.
\newblock {Navy Prototype Optical Interferometer Observations of the Double
  Stars Mizar A and Matar}.
\newblock \emph{\aj}, 116:\penalty0 2536--2548, Nov. 1998.
\newblock \doi{10.1086/300602}.

\bibitem[{Hummel} et~al.(2001){Hummel}, {Carquillat}, {Ginestet}, {Griffin},
  {Boden}, {Hajian}, {Mozurkewich}, and {Nordgren}]{Hummel01}
C.~A. {Hummel}, J.-M. {Carquillat}, N.~{Ginestet}, R.~F. {Griffin}, A.~F.
  {Boden}, A.~R. {Hajian}, D.~{Mozurkewich}, and T.~E. {Nordgren}.
\newblock {Orbital and Stellar Parameters of Omicron Leonis from Spectroscopy
  and Interferometry}.
\newblock \emph{\aj}, 121:\penalty0 1623--1635, Mar. 2001.
\newblock \doi{10.1086/319391}.

\bibitem[{Hummel} et~al.(2003){Hummel}, {Benson}, {Hutter}, {Johnston},
  {Mozurkewich}, {Armstrong}, {Hindsley}, {Gilbreath}, {Rickard}, and
  {White}]{Hummel03}
C.~A. {Hummel}, J.~A. {Benson}, D.~J. {Hutter}, K.~J. {Johnston},
  D.~{Mozurkewich}, J.~T. {Armstrong}, R.~B. {Hindsley}, G.~C. {Gilbreath},
  L.~J. {Rickard}, and N.~M. {White}.
\newblock {First Observations with a Co-phased Six-Station Optical
  Long-Baseline Array: Application to the Triple Star {$\eta$} Virginis}.
\newblock \emph{\aj}, 125:\penalty0 2630--2644, May 2003.
\newblock \doi{10.1086/374572}.

\bibitem[{Hummel} et~al.(2013){Hummel}, {Rivinius}, {Nieva}, {Stahl}, {van
  Belle}, and {Zavala}]{Hummel13}
C.~A. {Hummel}, T.~{Rivinius}, M.-F. {Nieva}, O.~{Stahl}, G.~{van Belle}, and
  R.~T. {Zavala}.
\newblock {Dynamical mass of the O-type supergiant in {$\zeta$} Orionis A}.
\newblock \emph{\aap}, 554:\penalty0 A52, June 2013.
\newblock \doi{10.1051/0004-6361/201321434}.

\bibitem[{Ireland} et~al.(2008){Ireland}, {M{\'e}rand}, {ten Brummelaar},
  {Tuthill}, {Schaefer}, {Turner}, {Sturmann}, {Sturmann}, and
  {McAlister}]{PAVO08}
M.~J. {Ireland}, A.~{M{\'e}rand}, T.~A. {ten Brummelaar}, P.~G. {Tuthill},
  G.~H. {Schaefer}, N.~H. {Turner}, J.~{Sturmann}, L.~{Sturmann}, and H.~A.
  {McAlister}.
\newblock {Sensitive visible interferometry with PAVO}.
\newblock In \emph{Society of Photo-Optical Instrumentation Engineers (SPIE)
  Conference Series}, volume 7013 of \emph{Society of Photo-Optical
  Instrumentation Engineers (SPIE) Conference Series}, page~24, July 2008.
\newblock \doi{10.1117/12.788386}.

\bibitem[{Jamialahmadi} et~al.(2015){Jamialahmadi}, {Berio}, {Meilland},
  {Perraut}, {Mourard}, {Lopez}, {Stee}, {Nardetto}, {Pichon}, {Clausse},
  {Spang}, {McAlister}, {ten Brummelaar}, {Sturmann}, {Sturmann}, {Turner},
  {Farrington}, {Vargas}, and {Scott}]{Jamial15}
N.~{Jamialahmadi}, P.~{Berio}, A.~{Meilland}, K.~{Perraut}, D.~{Mourard},
  B.~{Lopez}, P.~{Stee}, N.~{Nardetto}, B.~{Pichon}, J.~M. {Clausse},
  A.~{Spang}, H.~{McAlister}, T.~{ten Brummelaar}, J.~{Sturmann},
  L.~{Sturmann}, N.~{Turner}, C.~{Farrington}, N.~{Vargas}, and N.~{Scott}.
\newblock {The peculiar fast-rotating star 51 Ophiuchi probed by VEGA/CHARA}.
\newblock \emph{\aap}, 579:\penalty0 A81, July 2015.
\newblock \doi{10.1051/0004-6361/201425473}.

\bibitem[{Jorgensen} et~al.(2006){Jorgensen}, {Mozurkewich}, {Schmitt},
  {Armstrong}, {Gilbreath}, {Hindsley}, {Pauls}, and {Peterson}]{Jorg06}
A.~M. {Jorgensen}, D.~{Mozurkewich}, H.~{Schmitt}, J.~T. {Armstrong}, G.~C.
  {Gilbreath}, R.~{Hindsley}, T.~A. {Pauls}, and D.~M. {Peterson}.
\newblock {Coherent integrations, fringe modeling, and bootstrapping with the
  NPOI}.
\newblock In \emph{Society of Photo-Optical Instrumentation Engineers (SPIE)
  Conference Series}, volume 6268 of \emph{Society of Photo-Optical
  Instrumentation Engineers (SPIE) Conference Series}, page~1, June 2006.
\newblock \doi{10.1117/12.672829}.

\bibitem[{Jorgensen} et~al.(2014){Jorgensen}, {Mozurkewich}, and
  {Hutter}]{Jorg14}
A.~M. {Jorgensen}, D.~{Mozurkewich}, and D.~{Hutter}.
\newblock {Extremely Precise Measurements of Stellar Diameters and Binary Stars
  with Coherent Integration}.
\newblock In M.~J. {Creech-Eakman}, J.~A. {Guzik}, and R.~E. {Stencel},
  editors, \emph{Resolving The Future Of Astronomy With Long-Baseline
  Interferometry}, volume 487 of \emph{Astronomical Society of the Pacific
  Conference Series}, page 303, Sept. 2014.

\bibitem[{Kloppenborg} et~al.(2015){Kloppenborg}, {Stencel}, {Monnier},
  {Schaefer}, {Baron}, {Tycner}, {Zavala}, {Hutter}, {Zhao}, {Che}, {ten
  Brummelaar}, {Farrington}, {Parks}, {McAlister}, {Sturmann}, {Sturmann},
  {Sallave-Goldfinger}, {Turner}, {Pedretti}, and {Thureau}]{Klop15}
B.~{Kloppenborg}, R.~{Stencel}, J.~D. {Monnier}, G.~{Schaefer}, F.~{Baron},
  C.~{Tycner}, R.~T. {Zavala}, D.~{Hutter}, M.~{Zhao}, X.~{Che}, T.~{ten
  Brummelaar}, C.~{Farrington}, R.~{Parks}, H.~{McAlister}, J.~{Sturmann},
  L.~{Sturmann}, P.~J. {Sallave-Goldfinger}, N.~{Turner}, E.~{Pedretti}, and
  N.~{Thureau}.
\newblock {Interferometry of \$$\backslash$epsilon\$ Aurigae: Characterization
  of the asymmetric eclipsing disk}.
\newblock \emph{ArXiv e-prints}, Aug. 2015.

\bibitem[{Kok} et~al.(2012){Kok}, {Ireland}, {Tuthill}, {Robertson},
  {Warrington}, and {Tango}]{Kok12}
Y.~{Kok}, M.~J. {Ireland}, P.~G. {Tuthill}, J.~G. {Robertson}, B.~A.
  {Warrington}, and W.~J. {Tango}.
\newblock {Self-phase-referencing interferometry with SUSI}.
\newblock In \emph{Society of Photo-Optical Instrumentation Engineers (SPIE)
  Conference Series}, volume 8445 of \emph{Society of Photo-Optical
  Instrumentation Engineers (SPIE) Conference Series}, page~21, July 2012.
\newblock \doi{10.1117/12.925238}.

\bibitem[{Landavazo} et~al.(2014){Landavazo}, {Jorgensen}, {Sun}, {Newman},
  {Mozurkewich}, {van Belle}, {Hutter}, {Schmitt}, {Armstrong}, {Baines}, and
  {Restaino}]{Landavazo14}
M.~I. {Landavazo}, A.~M. {Jorgensen}, B.~{Sun}, K.~{Newman}, D.~{Mozurkewich},
  G.~T. {van Belle}, D.~J. {Hutter}, H.~R. {Schmitt}, J.~T. {Armstrong}, E.~K.
  {Baines}, and S.~R. {Restaino}.
\newblock {6-station, 5-baseline fringe tracking with the new classic data
  acquisition system at the Navy Precision Optical Interferometer}.
\newblock In \emph{Society of Photo-Optical Instrumentation Engineers (SPIE)
  Conference Series}, volume 9146 of \emph{Society of Photo-Optical
  Instrumentation Engineers (SPIE) Conference Series}, page~21, July 2014.
\newblock \doi{10.1117/12.2057278}.

\bibitem[{Maestro} et~al.(2012){Maestro}, {Kok}, {Huber}, {Ireland}, {Tuthill},
  {White}, {Schaefer}, {ten Brummelaar}, {McAlister}, {Turner}, {Farrington},
  and {Goldfinger}]{PAVO12}
V.~{Maestro}, Y.~{Kok}, D.~{Huber}, M.~J. {Ireland}, P.~G. {Tuthill},
  T.~{White}, G.~{Schaefer}, T.~A. {ten Brummelaar}, H.~A. {McAlister},
  N.~{Turner}, C.~D. {Farrington}, and P.~J. {Goldfinger}.
\newblock {Imaging rapid rotators with the PAVO beam combiner at CHARA}.
\newblock In \emph{Society of Photo-Optical Instrumentation Engineers (SPIE)
  Conference Series}, volume 8445 of \emph{Society of Photo-Optical
  Instrumentation Engineers (SPIE) Conference Series}, page~0, July 2012.
\newblock \doi{10.1117/12.926881}.

\bibitem[{Maestro} et~al.(2013){Maestro}, {Che}, {Huber}, {Ireland}, {Monnier},
  {White}, {Kok}, {Robertson}, {Schaefer}, {ten Brummelaar}, and
  {Tuthill}]{PAVO13}
V.~{Maestro}, X.~{Che}, D.~{Huber}, M.~J. {Ireland}, J.~D. {Monnier}, T.~R.
  {White}, Y.~{Kok}, J.~G. {Robertson}, G.~H. {Schaefer}, T.~A. {ten
  Brummelaar}, and P.~G. {Tuthill}.
\newblock {Optical interferometry of early-type stars with PAVO@CHARA - I.
  Fundamental stellar properties}.
\newblock \emph{\mnras}, 434:\penalty0 1321--1331, Sept. 2013.
\newblock \doi{10.1093/mnras/stt1092}.

\bibitem[{Markwardt}(2009)]{Markwardt09}
C.~B. {Markwardt}.
\newblock {Non-linear Least-squares Fitting in IDL with MPFIT}.
\newblock In D.~A. {Bohlender}, D.~{Durand}, and P.~{Dowler}, editors,
  \emph{Astronomical Data Analysis Software and Systems XVIII}, volume 411 of
  \emph{Astronomical Society of the Pacific Conference Series}, page 251, Sept.
  2009.

\bibitem[{Monnier} et~al.(2004){Monnier}, {Berger}, {Millan-Gabet}, and {ten
  Brummelaar}]{Monnier04}
J.~D. {Monnier}, J.-P. {Berger}, R.~{Millan-Gabet}, and T.~A. {ten Brummelaar}.
\newblock {The Michigan Infrared Combiner (MIRC): IR imaging with the CHARA
  Array}.
\newblock In W.~A. {Traub}, editor, \emph{New Frontiers in Stellar
  Interferometry}, volume 5491 of \emph{Society of Photo-Optical
  Instrumentation Engineers (SPIE) Conference Series}, page 1370, Oct. 2004.

\bibitem[{Monnier} et~al.(2006){Monnier}, {Pedretti}, {Thureau}, {Berger},
  {Millan-Gabet}, {ten Brummelaar}, {McAlister}, {Sturmann}, {Sturmann},
  {Muirhead}, {Tannirkulam}, {Webster}, and {Zhao}]{Monnier06}
J.~D. {Monnier}, E.~{Pedretti}, N.~{Thureau}, J.-P. {Berger},
  R.~{Millan-Gabet}, T.~{ten Brummelaar}, H.~{McAlister}, J.~{Sturmann},
  L.~{Sturmann}, P.~{Muirhead}, A.~{Tannirkulam}, S.~{Webster}, and M.~{Zhao}.
\newblock {Michigan Infrared Combiner (MIRC): commissioning results at the
  CHARA Array}.
\newblock In \emph{Society of Photo-Optical Instrumentation Engineers (SPIE)
  Conference Series}, volume 6268 of \emph{Society of Photo-Optical
  Instrumentation Engineers (SPIE) Conference Series}, page~1, June 2006.
\newblock \doi{10.1117/12.671982}.

\bibitem[{Monnier} et~al.(2007){Monnier}, {Zhao}, {Pedretti}, {Thureau},
  {Ireland}, {Muirhead}, {Berger}, {Millan-Gabet}, {Van Belle}, {ten
  Brummelaar}, {McAlister}, {Ridgway}, {Turner}, {Sturmann}, {Sturmann}, and
  {Berger}]{Monnier07}
J.~D. {Monnier}, M.~{Zhao}, E.~{Pedretti}, N.~{Thureau}, M.~{Ireland},
  P.~{Muirhead}, J.-P. {Berger}, R.~{Millan-Gabet}, G.~{Van Belle}, T.~{ten
  Brummelaar}, H.~{McAlister}, S.~{Ridgway}, N.~{Turner}, L.~{Sturmann},
  J.~{Sturmann}, and D.~{Berger}.
\newblock {Imaging the Surface of Altair}.
\newblock \emph{Science}, 317:\penalty0 342--, July 2007.
\newblock \doi{10.1126/science.1143205}.

\bibitem[{Monnier} et~al.(2012){Monnier}, {Che}, {Zhao}, {Ekstr{\"o}m},
  {Maestro}, {Aufdenberg}, {Baron}, {Georgy}, {Kraus}, {McAlister}, {Pedretti},
  {Ridgway}, {Sturmann}, {Sturmann}, {ten Brummelaar}, {Thureau}, {Turner}, and
  {Tuthill}]{Monnier12}
J.~D. {Monnier}, X.~{Che}, M.~{Zhao}, S.~{Ekstr{\"o}m}, V.~{Maestro},
  J.~{Aufdenberg}, F.~{Baron}, C.~{Georgy}, S.~{Kraus}, H.~{McAlister},
  E.~{Pedretti}, S.~{Ridgway}, J.~{Sturmann}, L.~{Sturmann}, T.~{ten
  Brummelaar}, N.~{Thureau}, N.~{Turner}, and P.~G. {Tuthill}.
\newblock {Resolving Vega and the Inclination Controversy with CHARA/MIRC}.
\newblock \emph{\apjl}, 761:\penalty0 L3, Dec. 2012.
\newblock \doi{10.1088/2041-8205/761/1/L3}.

\bibitem[{Mourard} et~al.(2008){Mourard}, {Perraut}, {Bonneau}, {Clausse},
  {Stee}, {Tallon-Bosc}, {Kervella}, {Hughes}, {Marcotto}, {Blazit},
  {Chesneau}, {Domiciano de Souza}, {Foy}, {H{\'e}nault}, {Mattei}, {Merlin},
  {Roussel}, {Tallon}, {Thiebaut}, {McAlister}, {ten Brummelaar}, {Sturmann},
  {Sturmann}, {Turner}, {Farrington}, and {Goldfinger}]{Vega08}
D.~{Mourard}, K.~{Perraut}, D.~{Bonneau}, J.~M. {Clausse}, P.~{Stee},
  I.~{Tallon-Bosc}, P.~{Kervella}, Y.~{Hughes}, A.~{Marcotto}, A.~{Blazit},
  O.~{Chesneau}, A.~{Domiciano de Souza}, R.~{Foy}, F.~{H{\'e}nault},
  D.~{Mattei}, G.~{Merlin}, A.~{Roussel}, M.~{Tallon}, E.~{Thiebaut},
  H.~{McAlister}, T.~{ten Brummelaar}, J.~{Sturmann}, L.~{Sturmann},
  N.~{Turner}, C.~{Farrington}, and P.~J. {Goldfinger}.
\newblock {VEGA: a new visible spectrograph and polarimeter on the CHARA
  Array}.
\newblock In \emph{Society of Photo-Optical Instrumentation Engineers (SPIE)
  Conference Series}, volume 7013 of \emph{Society of Photo-Optical
  Instrumentation Engineers (SPIE) Conference Series}, page~23, July 2008.
\newblock \doi{10.1117/12.787505}.

\bibitem[{Mourard} et~al.(2009){Mourard}, {Clausse}, {Marcotto}, {Perraut},
  {Tallon-Bosc}, {B{\'e}rio}, {Blazit}, {Bonneau}, {Bosio}, {Bresson},
  {Chesneau}, {Delaa}, {H{\'e}nault}, {Hughes}, {Lagarde}, {Merlin}, {Roussel},
  {Spang}, {Stee}, {Tallon}, {Antonelli}, {Foy}, {Kervella}, {Petrov},
  {Thiebaut}, {Vakili}, {McAlister}, {ten Brummelaar}, {Sturmann}, {Sturmann},
  {Turner}, {Farrington}, and {Goldfinger}]{Vega09}
D.~{Mourard}, J.~M. {Clausse}, A.~{Marcotto}, K.~{Perraut}, I.~{Tallon-Bosc},
  P.~{B{\'e}rio}, A.~{Blazit}, D.~{Bonneau}, S.~{Bosio}, Y.~{Bresson},
  O.~{Chesneau}, O.~{Delaa}, F.~{H{\'e}nault}, Y.~{Hughes}, S.~{Lagarde},
  G.~{Merlin}, A.~{Roussel}, A.~{Spang}, P.~{Stee}, M.~{Tallon},
  P.~{Antonelli}, R.~{Foy}, P.~{Kervella}, R.~{Petrov}, E.~{Thiebaut},
  F.~{Vakili}, H.~{McAlister}, T.~{ten Brummelaar}, J.~{Sturmann},
  L.~{Sturmann}, N.~{Turner}, C.~{Farrington}, and P.~J. {Goldfinger}.
\newblock {VEGA: Visible spEctroGraph and polArimeter for the CHARA array:
  principle and performance}.
\newblock \emph{\aap}, 508:\penalty0 1073--1083, Dec. 2009.
\newblock \doi{10.1051/0004-6361/200913016}.

\bibitem[{Mourard} et~al.(2011){Mourard}, {B{\'e}rio}, {Perraut}, {Ligi},
  {Blazit}, {Clausse}, {Nardetto}, {Spang}, {Tallon-Bosc}, {Bonneau},
  {Chesneau}, {Delaa}, {Millour}, {Stee}, {Le Bouquin}, {ten Brummelaar},
  {Farrington}, {Goldfinger}, and {Monnier}]{Vega11}
D.~{Mourard}, P.~{B{\'e}rio}, K.~{Perraut}, R.~{Ligi}, A.~{Blazit}, J.~M.
  {Clausse}, N.~{Nardetto}, A.~{Spang}, I.~{Tallon-Bosc}, D.~{Bonneau},
  O.~{Chesneau}, O.~{Delaa}, F.~{Millour}, P.~{Stee}, J.~B. {Le Bouquin},
  T.~{ten Brummelaar}, C.~{Farrington}, P.~J. {Goldfinger}, and J.~D.
  {Monnier}.
\newblock {Spatio-spectral encoding of fringes in optical long-baseline
  interferometry. Example of the 3T and 4T recombining mode of VEGA/CHARA}.
\newblock \emph{\aap}, 531:\penalty0 A110, July 2011.
\newblock \doi{10.1051/0004-6361/201116976}.

\bibitem[{Mourard} et~al.(2012){Mourard}, {Challouf}, {Ligi}, {B{\'e}rio},
  {Clausse}, {Gerakis}, {Bourges}, {Nardetto}, {Perraut}, {Tallon-Bosc},
  {McAlister}, {ten Brummelaar}, {Ridgway}, {Sturmann}, {Sturmann}, {Turner},
  {Farrington}, and {Goldfinger}]{Vega12}
D.~{Mourard}, M.~{Challouf}, R.~{Ligi}, P.~{B{\'e}rio}, J.-M. {Clausse},
  J.~{Gerakis}, L.~{Bourges}, N.~{Nardetto}, K.~{Perraut}, I.~{Tallon-Bosc},
  H.~{McAlister}, T.~{ten Brummelaar}, S.~{Ridgway}, J.~{Sturmann},
  L.~{Sturmann}, N.~{Turner}, C.~{Farrington}, and P.~J. {Goldfinger}.
\newblock {Performance, results, and prospects of the visible spectrograph VEGA
  on CHARA}.
\newblock In \emph{Society of Photo-Optical Instrumentation Engineers (SPIE)
  Conference Series}, volume 8445 of \emph{Society of Photo-Optical
  Instrumentation Engineers (SPIE) Conference Series}, page~0, July 2012.
\newblock \doi{10.1117/12.925223}.

\bibitem[{Mozurkewich}(1994)]{Mozurk94}
D.~{Mozurkewich}.
\newblock {Hybrid design for a six-way beam combiner}.
\newblock In J.~B. {Breckinridge}, editor, \emph{Amplitude and Intensity
  Spatial Interferometry II}, volume 2200 of \emph{Society of Photo-Optical
  Instrumentation Engineers (SPIE) Conference Series}, pages 76--80, June 1994.

\bibitem[{Muterspaugh} et~al.(2005){Muterspaugh}, {Lane}, {Konacki}, {Burke},
  {Colavita}, {Kulkarni}, and {Shao}]{Muterspaugh05}
M.~W. {Muterspaugh}, B.~F. {Lane}, M.~{Konacki}, B.~F. {Burke}, M.~M.
  {Colavita}, S.~R. {Kulkarni}, and M.~{Shao}.
\newblock {PHASES High-Precision Differential Astrometry of {$\delta$}
  Equulei}.
\newblock \emph{\aj}, 130:\penalty0 2866--2875, Dec. 2005.
\newblock \doi{10.1086/497035}.

\bibitem[{Muterspaugh} et~al.(2006a){Muterspaugh}, {Lane}, {Konacki}, {Burke},
  {Colavita}, {Kulkarni}, and {Shao}]{Muterspaugh06a}
M.~W. {Muterspaugh}, B.~F. {Lane}, M.~{Konacki}, B.~F. {Burke}, M.~M.
  {Colavita}, S.~R. {Kulkarni}, and M.~{Shao}.
\newblock {PHASES differential astrometry and the mutual inclination of the
  V819 Herculis triple star system}.
\newblock \emph{\aap}, 446:\penalty0 723--732, Feb. 2006a.
\newblock \doi{10.1051/0004-6361:20053749}.

\bibitem[{Muterspaugh} et~al.(2006b){Muterspaugh}, {Lane}, {Kulkarni}, {Burke},
  {Colavita}, and {Shao}]{Muterspaugh06b}
M.~W. {Muterspaugh}, B.~F. {Lane}, S.~R. {Kulkarni}, B.~F. {Burke}, M.~M.
  {Colavita}, and M.~{Shao}.
\newblock {Limits to Tertiary Astrometric Companions in Binary Systems}.
\newblock \emph{\apj}, 653:\penalty0 1469--1479, Dec. 2006b.
\newblock \doi{10.1086/508743}.

\bibitem[{Muterspaugh} et~al.(2006c){Muterspaugh}, {Lane}, {Konacki},
  {Wiktorowicz}, {Burke}, {Colavita}, {Kulkarni}, and {Shao}]{Muterspaugh06c}
M.~W. {Muterspaugh}, B.~F. {Lane}, M.~{Konacki}, S.~{Wiktorowicz}, B.~F.
  {Burke}, M.~M. {Colavita}, S.~R. {Kulkarni}, and M.~{Shao}.
\newblock {PHASES Differential Astrometry and Iodine Cell Radial Velocities of
  the {$\kappa$} Pegasi Triple Star System}.
\newblock \emph{\apj}, 636:\penalty0 1020--1032, Jan. 2006c.
\newblock \doi{10.1086/498209}.

\bibitem[{Muterspaugh} et~al.(2008){Muterspaugh}, {Lane}, {Fekel}, {Konacki},
  {Burke}, {Kulkarni}, {Colavita}, {Shao}, and {Wiktorowicz}]{Muterspaugh08}
M.~W. {Muterspaugh}, B.~F. {Lane}, F.~C. {Fekel}, M.~{Konacki}, B.~F. {Burke},
  S.~R. {Kulkarni}, M.~M. {Colavita}, M.~{Shao}, and S.~J. {Wiktorowicz}.
\newblock {Masses, Luminosities, and Orbital Coplanarities of the {$\mu$}
  Orionis Quadruple-Star System from Phases Differential Astrometry}.
\newblock \emph{\aj}, 135:\penalty0 766--776, Mar. 2008.
\newblock \doi{10.1088/0004-6256/135/3/766}.

\bibitem[{Muterspaugh} et~al.(2010){Muterspaugh}, {Fekel}, {Lane}, {Hartkopf},
  {Kulkarni}, {Konacki}, {Burke}, {Colavita}, {Shao}, and
  {Williamson}]{Muterspaugh10}
M.~W. {Muterspaugh}, F.~C. {Fekel}, B.~F. {Lane}, W.~I. {Hartkopf}, S.~R.
  {Kulkarni}, M.~{Konacki}, B.~F. {Burke}, M.~M. {Colavita}, M.~{Shao}, and
  M.~{Williamson}.
\newblock {The Phases Differential Astrometry Data Archive. IV. The Triple Star
  Systems 63 Gem A and HR 2896}.
\newblock \emph{\aj}, 140:\penalty0 1646--1656, Dec. 2010.
\newblock \doi{10.1088/0004-6256/140/6/1646}.

\bibitem[{North} et~al.(2007){North}, {Tuthill}, {Tango}, and {Davis}]{North07}
J.~R. {North}, P.~G. {Tuthill}, W.~J. {Tango}, and J.~{Davis}.
\newblock {{$\gamma$}$^{2}$ Velorum: orbital solution and fundamental parameter
  determination with SUSI}.
\newblock \emph{\mnras}, 377:\penalty0 415--424, May 2007.
\newblock \doi{10.1111/j.1365-2966.2007.11608.x}.

\bibitem[{North} et~al.(2009){North}, {Davis}, {Robertson}, {Bedding},
  {Bruntt}, {Ireland}, {Jacob}, {Lacour}, {O'Byrne}, {Owens}, {Stello},
  {Tango}, and {Tuthill}]{North09}
J.~R. {North}, J.~{Davis}, J.~G. {Robertson}, T.~R. {Bedding}, H.~{Bruntt},
  M.~J. {Ireland}, A.~P. {Jacob}, S.~{Lacour}, J.~W. {O'Byrne}, S.~M. {Owens},
  D.~{Stello}, W.~J. {Tango}, and P.~G. {Tuthill}.
\newblock {The radius and other fundamental parameters of the F9V star
  {$\beta$} Virginis}.
\newblock \emph{\mnras}, 393:\penalty0 245--252, Feb. 2009.
\newblock \doi{10.1111/j.1365-2966.2008.14216.x}.

\bibitem[{Ohishi} et~al.(2004){Ohishi}, {Nordgren}, and {Hutter}]{Ohishi04}
N.~{Ohishi}, T.~E. {Nordgren}, and D.~J. {Hutter}.
\newblock {Asymmetric Surface Brightness Distribution of Altair Observed with
  the Navy Prototype Optical Interferometer}.
\newblock \emph{\apj}, 612:\penalty0 463--471, Sept. 2004.
\newblock \doi{10.1086/422422}.

\bibitem[{Patience} et~al.(2008){Patience}, {Zavala}, {Prato}, {Franz},
  {Wasserman}, {Tycner}, {Hutter}, and {Hummel}]{Patience08}
J.~{Patience}, R.~T. {Zavala}, L.~{Prato}, O.~{Franz}, L.~{Wasserman},
  C.~{Tycner}, D.~J. {Hutter}, and C.~A. {Hummel}.
\newblock {Optical Interferometric Observations of ${\theta}^{1}$ Orionis C
  from NPOI and Implications for the System Orbit}.
\newblock \emph{\apjl}, 674:\penalty0 L97--L100, Feb. 2008.
\newblock \doi{10.1086/529041}.

\bibitem[{Peterson} et~al.(2006){Peterson}, {Hummel}, {Pauls}, {Armstrong},
  {Benson}, {Gilbreath}, {Hindsley}, {Hutter}, {Johnston}, {Mozurkewich}, and
  {Schmitt}]{Peterson06}
D.~M. {Peterson}, C.~A. {Hummel}, T.~A. {Pauls}, J.~T. {Armstrong}, J.~A.
  {Benson}, G.~C. {Gilbreath}, R.~B. {Hindsley}, D.~J. {Hutter}, K.~J.
  {Johnston}, D.~{Mozurkewich}, and H.~R. {Schmitt}.
\newblock {Vega is a rapidly rotating star}.
\newblock \emph{\nat}, 440:\penalty0 896--899, Apr. 2006.
\newblock \doi{10.1038/nature04661}.

\bibitem[{Petrov} et~al.(2007){Petrov}, {Malbet}, {Weigelt}, {Antonelli},
  {Beckmann}, {Bresson}, {Chelli}, {Dugu{\'e}}, {Duvert}, {Gennari},
  {Gl{\"u}ck}, {Kern}, {Lagarde}, {Le Coarer}, {Lisi}, {Millour}, {Perraut},
  {Puget}, {Rantakyr{\"o}}, {Robbe-Dubois}, {Roussel}, {Salinari}, {Tatulli},
  {Zins}, {Accardo}, {Acke}, {Agabi}, {Altariba}, {Arezki}, {Aristidi},
  {Baffa}, {Behrend}, {Bl{\"o}cker}, {Bonhomme}, {Busoni}, {Cassaing},
  {Clausse}, {Colin}, {Connot}, {Delboulb{\'e}}, {Domiciano de Souza},
  {Driebe}, {Feautrier}, {Ferruzzi}, {Forveille}, {Fossat}, {Foy},
  {Fraix-Burnet}, {Gallardo}, {Giani}, {Gil}, {Glentzlin}, {Heiden},
  {Heininger}, {Hernandez Utrera}, {Hofmann}, {Kamm}, {Kiekebusch}, {Kraus},
  {Le Contel}, {Le Contel}, {Lesourd}, {Lopez}, {Lopez}, {Magnard}, {Marconi},
  {Mars}, {Martinot-Lagarde}, {Mathias}, {M{\`e}ge}, {Monin}, {Mouillet},
  {Mourard}, {Nussbaum}, {Ohnaka}, {Pacheco}, {Perrier}, {Rabbia}, {Rebattu},
  {Reynaud}, {Richichi}, {Robini}, {Sacchettini}, {Schertl}, {Sch{\"o}ller},
  {Solscheid}, {Spang}, {Stee}, {Stefanini}, {Tallon}, {Tallon-Bosc}, {Tasso},
  {Testi}, {Vakili}, {von der L{\"u}he}, {Valtier}, {Vannier}, and
  {Ventura}]{Petrov07}
R.~G. {Petrov}, F.~{Malbet}, G.~{Weigelt}, P.~{Antonelli}, U.~{Beckmann},
  Y.~{Bresson}, A.~{Chelli}, M.~{Dugu{\'e}}, G.~{Duvert}, S.~{Gennari},
  L.~{Gl{\"u}ck}, P.~{Kern}, S.~{Lagarde}, E.~{Le Coarer}, F.~{Lisi},
  F.~{Millour}, K.~{Perraut}, P.~{Puget}, F.~{Rantakyr{\"o}},
  S.~{Robbe-Dubois}, A.~{Roussel}, P.~{Salinari}, E.~{Tatulli}, G.~{Zins},
  M.~{Accardo}, B.~{Acke}, K.~{Agabi}, E.~{Altariba}, B.~{Arezki},
  E.~{Aristidi}, C.~{Baffa}, J.~{Behrend}, T.~{Bl{\"o}cker}, S.~{Bonhomme},
  S.~{Busoni}, F.~{Cassaing}, J.-M. {Clausse}, J.~{Colin}, C.~{Connot},
  A.~{Delboulb{\'e}}, A.~{Domiciano de Souza}, T.~{Driebe}, P.~{Feautrier},
  D.~{Ferruzzi}, T.~{Forveille}, E.~{Fossat}, R.~{Foy}, D.~{Fraix-Burnet},
  A.~{Gallardo}, E.~{Giani}, C.~{Gil}, A.~{Glentzlin}, M.~{Heiden},
  M.~{Heininger}, O.~{Hernandez Utrera}, K.-H. {Hofmann}, D.~{Kamm},
  M.~{Kiekebusch}, S.~{Kraus}, D.~{Le Contel}, J.-M. {Le Contel}, T.~{Lesourd},
  B.~{Lopez}, M.~{Lopez}, Y.~{Magnard}, A.~{Marconi}, G.~{Mars},
  G.~{Martinot-Lagarde}, P.~{Mathias}, P.~{M{\`e}ge}, J.-L. {Monin},
  D.~{Mouillet}, D.~{Mourard}, E.~{Nussbaum}, K.~{Ohnaka}, J.~{Pacheco},
  C.~{Perrier}, Y.~{Rabbia}, S.~{Rebattu}, F.~{Reynaud}, A.~{Richichi},
  A.~{Robini}, M.~{Sacchettini}, D.~{Schertl}, M.~{Sch{\"o}ller},
  W.~{Solscheid}, A.~{Spang}, P.~{Stee}, P.~{Stefanini}, M.~{Tallon},
  I.~{Tallon-Bosc}, D.~{Tasso}, L.~{Testi}, F.~{Vakili}, O.~{von der L{\"u}he},
  J.-C. {Valtier}, M.~{Vannier}, and N.~{Ventura}.
\newblock {AMBER, the near-infrared spectro-interferometric three-telescope
  VLTI instrument}.
\newblock \emph{\aap}, 464:\penalty0 1--12, Mar. 2007.
\newblock \doi{10.1051/0004-6361:20066496}.

\bibitem[{Press} et~al.(2002){Press}, {Teukolsky}, {Vetterling}, and
  {Flannery}]{NR}
W.~H. {Press}, S.~A. {Teukolsky}, W.~T. {Vetterling}, and B.~P. {Flannery}.
\newblock \emph{{Numerical recipes in C++ : the art of scientific computing}}.
\newblock 2002.

\bibitem[{Ragland} et~al.(2004){Ragland}, {Traub}, {Berger}, {Millan-Gabet},
  {Monnier}, {Pedretti}, {Schloerb}, {Carleton}, {Haguenauer}, {Kern},
  {Labeye}, {Lacasse}, {Malbet}, and {Rousselet-Perraut}]{Ragland04}
S.~{Ragland}, W.~A. {Traub}, J.-P. {Berger}, R.~{Millan-Gabet}, J.~D.
  {Monnier}, E.~{Pedretti}, F.~P. {Schloerb}, N.~P. {Carleton},
  P.~{Haguenauer}, P.~Y. {Kern}, P.~R. {Labeye}, M.~G. {Lacasse}, F.~{Malbet},
  and K.~{Rousselet-Perraut}.
\newblock {Characterizing closure-phase measurements at IOTA}.
\newblock In W.~A. {Traub}, editor, \emph{New Frontiers in Stellar
  Interferometry}, volume 5491 of \emph{Society of Photo-Optical
  Instrumentation Engineers (SPIE) Conference Series}, page 1390, Oct. 2004.

\bibitem[{Roettenbacher} et~al.(2015a){Roettenbacher}, {Monnier}, {Henry},
  {Fekel}, {Williamson}, {Pourbaix}, {Latham}, {Latham}, {Torres}, {Baron},
  {Che}, {Kraus}, {Schaefer}, {Aarnio}, {Korhonen}, {Harmon}, {ten Brummelaar},
  {Sturmann}, {Sturmann}, and {Turner}]{Roet15a}
R.~M. {Roettenbacher}, J.~D. {Monnier}, G.~W. {Henry}, F.~C. {Fekel}, M.~H.
  {Williamson}, D.~{Pourbaix}, D.~W. {Latham}, C.~A. {Latham}, G.~{Torres},
  F.~{Baron}, X.~{Che}, S.~{Kraus}, G.~H. {Schaefer}, A.~N. {Aarnio},
  H.~{Korhonen}, R.~O. {Harmon}, T.~A. {ten Brummelaar}, J.~{Sturmann},
  L.~{Sturmann}, and N.~H. {Turner}.
\newblock {Detecting the Companions and Ellipsoidal Variations of RS CVn
  Primaries. I. {$\sigma$}Geminorum}.
\newblock \emph{\apj}, 807:\penalty0 23, July 2015a.
\newblock \doi{10.1088/0004-637X/807/1/23}.

\bibitem[{Roettenbacher} et~al.(2015b){Roettenbacher}, {Monnier}, {Fekel},
  {Henry}, {Korhonen}, {Latham}, {Muterspaugh}, {Williamson}, {Baron}, {ten
  Brummelaar}, {Che}, {Harmon}, {Schaefer}, {Scott}, {Sturmann}, {Sturmann},
  and {Turner}]{Roet15b}
R.~M. {Roettenbacher}, J.~D. {Monnier}, F.~C. {Fekel}, G.~W. {Henry},
  H.~{Korhonen}, D.~W. {Latham}, M.~W. {Muterspaugh}, M.~H. {Williamson},
  F.~{Baron}, T.~A. {ten Brummelaar}, X.~{Che}, R.~O. {Harmon}, G.~H.
  {Schaefer}, N.~J. {Scott}, J.~{Sturmann}, L.~{Sturmann}, and N.~H. {Turner}.
\newblock {Detecting the Companions and Ellipsoidal Variations of RS CVn
  Primaries. II. o Draconis, a Candidate for Recent Low-mass Companion
  Ingestion}.
\newblock \emph{\apj}, 809:\penalty0 159, Aug. 2015b.
\newblock \doi{10.1088/0004-637X/809/2/159}.

\bibitem[{Rousselet-Perraut} et~al.(1999){Rousselet-Perraut}, {Stadler},
  {Feautrier}, {Le Coarer}, {Petmezakis}, {Haguenauer}, {Kern}, {Malbet},
  {Berger}, {Schanen-Duport}, {Benech}, and {Delage}]{Rouss99}
K.~{Rousselet-Perraut}, E.~{Stadler}, P.~{Feautrier}, E.~{Le Coarer},
  P.~{Petmezakis}, P.~{Haguenauer}, P.~{Kern}, F.~{Malbet}, J.-P. {Berger},
  I.~{Schanen-Duport}, P.~{Benech}, and L.~{Delage}.
\newblock {The Integrated Optics Near-infrared Interferometric Camera (IONIC)}.
\newblock In S.~{Unwin} and R.~{Stachnik}, editors, \emph{Working on the
  Fringe: Optical and IR Interferometry from Ground and Space}, volume 194 of
  \emph{Astronomical Society of the Pacific Conference Series}, page 344, 1999.

\bibitem[{Rousselet-Perraut} et~al.(2000){Rousselet-Perraut}, {Haguenauer},
  {Petmezakis}, {Berger}, {Mourard}, {Ragland}, {Huss}, {Reynaud}, {Le Coarer},
  {Kern}, and {Malbet}]{Rouss00}
K.~{Rousselet-Perraut}, P.~{Haguenauer}, P.~{Petmezakis}, J.-P. {Berger},
  D.~{Mourard}, S.~D. {Ragland}, G.~{Huss}, F.~{Reynaud}, E.~{Le Coarer}, P.~Y.
  {Kern}, and F.~{Malbet}.
\newblock {Qualification of IONIC (integrated optics near-infrared
  interferometric camera)}.
\newblock In P.~{L{\'e}na} and A.~{Quirrenbach}, editors, \emph{Interferometry
  in Optical Astronomy}, volume 4006 of \emph{Society of Photo-Optical
  Instrumentation Engineers (SPIE) Conference Series}, pages 1042--1051, July
  2000.

\bibitem[{Schmitt} et~al.(2009){Schmitt}, {Pauls}, {Tycner}, {Armstrong},
  {Zavala}, {Benson}, {Gilbreath}, {Hindsley}, {Hutter}, {Johnston},
  {Jorgensen}, and {Mozurkewich}]{Schmitt09}
H.~R. {Schmitt}, T.~A. {Pauls}, C.~{Tycner}, J.~T. {Armstrong}, R.~T. {Zavala},
  J.~A. {Benson}, G.~C. {Gilbreath}, R.~B. {Hindsley}, D.~J. {Hutter}, K.~J.
  {Johnston}, A.~M. {Jorgensen}, and D.~{Mozurkewich}.
\newblock {Navy Prototype Optical Interferometer Imaging of Line Emission
  Regions of {$\beta$} Lyrae Using Differential Phase Referencing}.
\newblock \emph{\apj}, 691:\penalty0 984--996, Feb. 2009.
\newblock \doi{10.1088/0004-637X/691/2/984}.

\bibitem[{Shaklan} and {Roddier}(1988)]{Shaklan88}
S.~{Shaklan} and F.~{Roddier}.
\newblock {Coupling starlight into single-mode fiber optics}.
\newblock \emph{\ao}, 27:\penalty0 2334--2338, June 1988.
\newblock \doi{10.1364/AO.27.002334}.

\bibitem[{Shaklan} et~al.(1992){Shaklan}, {Colavita}, and {Shao}]{Shaklan92}
S.~B. {Shaklan}, M.~M. {Colavita}, and M.~{Shao}.
\newblock {Visibility calibration using single mode fibers in a long-baseline
  interferometer.}
\newblock In J.~M. {Beckers} and F.~{Merkle}, editors, \emph{European Southern
  Observatory Conference and Workshop Proceedings}, volume~39 of \emph{European
  Southern Observatory Conference and Workshop Proceedings}, pages 1271--1283,
  Mar. 1992.

\bibitem[{Sun} et~al.(2014){Sun}, {Jorgensen}, {Landavazo}, {Hutter}, {van
  Belle}, {Mozurkewich}, {Armstrong}, {Schmitt}, {Baines}, and
  {Restaino}]{Sun14}
B.~{Sun}, A.~M. {Jorgensen}, M.~{Landavazo}, D.~J. {Hutter}, G.~T. {van Belle},
  D.~{Mozurkewich}, J.~T. {Armstrong}, H.~R. {Schmitt}, E.~K. {Baines}, and
  S.~R. {Restaino}.
\newblock {The new classic data acquisition system for NPOI}.
\newblock In \emph{Society of Photo-Optical Instrumentation Engineers (SPIE)
  Conference Series}, volume 9146 of \emph{Society of Photo-Optical
  Instrumentation Engineers (SPIE) Conference Series}, page~20, July 2014.
\newblock \doi{10.1117/12.2057013}.

\bibitem[{Tango} et~al.(2009){Tango}, {Davis}, {Jacob}, {Mendez}, {North},
  {O'Byrne}, {Seneta}, and {Tuthill}]{Tango09}
W.~J. {Tango}, J.~{Davis}, A.~P. {Jacob}, A.~{Mendez}, J.~R. {North}, J.~W.
  {O'Byrne}, E.~B. {Seneta}, and P.~G. {Tuthill}.
\newblock {A new determination of the orbit and masses of the Be binary system
  {$\delta$} Scorpii}.
\newblock \emph{\mnras}, 396:\penalty0 842--848, June 2009.
\newblock \doi{10.1111/j.1365-2966.2009.14272.x}.

\bibitem[{ten Brummelaar} et~al.(2005){ten Brummelaar}, {McAlister}, {Ridgway},
  {Bagnuolo}, {Turner}, {Sturmann}, {Sturmann}, {Berger}, {Ogden}, {Cadman},
  {Hartkopf}, {Hopper}, and {Shure}]{tenBrummelaar05}
T.~A. {ten Brummelaar}, H.~A. {McAlister}, S.~T. {Ridgway}, W.~G. {Bagnuolo},
  Jr., N.~H. {Turner}, L.~{Sturmann}, J.~{Sturmann}, D.~H. {Berger}, C.~E.
  {Ogden}, R.~{Cadman}, W.~I. {Hartkopf}, C.~H. {Hopper}, and M.~A. {Shure}.
\newblock {First Results from the CHARA Array. II. A Description of the
  Instrument}.
\newblock \emph{\apj}, 628:\penalty0 453--465, July 2005.
\newblock \doi{10.1086/430729}.

\bibitem[{Traub} et~al.(2004){Traub}, {Berger}, {Brewer}, {Carleton}, {Kern},
  {Kraus}, {Lacasse}, {McGonagle}, {Millan-Gabet}, {Monnier}, {Pedretti},
  {Ragland}, {Reich}, {Schloerb}, {Schuller}, {Souccar}, and
  {Wallace}]{Traub04}
W.~A. {Traub}, J.-P. {Berger}, M.~K. {Brewer}, N.~P. {Carleton}, P.~Y. {Kern},
  S.~{Kraus}, M.~G. {Lacasse}, W.~H. {McGonagle}, R.~{Millan-Gabet}, J.~D.
  {Monnier}, E.~{Pedretti}, S.~{Ragland}, R.~K. {Reich}, F.~P. {Schloerb},
  P.~A. {Schuller}, K.~{Souccar}, and G.~{Wallace}.
\newblock {IOTA: recent technology and science}.
\newblock In W.~A. {Traub}, editor, \emph{New Frontiers in Stellar
  Interferometry}, volume 5491 of \emph{Society of Photo-Optical
  Instrumentation Engineers (SPIE) Conference Series}, page 482, Oct. 2004.

\bibitem[{van Belle}(2012)]{vanBelle12}
G.~T. {van Belle}.
\newblock {Interferometric observations of rapidly rotating stars}.
\newblock \emph{\aapr}, 20:\penalty0 51, Mar. 2012.
\newblock \doi{10.1007/s00159-012-0051-2}.

\bibitem[{Wang} et~al.(2015){Wang}, {Hummel}, {Ren}, and {Fu}]{Wang15}
X.~{Wang}, C.~A. {Hummel}, S.~{Ren}, and Y.~{Fu}.
\newblock {The Three-dimensional Orbit and Physical Parameters of 47 Oph}.
\newblock \emph{\aj}, 149:\penalty0 110, Mar. 2015.
\newblock \doi{10.1088/0004-6256/149/3/110}.

\bibitem[{Wenger} et~al.(2000){Wenger}, {Ochsenbein}, {Egret}, {Dubois},
  {Bonnarel}, {Borde}, {Genova}, {Jasniewicz}, {Lalo{\"e}}, {Lesteven}, and
  {Monier}]{Wenger00}
M.~{Wenger}, F.~{Ochsenbein}, D.~{Egret}, P.~{Dubois}, F.~{Bonnarel},
  S.~{Borde}, F.~{Genova}, G.~{Jasniewicz}, S.~{Lalo{\"e}}, S.~{Lesteven}, and
  R.~{Monier}.
\newblock {The SIMBAD astronomical database. The CDS reference database for
  astronomical objects}.
\newblock \emph{\aaps}, 143:\penalty0 9--22, Apr. 2000.
\newblock \doi{10.1051/aas:2000332}.

\bibitem[{White} et~al.(2013){White}, {Huber}, {Maestro}, {Bedding}, {Ireland},
  {Baron}, {Boyajian}, {Che}, {Monnier}, {Pope}, {Roettenbacher}, {Stello},
  {Tuthill}, {Farrington}, {Goldfinger}, {McAlister}, {Schaefer}, {Sturmann},
  {Sturmann}, {ten Brummelaar}, and {Turner}]{White13}
T.~R. {White}, D.~{Huber}, V.~{Maestro}, T.~R. {Bedding}, M.~J. {Ireland},
  F.~{Baron}, T.~S. {Boyajian}, X.~{Che}, J.~D. {Monnier}, B.~J.~S. {Pope},
  R.~M. {Roettenbacher}, D.~{Stello}, P.~G. {Tuthill}, C.~D. {Farrington},
  P.~J. {Goldfinger}, H.~A. {McAlister}, G.~H. {Schaefer}, J.~{Sturmann},
  L.~{Sturmann}, T.~A. {ten Brummelaar}, and N.~H. {Turner}.
\newblock {Interferometric radii of bright Kepler stars with the CHARA Array:
  $\theta$ Cygni and 16 Cygni A and B}.
\newblock \emph{\mnras}, 433:\penalty0 1262--1270, Aug. 2013.
\newblock \doi{10.1093/mnras/stt802}.

\bibitem[{Wirnitzer}(1985)]{wirnitzer85}
B.~{Wirnitzer}.
\newblock {Bispectral analysis at low light levels and astronomical speckle
  masking}.
\newblock \emph{Journal of the Optical Society of America A}, 2:\penalty0
  14--21, Jan. 1985.
\newblock \doi{10.1364/JOSAA.2.000014}.

\bibitem[{Wittkowski} et~al.(2006){Wittkowski}, {Hummel}, {Aufdenberg}, and
  {Roccatagliata}]{Witt06}
M.~{Wittkowski}, C.~A. {Hummel}, J.~P. {Aufdenberg}, and V.~{Roccatagliata}.
\newblock {Tests of stellar model atmospheres by optical interferometry. III.
  NPOI and VINCI interferometry of the M0 giant {$\gamma$} Sagittae covering
  0.5-2.2 {$\mu$}m}.
\newblock \emph{\aap}, 460:\penalty0 843--853, Dec. 2006.
\newblock \doi{10.1051/0004-6361:20065853}.

\bibitem[{Zavala} et~al.(2010){Zavala}, {Hummel}, {Boboltz}, {Ojha}, {Shaffer},
  {Tycner}, {Richards}, and {Hutter}]{Zavala10}
R.~T. {Zavala}, C.~A. {Hummel}, D.~A. {Boboltz}, R.~{Ojha}, D.~B. {Shaffer},
  C.~{Tycner}, M.~T. {Richards}, and D.~J. {Hutter}.
\newblock {The Algol Triple System Spatially Resolved at Optical Wavelengths}.
\newblock \emph{\apjl}, 715:\penalty0 L44--L48, May 2010.
\newblock \doi{10.1088/2041-8205/715/1/L44}.

\bibitem[{Zhang} et~al.(2006){Zhang}, {Armstrong}, {Clark}, {Gilbreath},
  {Lucke}, {Restaino}, {Mozurkewich}, {Benson}, {Hutter}, {White}, {Schmitt},
  and {Walton}]{Zhang06}
X.~{Zhang}, J.~T. {Armstrong}, J.~A. {Clark}, III, G.~C. {Gilbreath},
  R.~{Lucke}, S.~{Restaino}, D.~{Mozurkewich}, J.~A. {Benson}, D.~J. {Hutter},
  N.~{White}, H.~{Schmitt}, and J.~P. {Walton}.
\newblock {Characterization of the optical throughput performance of the Navy
  Prototype Optical Interferometer (NPOI)}.
\newblock In \emph{Society of Photo-Optical Instrumentation Engineers (SPIE)
  Conference Series}, volume 6268 of \emph{Society of Photo-Optical
  Instrumentation Engineers (SPIE) Conference Series}, page~3, June 2006.
\newblock \doi{10.1117/12.670001}.

\bibitem[{Zhao} et~al.(2008){Zhao}, {Gies}, {Monnier}, {Thureau}, {Pedretti},
  {Baron}, {Merand}, {ten Brummelaar}, {McAlister}, {Ridgway}, {Turner},
  {Sturmann}, {Sturmann}, {Farrington}, and {Goldfinger}]{Zhao08}
M.~{Zhao}, D.~{Gies}, J.~D. {Monnier}, N.~{Thureau}, E.~{Pedretti}, F.~{Baron},
  A.~{Merand}, T.~{ten Brummelaar}, H.~{McAlister}, S.~T. {Ridgway},
  N.~{Turner}, J.~{Sturmann}, L.~{Sturmann}, C.~{Farrington}, and P.~J.
  {Goldfinger}.
\newblock {First Resolved Images of the Eclipsing and Interacting Binary
  {$\beta$} Lyrae}.
\newblock \emph{\apjl}, 684:\penalty0 L95--L98, Sept. 2008.
\newblock \doi{10.1086/592146}.

\end{thebibliography}
\clearpage


\begin{table}[htbp]
 \begin{center}
Table \ref{table:frngfreq}\\
Laser Fringe Results\\
 \begin{tabular}{lll}
\hline
\hline
Beam & Pixels Per & Fitted \\
Pair & Fringe & Visibility \\
\hline 
1-3 & $39.38$ & $0.968\pm0.006$ \\
2-4 & $16.52$ & $0.969\pm0.010$ \\
1-4 & $13.89$ & $0.939\pm0.009$ \\
2-5 & $11.91$ & $0.949\pm0.019$ \\
3-4 & $10.45$ & $0.954\pm0.010$ \\
1-2 & $7.64$ & $0.922\pm0.005$ \\
4-5 & $7.01$ & $0.884\pm0.018$ \\
2-3 & $6.48$ & $0.928\pm0.008$ \\
1-5 & $4.65$ & $0.858\pm0.015$ \\
3-5 & $4.20$ & $0.867\pm0.014$ \\
 \end{tabular}
 \end{center}
 \caption{Pixels per fringe and raw visibility derived from the model fits to the 632.8 nm HeNe laser fringes in 
 Figure~\ref{fig:labvis}.}
 \label{table:frngfreq}
 \end{table}

\begin{table}[htbp]
 \begin{center}
Table \ref{table:obsseq}\\
Observing Sequence\\
 \begin{tabular}{ll}
\hline
Align Fibers & $120$ s \\
Take Darks & $30$ s \\
Fringe Search & $30$ s \\
Record Fringes & $720$ s \\
Record Foreground & $60$ s \\
Record Single Beams 1$-$6 & $45$ s per beam \\
Total & 1230 s \\ 
\hline
 \end{tabular}
 \end{center}
 \caption{The observing sequence for each target or calibrator star. 
Raw interferograms are recorded while fringe tracking. The fibers only 
need to be aligned a couple of times per night. The 
rest of the data obtained are used to calibrate these raw interferograms: 
``Dark" frames are used to subtract off the bias counts of the EMCCDs. 
``Foreground'' frames  are observations of the star with incoherent flux (no fringes). 
These data are used to characterize the bias 
in power spectrum and triple amplitudes. ``Single Beam" frames are the flux measurements of each beam individually, which 
is used to measure how the polarizing beam splitter 
splits light between the photometric and fringing cameras at each wavelength. 
 }
 \label{table:obsseq}
 \end{table}

\begin{table}[htbp]
 \begin{center}
Table \ref{table:throughput}\\
Throughput\\
 \begin{tabular}{ll}
 \hline
 \hline
 Reflection      &  \\
 \hline
Feed System & \tpfeedsystem \\
Delay Line Carts & \tpdelayline \\
$\sfrac{70}{20}$ Beam Splitter, ($20$\% to Tip/Tilt)  & $70$\%\\
3 Inch Flat Mirror (Protected silver coating) & \tpthreeinchflatmirror \\
2 Inch Flat Mirror (Silver coating) & \tptwoinchflatmirror \\
Off Axis Parabola (Silver coating) & \tpoap \\
Fiber Coupling Efficiency ($r_{0}=9$ cm) & \tpfibercoupling \\
Fiber Fresnel Air-to-Glass & \tpfiberfresnel \\
Lenslet Array at output of Fibers & \tplenslet \\
\\
\multicolumn{1}{c}{Light to Fringing Camera} \\ \hline
Polarizing Beam Splitter $S$ mode & 50$\%$\\
Fringe Focusing Lens (AR coating) & \tpfrngfocuslens \\
Cylindrical Lens (AR coating) & \tpcylindricallens \\
$R=200$ Fringing Spectrograph Filter & \tpspecfrng \\
Quantum Efficiency of Andor Ixon DU 860 (8\% loss from air to chip) & \tpqefrng \\
{\bf Throughput Fringing Camera (theoretical)} & $\approx${\bf \tpfrngtheo} \\
{\bf Throughput Fringing Camera (observed)} & $\approx${\bf \tpfrngobs} \\
\\
\multicolumn{1}{c}{Light to Photometric Camera} \\ \hline
Lenslet Array at input of Multi-mode Fibers & \tplensletphot \\
Polarizing Beam Splitter $P$ mode & 50$\%$\\
$R=200$ Photometric Spectrograph Filter & \tpspecphot \\
Quantum Efficiency of Andor Ixon DU 860 (8\% loss from air to chip)  & \tpqephot \\
{\bf Throughput Photometric Camera (theoretical)}  & $\approx${\bf \tpphottheo} \\
{\bf Throughput Photometric Camera (observed)}  & $\approx${\bf \tpphotobs} \\
 \end{tabular}
 \end{center}
 \caption[Throughput]
 { \label{table:throughput}
List of throughput for the VISION fringing and photometric cameras. Most VISION mirrors are 
 silver-coated. The biggest single contributor to the loss of light is coupling to 
 the single-model fibers. The theoretical 
 fiber coupling efficiency of \tpfibercoupling~is likely greatly over estimated 
 due to feed alignment and delay line cart optics misalignments detailed in \S\ref{sec:throughput}. 
 }
 \end{table}

\begin{table}[htbp]
 \begin{center}
Table \ref{table:fringesearchtrack}\\
Fringe Searching and Tracking Parameters\\
 \begin{tabular}{ll}
\hline
\hline
EMCCD Exposure Time & 6 ms \\
Effective Coherent Exposure Time & 12 ms \\
Number of Coherent Co-adds & 2 \\
Effective In-coherent Exposure Time & 360 ms \\
Number of Incoherent Co-adds & 30 \\
Search Step Size & 12.5~$\mu$m \\ 
 \hline
 \hline
 \end{tabular}
 \end{center}
 \caption{Fringe searching and tracking parameters detailed in \S\ref{sec:frngsearch} and \S\ref{sec:frngtrack}. These optimal parameters were determined using on-sky observations in median-seeing conditions.  }
 \label{table:fringesearchtrack}
 \end{table}

\begin{table}[htbp]
 \begin{center}
Table \ref{table:orbit}\\
Observations of $\zeta$ Orionis\\
 \begin{tabular}{lll}
\hline
\hline
HJD & \multicolumn{2}{c}{$24557098.122$}\\
UT Date & \multicolumn{2}{c}{2015$-$03$-$17} \\
Telescopes Used & \multicolumn{2}{c}{AC$-$AE$-$N3} \\
Baseline Length Range & \multicolumn{2}{c}{$\baserange$~m} \\
Wavelength Range & \multicolumn{2}{c}{570-750 nm}\\
$\#$ Closure Phases & \multicolumn{2}{c}{$35$}\\
$\#$ $V^{2}$ & \multicolumn{2}{c}{$90$} \\
Median Closure Phase Error & \multicolumn{2}{c}{$\cphaseerr^{\circ}$}  \\
Median $V^{2}$ Error & \multicolumn{2}{c}{$\vsqrerr$\%} \\
& This Work & \cite{Hummel13} \\
Separation (mas) & $\zetasep$ & $\hummelsep^1$ \\
Position Angle (deg) & $\zetapa^{\circ}$ & $\hummelpa^{\circ~1}$ \\
$\Delta {\rm Mag}$(570-750 nm) & $\zetafratio$ & $\hummelfratio^1$ \\
 \hline
 \hline
 \end{tabular}
 \end{center}
 \label{table:orbit}
 \end{table}

\clearpage

\begin{figure}[ht]
\begin{center}
\includegraphics[width=\textwidth]{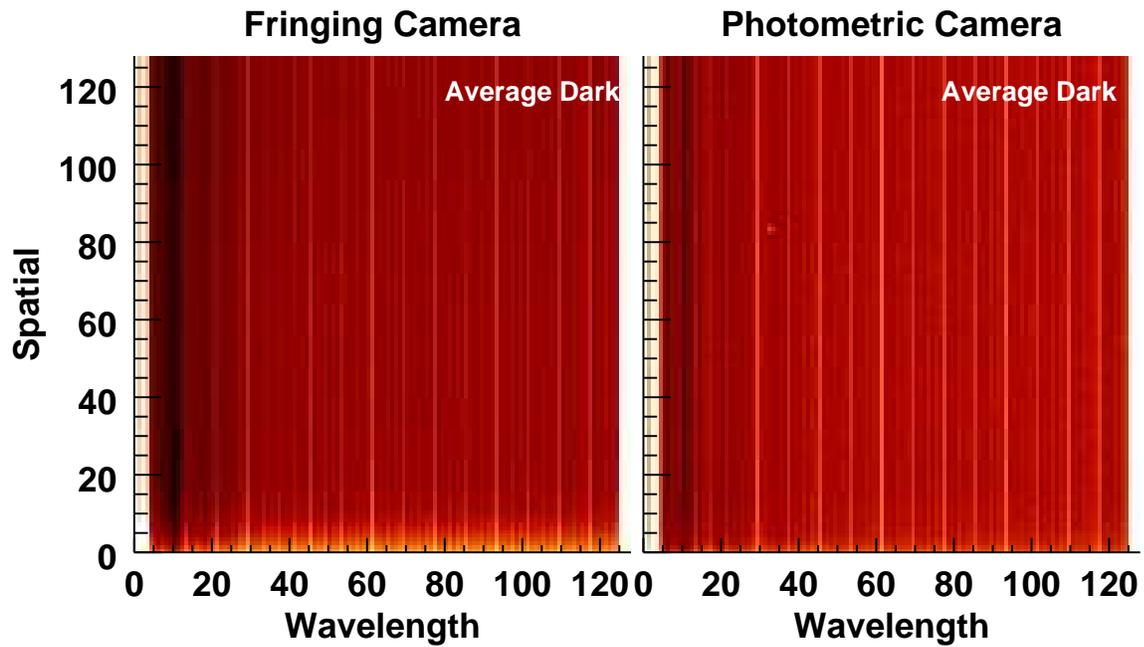}  
\end{center}
\caption{\label{fig:masterdark}
Thirty second time-averaged darks on the fringing and photometric cameras. There are variable bias counts across each EMCCD that must be subtracted off for each VISION data set. The maximum, median, and minimum for the average fringing camera dark are $\approx87$, $\approx91$ and $\approx107$ counts respectively.  The maximum, median, and minimum for the average photometric camera dark are $\approx97$, $\approx99$ and $\approx109$ counts respectively.
}
\end{figure}

\begin{figure}[ht]
\begin{center}
\includegraphics[angle=90.0,width=\textwidth]{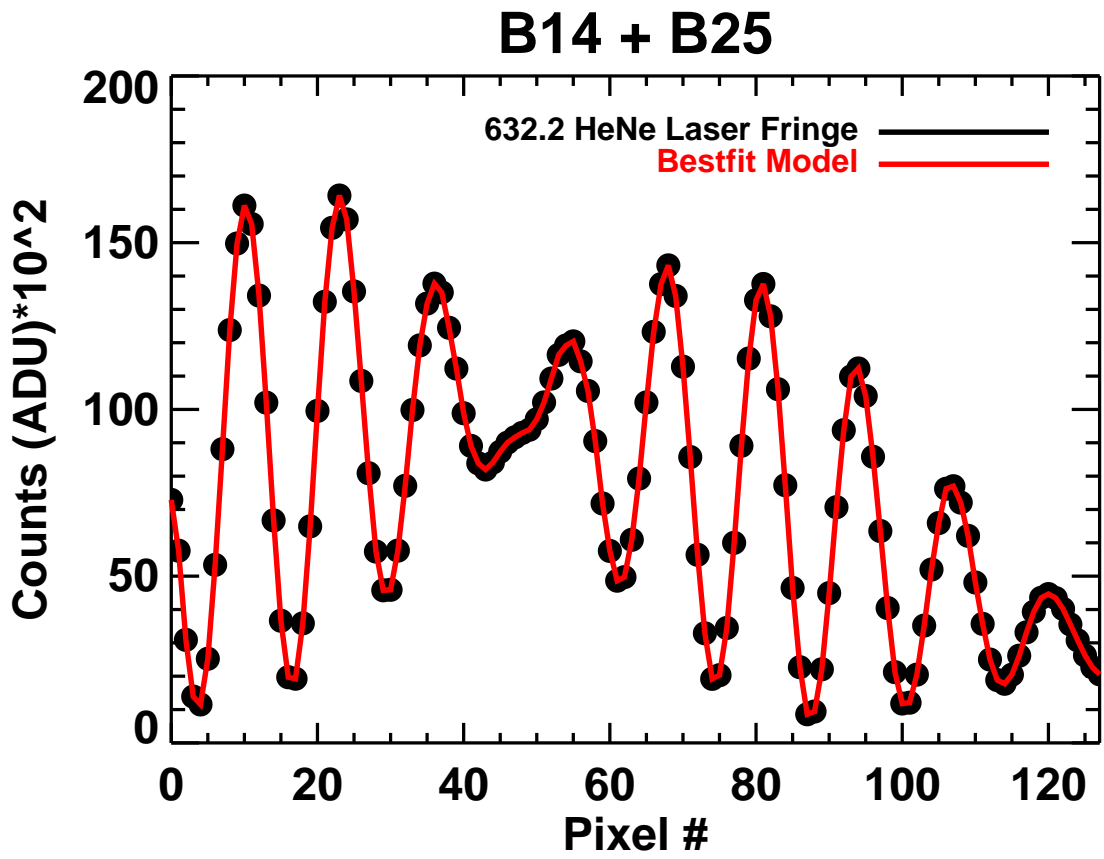}
\end{center}
\caption{\label{fig:cross} 
Laser fringe data from beam pair 1-4 and 2-5 added together on the detector, with the corresponding 
best fit model using Equation~\ref{eqn:frngmulti} in \S\ref{sec:crosstalk}.  
}
\end{figure}

\end{document}